\documentclass[showpacs,preprintnumbers,amsmath,amssymb,floatfix]{revtex4}

\usepackage{graphicx}
\usepackage{epsfig}
\usepackage{amsfonts}
\usepackage{amssymb}
\usepackage{epsf}
\newcommand{\insertplot}[5]{\begin{figure}
 \hfill\hbox to 0.05in{\vbox to #5in{\vfill
 \inputplot{#1}{#4}{#5}}\hfill}
 \hfill\vspace{-.1in}
 \caption{#2}\label{#3}
 \end{figure}}
\newcommand{\inputplot}[3]{
 \special{ps: plotfile #1}
}

\newcounter{fig}   \newcommand{\lbfig}[1]{\refstepcounter{fig}
\label{#1} }

\newcommand{\vphi}{\varphi}

\begin{document}

\title{Stable Lorentzian
Wormholes in Dilatonic Einstein-Gauss-Bonnet Theory}

\author{ 
{\bf Panagiota Kanti}
}
\affiliation{ Division of Theoretical Physics, Department of Physics,\\
University of Ioannina, Ioannina GR-45110, Greece}

\author{
{\bf Burkhard Kleihaus, Jutta Kunz}
}
\affiliation{
{Institut f\"ur Physik, Universit\"at Oldenburg,
D-26111 Oldenburg, Germany}
}
\date{\today}
\begin{abstract}
We discuss the properties
of Lorentzian wormholes in dilatonic Einstein-Gauss-Bonnet
theory in four spacetime dimensions.
These wormholes do not need any form of exotic matter for their existence.
A subset of these wormholes is shown to be linearly stable 
with respect to radial perturbations.
We perform a comprehensive study of their domain of existence,
and derive a generalised Smarr relation for these wormholes.
We also investigate their geodesics determining all possible particle
trajectories, and perform a study of the acceleration and tidal forces
that a traveler crossing the wormhole would feel.
\end{abstract}
\maketitle

\section{Introduction}

Black holes are by far the most celebrated class of solutions derived
from Einstein's field equations. Being among the first types of solutions
to be found almost a century ago, they have undergone an extensive investigation
over the years. Their existence conditions, forms of solutions and set of
properties have been studied in the context of both traditional General Relativity
and generalized gravitational theories admitting either four or more dimensions.

The most well-known example of such a generalized gravitational theory in four
dimensions is provided by the low-energy heterotic string effective theory 
\cite{Gross:1986mw,Metsaev:1987zx}. In this theory, the scalar curvature term
of Einstein's theory is only one part of a more complex action functional where 
higher-curvature gravitational terms as well as kinetic and interaction terms
of a variety of additional fields (axions, fermions and gauge fields) make their
appearance. The dilatonic Einstein-Gauss-Bonnet theory is a minimal version
of the aforementioned theory and contains the scalar curvature term $R$, a
scalar field called the dilaton, and a quadratic curvature term, the Gauss-Bonnet
(GB) term, given by $R^2_{GB}=R_{\mu\nu\rho\sigma} R^{\mu\nu\rho\sigma}
- 4 R_{\mu\nu} R^{\mu\nu} + R^2$.  The GB term in four dimensions can be
expressed as a total derivative term and, normally, makes no contribution to
the field equations; however, the exponential coupling to the dilaton field, that
emerges in the context of the dilatonic Einstein-Gauss-Bonnet theory, ensures
that the GB term is kept in the theory. 

In the framework of the dilatonic Einstein-Gauss-Bonnet theory, new types of
black hole solutions emerged that are endowed with non-trivial dilaton hair
(for an indicative list of works on this topic, see \cite{Boulware, Kanti:1995vq,
Guo, Ohta, Maeda,Kleihaus:2011tg}). 
The presence of the GB term in the theory
caused the circumvention of the traditional no-hair theorems as it bypassed 
the conditions for their validity. In reality, the
existence of the dilatonic black holes violated only the `letter' of the
no-hair theorems, and not the `essence' of them since the dilaton charge
was of a `secondary' type.
However, it became
evident that in the context of this type of generalized gravitational theories
solutions with a much richer structure and a modified set of properties, 
compared to the ones in General Relativity, can emerge.

Another class of gravitational solutions whose properties are strictly set by
the General Theory of Relativity are wormholes. They were first discovered
in 1935 as a feature of Schwarzschild geometry \cite{Einstein:1935tc} and
named the `Einstein-Rosen bridge' as they connect two different universes. 
Their importance was realised in full when Wheeler \cite{Wheeler:1957mu,Wheeler:1962} 
showed that such a bridge, or wormhole, can connect also two distant regions of
our own universe thus opening the way for fast interstellar travel. However,
it was soon demonstrated that this is not possible for the following reasons:
(i) the Schwarzschild wormhole is hidden inside the event horizon of the
corresponding black hole, therefore, is not static but evolves with time;
as a result, the circumference of its `throat' is not constant but opens and
closes so quickly that not even a light signal can pass through
\cite{Kruskal:1959vx,Fuller:1962zza}, (ii) even if a traveler could somehow
pass the throat, she would be bound to exit the wormhole through the past
horizon of the  Schwarzschild geometry; this horizon was shown to be
unstable against small perturbations and that it would change to a proper,
and thus non-traversable, horizon at the mere approaching of the traveler
\cite{Redmount:1985,Eardley:1974zz,Wald:1980nk}.

In \cite{Morris:1988cz} a new class of wormhole solutions was found that
possess no horizon and could in principle be traversable. However, some form
of exotic matter whose energy-momentum tensor had to violate all (null,
weak and strong) energy conditions was necessary in order to keep the throat
of the wormhole open. Several studies have considered a phantom field, i.e.
a scalar field with a reversed sign in front of its kinetic term, as a candidate
for such a form of matter
\cite{Ellis:1973yv,Ellis:1979bh,Bronnikov:1973fh,Kodama:1978dw,ArmendarizPicon:2002km}.

It was demonstrated in \cite{Kanti:1995vq} that the GB term leads to an
effective energy-momentum tensor that also violates the energy conditions.
It is in fact this violation that causes the circumvention of the no-hair
theorem forbidding the existence of 
regular black holes with non-trivial scalar hair. 
It was noted recently in \cite{Bronnikov:2009az}, too, 
that the presence of the GB term in the
context of a scalar-tensor theory has the property to evade the various
no-go theorems of General Relativity. 
Indeed, various wormhole solutions
were found in the context of gravitational
theories with higher curvature terms
\cite{Hochberg:1990is,Fukutaka:1989zb,Ghoroku:1992tz,Furey:2004rq}.
In the presence of the Gauss-Bonnet term in particular, wormhole solutions
were found in the context of higher-dimensional gravitational theories
\cite{Bhawal:1992sz,Dotti:2006cp,Gravanis,Giribet1,Dotti:2007az,Richarte,Maeda:2008nz,
Giribet2,Simeone}.

In this work, we will investigate the properties of wormhole solutions that arise
in the context of the four-dimensional dilatonic Einstein-Gauss-Bonnet theory,
first reported in \cite{Kanti:2011jz}. The presence of
the higher-curvature GB term, that follows naturally from the compactification
of the 10-dimensional heterotic superstring theory down to four dimensions,
suffices to support these types of solutions without the need for phantom
scalar fields or other forms of exotic matter.  

The outline of our paper is as follows: In section II, we present
the theoretical context of our model and discuss the asymptotic forms
of the sought-for wormhole solutions at the regions of radial infinity
and the regular throat. Based on the latter, we derive the embedding diagram
and study the violation of the energy conditions. 
We also present a Smarr relation for the wormhole solutions. In section III, we
present the results of our numerical analysis that reveal
the existence of wormhole solutions in the dilatonic Einstein-Gauss-Bonnet
theory, and discuss their properties. 
We demonstrate the stability with respect to radial perturbations
of a subset of these solutions in section IV.
In section V we discuss the junction conditions.
The geodesics in these wormhole spacetimes are presented in section VI.
We calculate the magnitude of the acceleration and tidal forces 
that a traveler traversing the wormhole would feel in section VII,
and conclude in section VIII.


\section{Einstein-Gauss-Bonnet-Dilaton Theory}

\subsection{Action}

We consider the following effective action 
\cite{Mignemi:1992nt,Kanti:1995vq,Chen:2009rv}
motivated by the low-energy heterotic string theory
\cite{Gross:1986mw,Metsaev:1987zx}
\begin{eqnarray}  
S=\frac{1}{16 \pi}\int d^4x \sqrt{-g} \left[R - \frac{1}{2}
 \partial_\mu \phi \,\partial^\mu \phi
 + \alpha  e^{-\gamma \phi} R^2_{\rm GB}   \right],
\label{act}
\end{eqnarray} 
where 
$\phi$ is the dilaton field
with coupling constant $\gamma$, $\alpha $ is a positive numerical
coefficient given in terms of the Regge slope parameter,
and
$R^2_{\rm GB} = R_{\mu\nu\rho\sigma} R^{\mu\nu\rho\sigma}
- 4 R_{\mu\nu} R^{\mu\nu} + R^2$ 
is the GB correction. 

The dilaton and Einstein equations are given by
\begin{eqnarray}
\nabla^2 \phi & = & \alpha \gamma  e^{-\gamma \phi}R^2_{\rm GB}
\label{dileq}\\
G_{\mu\nu} & = &
\frac{1}{2}\left[\nabla_\mu \phi \nabla_\nu \phi 
                 -\frac{1}{2}g_{\mu\nu}\nabla_\lambda \phi \nabla^\lambda\phi 
		 \right]
\nonumber\\
& &
-\alpha e^{-\gamma \phi} 
\left[	H_{\mu\nu}
  +4\left(\gamma^2\nabla^\rho \phi \nabla^\sigma \phi
           -\gamma \nabla^\rho\nabla^\sigma \phi\right)	P_{\mu\rho\nu\sigma}
		 \right]
\label{Einsteq}
\end{eqnarray}
with
\begin{eqnarray}
H_{\mu\nu} & = & 2\left[R R_{\mu\nu} -2 R_{\mu\rho}R^\rho_\nu
                        -2 R_{\mu\rho\nu\sigma}R^{\rho\sigma}
			+R_{\mu\rho\sigma\lambda}R_\nu^{\ \rho\sigma\lambda}
		   \right]
		   -\frac{1}{2}g_{\mu\nu}R^2_{\rm GB}	\ ,
\\
 P_{\mu\nu\rho\sigma} & = & 
R_{\mu\nu\rho\sigma}
+2 g_{\mu [ \sigma} R_{\rho ]\nu}
+2 g_{\nu [ \rho} R_{\sigma ]\mu}
+R g_{\mu [ \rho} g_{\sigma ]\nu} \ .
\label{HP}
\end{eqnarray}

\subsection{Ansatz and equations}

Throughout this paper we consider only static, spherically-symmetric
solutions of the above set of equations. 
Thus we can write the spacetime line-element
in the form \cite{Kanti:1995vq}
\begin{equation}
ds^2 = g_{\mu\nu}\,dx^\mu dx^\nu= -e^{\Gamma(r)}dt^2+e^{\Lambda(r)}dr^2
+r^2\left(d\theta^2+\sin^2\theta d\varphi^2 \right) \ .
\label{metricS} 
\end{equation}

As was demonstrated in \cite{Kanti:1995vq}, the dilatonic-Einstein-Gauss-Bonnet
(EGBd) theory admits black hole solutions whose gravitational background has the
line-element of Eq.~(\ref{metricS}). It was also shown that further classes of
solutions emerge in the context of the same theory. One of them, in particular,
possesses no curvature singularity and no proper horizon
(with $g_{tt}$ being regular over the whole radial regime). 
However, the $g_{rr}$ metric component
as well as the dilaton field showed some pathological behavior 
at a finite radius $r=r_0$,
as seen from the expansion near $r_0$ \cite{Kanti:1995vq} 
\begin{eqnarray}
e^{-\Lambda(r)} & = & \lambda_1 (r-r_0) + \cdots \ , 
\\
\Gamma'(r)      & = & \frac{\gamma_1}{\sqrt{r-r_0}} +  \cdots \ ,
\\
\phi(r)         & = & \phi_0 + \phi_1 \sqrt{r-r_0} +  \cdots \ .
\end{eqnarray}
The absence of any singular behaviour of the curvature invariants at $r_0$ signifies
that the aforementioned pathological behaviour is merely due to the particular
choice of the coordinate system.

In \cite{Kanti:2011jz} we have argued that this class of asymptotically flat
solutions can be brought to a regular form by employing the coordinate
transformation $r^2  = r_0^2 + l^2$. Then, the metric becomes
\begin{equation}
ds^2 =  -e^{2\nu(l)}dt^2+f(l)dl^2
+(l^2+r_0^2)\left(d\theta^2+\sin^2\theta d\varphi^2 \right) \ .
\label{metricL} 
\end{equation} 
The above form is regular and describes a wormhole solution, where $r_0$ is the
radius of the throat. Indeed, in terms of the new coordinate $l$, the expansion at
$l=0$ assumes the form
\begin{eqnarray}
f(l) & = & f_0 + f_1 l + \cdots \ ,
\label{expansion-throat1}
\\
e^{2\nu(l)} & = & e^{2\nu_0}(1 +\nu_1 l)  + \cdots\ ,
\label{expansion-throat2}
\\
\phi(l)         & = & \phi_0 + \phi_1 l +  \cdots \ ,
\label{expansion-throat3}
\end{eqnarray}
where $f_i$, $\nu_i$ and $\phi_i$ are constant coefficients, and shows no pathology.
Both metric functions and the dilaton field remain finite 
in this asymptotic regime. 
Thus these solutions possess no horizon. 
In addition, all curvature invariant quantities -- including the GB term
-- turn out to be finite at $l=0$, a result that demonstrates the absence
of any singularity. 

Substitution of the metric Eq.~(\ref{metricL}) 
into the dilaton equation (\ref{dileq})
and Einstein equations (\ref{Einsteq}) 
yields a coupled system of ordinary differential equations (ODE's) 
for the metric functions and the dilaton field
\begin{eqnarray}
f' + \frac{f (r^2 f + l^2 - 2 r^2)}{l r^2}
& = & 
\frac{r^2 f \phi'^2}{4 l}
 +2\alpha \gamma\frac{e^{-\gamma\phi}}{l r^2} \left\{
  2 (r^2 f - l^2)(\gamma\phi'^2-\phi'')
 + \phi'\left[\frac{f'}{f}\,(r^2 f-3 l^2) +\frac{4 l r_0^2}{r^2}\right]\right\}, \hspace*{0.3cm}\,\,\,
 \label{ode1} \\[1mm] && \nonumber\\
\nu' - \frac{r^2 f -l^2}{2 l r^2}
& = & 
\frac{\phi'^2 r^2}{8 l}
+2\alpha \gamma\frac{e^{-\gamma\phi}}{l r^2 f} 
\,\nu'\phi' (r^2 f- 3 l^2)\,,\label{ode2}  \\[1mm] && \nonumber
\\
 \nu''+\nu'^2 + \frac{\nu' (2 l f - r^2 f')}{2 r^2 f}
 & + &
  \frac{ 2 r_0^2 f - l  r^2 f'}{2 r^4 f}  \label{ode3} \\ 
& = & 
-\frac{\phi'^2}{4} 
+2\alpha \gamma\frac{e^{-\gamma\phi}}{r^2 f} \left\{
2 l \left[\nu' (\gamma\phi'^2-\phi'') -\phi' (\nu'^2+\nu'')\right]
+ \nu' \phi' \left(\frac{3lf'}{f}-\frac{2r_0^2}{r^2}\right)\right\},
\nonumber\\[1mm]
\phi'' + \nu'\phi' + \frac{\phi' (4 l f - r^2 f')}{2 r^2 f}
& = & 
4\alpha \gamma\frac{e^{-\gamma\phi}}{r^4 f} \left\{
-2 (r^2 f - l^2)(\nu'^2+\nu'') 
+\nu'\left[\frac{f'}{f}\,(r^2 f-3 l^2) +\frac{4 l r_0^2}{r^2}\right]\right\}.
\label{ode4}
\end{eqnarray}
Here $r^2=l^2+r_0^2$ and the prime denotes the derivative with respect to $l$.
Equations (\ref{ode1}), (\ref{ode2}) and (\ref{ode3}) follow from the 
$tt$, $ll$ and $\theta\theta$ components
of the Einstein equations, respectively, whereas the last equation (\ref{ode4}) follows
from the dilaton equation. 
For the numerical computation we `diagonalize' 
Eqs.~(\ref{ode1}), (\ref{ode2}) and (\ref{ode4}) with respect to 
$f'$, $\nu'$ and $\phi''$. 
The remaining equation, Eq.~(\ref{ode3}),
involving also second derivatives of $\nu$ and $\phi$,
is satisfied if the other three equations are fulfilled.
Thus a system of ODE's must be solved
that consists of two first-order equations
for the metric functions $f$ and $\nu$
and a second-order equation for the dilaton field.

\subsection{Expansions}

The expansion near the throat, 
Eqs.~(\ref{expansion-throat1})-(\ref{expansion-throat3}),
once substituted into the
set of equations leads to a number of recursive constraints that determine the
higher-order coefficients in terms of the lower ones. 
Looking at the lowest order, 
we observe that $f_0$, $\nu_0$ and $\phi_0$ are free parameters. 
Also,
the set of parameters of the theory includes the radius of the throat $r_0$ and
the value of $\alpha$. 
The value of the constant $\gamma$ will be later fixed to 1 for simplicity. 

However, not all of the above parameters are actually independent. To start with,
we observe that the field equations remain invariant under the
simultaneous changes
\begin{equation}
\label{sym1}
\phi \rightarrow \phi +\phi_*
\ , \hspace{1cm}
(r,l) \rightarrow (r,l)\ e^{-\gamma \phi_*/2} \,. 
\end{equation}
In addition, the following transformation
\begin{equation}
\label{sym2}
\alpha \rightarrow k \alpha 
\ , \hspace{1cm} \phi \rightarrow \phi + \frac{\ln k}{\gamma} \,.
\end{equation}
is also a symmetry of the equations. In the light of the above, we conclude
that only one parameter out of the set $(\alpha, r_0, \phi_0)$ is independent. 
We may therefore choose to have a zero asymptotic value of the dilaton field at
infinity which entails fixing the value of the dilaton field at the throat $\phi_0$.
The remaining two parameters can be combined to give a dimensionless
parameter $\alpha/r_0^2$ that will be used throughout our analysis. Finally,
among the two parameters associated with the metric functions,
($\nu_0$, $f_0$), again only the latter is independent -- it is only the derivatives
of the metric function $\nu$ that appear in the field equations of motion,
therefore, we may use this freedom to fix the value of $\nu_0$ in order to
ensure asymptotic flatness at radial infinity. 
Thus, our class of wormholes is a two-parameter family of solutions
that, in the context of our analysis, have been chosen to be 
$f_0$ and $\alpha/r_0^2$.

When the expansion of the metric functions and dilaton field near the throat 
(\ref{expansion-throat1})-(\ref{expansion-throat3}) are substituted into the
field equations, we obtain a set of constraints on the higher-order coefficients.   
The constraint on the value of the first derivative of the dilaton field at the
throat is particularly interesting and takes the form
\begin{equation}
\phi_1^2 = \frac{f_0(f_0-1)}{2\alpha\gamma^2 e^{-\gamma \phi_0}
\left[f_0-2(f_0-1)\frac{\alpha}{r_0^2} e^{-\gamma \phi_0}\right]}\ .
\label{constraint-phi1}
\end{equation}
As the left-hand-side of the above equation is positive-definite, the same
must hold for the right-hand-side. We may easily see that the expression
inside the square brackets in the denominator remains positive and has
no roots if 
\begin{equation}
\frac{\alpha}{r_0^2} < \frac{1}{2} e^{\gamma \phi_0} .
\label{constraint_1}
\end{equation}
This inequality is automatically satisfied for the set of wormhole solutions presented
in the next section, and gives a lower limit on the size of the throat of
the wormhole $r_0$. Then, the positivity of the right-hand-side of 
Eq. (\ref{constraint-phi1}) demands that $f_0 \geq 1$. The solutions satisfying $f_0=1$
comprise a boundary in the phase space of the wormhole solutions and their
physical significance will be discussed in the next section.

As already discussed above, at the asymptotic regime of radial infinity,
i.e. as $l \to \infty$, we demand asymptotic flatness for the two metric
functions and a vanishing value of the dilaton field. Then, the asymptotic
expansion at infinity takes the form
\begin{eqnarray}
\nu & \rightarrow & -\frac{M}{l} + \cdots \ ,
\nonumber \\
f   & \rightarrow & 1+ \frac{2M}{l} + \cdots \ ,
\\
\phi  & \rightarrow & -\frac{D}{l} + \cdots \ .
\nonumber
\end{eqnarray}
In the above, $M$ and $D$ are identified with the mass and dilaton charge of the
wormhole, respectively. We remind the reader that in the case of the black hole
solutions \cite{Kanti:1995vq}, the parameters $M$ and $D$ were related, and that
rendered the dilatonic hair as ``secondary''. However, in the case of the wormhole
solutions these parameters are not related; this result, together with the fact that
the number of independent parameters near the throat is also two, confirms the
classification of this group of solutions as a two-parameter class of solutions. 

\subsection{Wormhole geometry}

A general property of a wormhole is the existence of a throat,
i.~e.~a surface of minimal area (or minimal radius for spherically symmetric
spacetimes). Indeed, this property is implied by the form
of the line element (\ref{metricL}) above, with $f(0)$ and $\nu(0)$
finite.
To cast this condition in a coordinate independent way, we define the 
proper distance from the throat in the following way
\begin{equation}
\xi = \int_0^l \sqrt{g_{_{ll}}} dl' = \int_0^l \sqrt{f(l')} dl' \  .
\end{equation}
If we impose the condition for a minimal radius $R=\sqrt{l^2+r_0^2}$ at $l=0$, this 
translates to
\begin{equation}
\left.\frac{dR}{d\xi}\right|_{l=0} = 0 , \ \ \ 
\left.\frac{d^2R}{d\xi^2}\right|_{l=0} > 0\,.
\end{equation}
It is easily seen that the first condition is indeed satisfied.
For the second condition we find 
\begin{equation}
\left.\frac{d^2R}{d\xi^2}\right|_{l=0} = \frac{1}{r_0 f_0} > 0 \ .
\end{equation}
This gives a coordinate independent  meaning to the parameter $f_0$,
\begin{equation}
f_0 = \left[\left. R\frac{d^2R}{d\xi^2}\right|_{l=0}\right]^{-1} \ .
\end{equation}

A particularly useful concept with which we may examine the geometry of a given
manifold is the construction of the corresponding embedding diagram. In the
present case, we consider the isometric embedding of a plane passing through
the wormhole. Due to the spherical symmetry of the solutions, we may simplify
the analysis and choose $\theta=\pi/2$. Then, we set
\begin{equation}
f(l) dl^2 +(l^2+r_0^2) d\vphi^2 = dz^2 + d\eta^2 +\eta^2 d\vphi^2 \ ,
\label{eucemb}
\end{equation}
where $\{z,\eta,\varphi\}$ are a set of cylindrical coordinates 
in the three-dimensional Euclidean space $R^3$. 
Regarding $z$ and $\eta$ as functions of $l$, we find 
\begin{equation}
\eta(l)=\sqrt{l^2+r_0^2} \ , \ \ \ \ 
\left(\frac{dz}{dl}\right)^2+\left(\frac{d\eta}{dl}\right)^2 = f(l) \ .
\end{equation}
From the last equation it follows that 
\begin{equation}
z(l) = \pm \int_0^l \sqrt{f(l')-\frac{l'^2}{l'^2+r_0^2}} \ dl' \ .
\label{z_emb}
\end{equation}
Hence $\{\eta(l),z(l)\}$ is a parametric representation of a slice of the 
embedded $\theta=\pi/2$-plane for a fixed value of $\vphi$.
We observe that the curvature radius of the curve $\{\eta(l),z(l)\}$ at $l=0$ is 
given by $R_{0} = r_0 f_0$. Thus $f_0$ has a geometric
meaning: $f_0 = R_{0}/r_0$ is the ratio 
of the curvature radius to the radius of the throat.

\subsection{Energy conditions}

As was already mentioned in the Introduction, the existence of the wormhole solution
relies on the violation of the null energy condition. The null energy condition holds if
\begin{equation}
T_{\mu\nu} n^\mu n^\nu \geq 0
\end{equation}
for any null vector field $n^\mu$. In the particular case of spherically symmetric
solutions, and if we employ the Einstein equations, this condition can be expressed as
\begin{equation}
-G_0^0+G_l^l \geq  0 \ , \ \ \ {\rm and } \ \ \ -G_0^0+G_\theta^\theta\geq  0 \ .
\label{Nulleng}
\end{equation}
If one or both of the above conditions do not hold in some region of spacetime, then the
null energy condition is violated. By making use of the expansion of the fields near the
throat (\ref{expansion-throat1})-(\ref{expansion-throat3}), we find that, close to $r_0$,
\begin{equation}
\left[-G_0^0+G_l^l\right]_{l=0} = -\frac{2}{f_0 r_0^2} < 0 \ , 
\label{eng_pert_cond}
\end{equation}
provided $e^{2\nu(0)}\neq 0$, i.e.~in the absence of a horizon.
Thus for the wormhole solutions there is always a region close to the throat 
where the  null energy condition is violated.

On the other hand, from the asymptotic expansion of the solutions at infinity,
we find
\begin{eqnarray}
-G_0^0+G_l^l & \rightarrow & \frac{D^2}{2} \frac{1}{l^4} + {\cal O}(l^{-5}) \ , 
\label{term1_as}\\
-G_0^0+G_\theta^\theta & \rightarrow & 20 \alpha M D \frac{1}{l^6} + {\cal O}(l^{-7}) \ . 
\label{term2_as}
\end{eqnarray}
We observe that, for solutions with a positive dilaton charge $D$ (and a positive mass
$M$) the null energy condition is satisfied in the asymptotic region. However, if the
dilaton charge is negative, the null energy condition is violated also in that
asymptotic region. Note that $D \geq 0$ for the black hole solutions, where it
was given by a positive-definite combination of $(\alpha, M, \phi_\infty)$
\cite{Kanti:1995vq}. 
However, $D \geq 0$ does not necessarily hold for 
all of the wormhole solutions.

\subsection{Smarr relation}

To derive the Smarr-like mass formula 
for the wormhole solutions we start with the definition of the Komar mass
\begin{equation}
M = M_{\rm th} +\frac{1}{4\pi} \int_\Sigma{ R_{\mu\nu} \xi^\mu n^\nu} dV
  = M_{\rm th} -\frac{1}{4\pi} \int{ R^0_{0}\sqrt{-g}}d^3x \,, 
\label{M_adm}
\end{equation}
where $\xi^\mu$ is the timelike Killing vector field, 
$\Sigma$ is a spacelike hypersurface,
$n^\nu$ is a normal vector on $\Sigma$ 
and $dV$ is the natural volume element on $\Sigma$.
Here $M_{\rm th}$ denotes the contribution of the throat,
\begin{equation}
M_{\rm th} = \frac{1}{2} A_{\rm th} \frac{\kappa}{2\pi} \,,
\label{m_th}
\end{equation}
where  $A_{\rm th}$ is the area of the throat, 
and $\kappa$ is the surface gravity at the throat
\begin{equation}
\kappa^2 = -1/2 (\nabla_\mu \xi_\nu)(\nabla^\mu \xi^\nu)
\,, \quad
\kappa =  \frac{e^{\nu_0}}{\sqrt{f_0}}\nu'(0) \,.
\label{kap}
\end{equation}

To obtain the mass formula we express $ R^0_{0}$ in terms of the effective
stress energy tensor
\begin{equation}
R^0_{0} = T^0_{0}-\frac{1}{2} T^\mu_\mu 
= T^0_{0}-\frac{1}{2} T^\mu_\mu 
+\frac{1}{2\gamma}\left[ \nabla^2 \phi-\gamma\alpha e^{-\gamma\phi}R^2_{\rm GB} \right]\,,
\label{r00_th}
\end{equation}
where we have added a zero in the form of the dilaton equation.
Multiplication by $\sqrt{-g}$ yields
\begin{equation}
R^0_{0}\sqrt{-g} = \left( T^0_{0}-\frac{1}{2} T^\mu_\mu
                    -\frac{1}{2}\alpha e^{-\gamma\phi}R^2_{\rm GB}\right)\sqrt{-g}
	  +\frac{1}{2\gamma}\partial_\mu\left(\sqrt{-g} \partial^\mu \phi\right).
\label{r00g_th}
\end{equation}
Substitution of the metric yields a total derivative for the right-hand-side of the above equation.
Thus, after integration, we find
\begin{equation}
-\frac{1}{4\pi} \int{R^0_{0}\sqrt{-g}}d^3x = -\frac{D}{2\gamma}
+4\frac{e^{\nu_0}}{\sqrt{f_0}}\alpha e^{-\gamma\phi_0}\nu'(0)
+\frac{1}{2\gamma} \frac{e^{\nu_0}}{\sqrt{f_0}}
   r_0^2 \phi'(0)\left(1+4\frac{\alpha\gamma^2}{r_0^2} e^{-\gamma\phi_0}\right).
\label{int_r00}
\end{equation}

Now substitution into Eq.~(\ref{M_adm}) gives the			   
Smarr-like formula 
\begin{equation}
M = 2 S_{\rm th} \frac{\kappa}{2\pi} -\frac{D}{2\gamma} 
     +\frac{1}{8\pi\gamma}\int{\sqrt{-g}g^{ll}\frac{d\phi}{dl}
             \left(1+2\alpha\gamma^2 e^{-\gamma\phi} \tilde{R}\right)}d^2x ,  
\end{equation}
with
\begin{equation}
 S_{\rm th} = \frac{1}{4}\int{ \sqrt{h}\left(1+2\alpha e^{-\gamma\phi} 
  \tilde{R}\right) d^2x }\,. 
\end{equation}
Here $h$ is the induced spatial metric on the throat, $\tilde{R}$ is the scalar 
curvature of $h$, and the integral is evaluated at $l=0$.
Defining the normal vector $n_0^\mu$ on the surface $l=0$ the mass formula
becomes
\begin{equation}
M = 2 S_{\rm th} \frac{\kappa}{2\pi} -\frac{D}{2\gamma} 
     +\frac{D_{\rm th}}{2\gamma}\ ,  
\label{smarr}
\end{equation}
with 
\begin{equation}
D_{\rm th} =\frac{1}{4\pi}\int{\sqrt{h}e^{\nu_0} n_0^\mu \partial_\mu \phi
                    \left(1+2\alpha\gamma^2 e^{-\gamma\phi} \tilde{R}\right)}d^2x \ .
\end{equation}
According to the above, the Smarr-like mass formula for wormholes
is obtained by replacing the horizon properties
by the corresponding throat properties in
the known EGBd mass formula for black holes 
\cite{Kanti:1995vq,Kleihaus:2011tg}. In addition, an extra contribution
appears that may be interpreted as a modified throat dilaton charge, 
while the GB modification is of the same type as the GB modification
of the area (or entropy in the case of black holes).


\section{Numerical wormhole solutions}

For the numerical calculations we use the line-element (\ref{metricL}),
since the functions are well behaved at $r_0$. For the representation of 
the numerical results we employ the metric (\ref{metricS}) as well.

To solve the ODEs numerically, 
we introduce the compactified coordinate $x=l/(1+l)$,
thus mapping the semi-infinite range of $l$ to the finite range of $x$, 
i.e.~$0\leq x\leq 1$.
We cover the parameter range $0.002 \leq \alpha \leq 0.128$
and $1.0001 \leq f_0 \leq 20.0$,
keeping $r_0=1$ fixed. Also we set $\gamma = 1$.

\subsection{Metric and dilaton functions}

Let us start the discussion of the solutions by recalling
the boundary conditions for the system of equations, consisting of
two first order and one second order ODEs.
At the throat $l=0$ regularity requires
\begin{equation}
\left\{\phi'^2 - \frac{f(f-1)}{2\alpha\gamma^2 e^{-\gamma \phi}
                              \left[f-2(f-1)\frac{\alpha}{r_0^2} e^{-\gamma \phi}\right]}
\right\}_{l=0} = 0\,.
\label{BC_phi}
\end{equation}
This boundary condition has to be supplemented by 
$f(0) = f_0$ or $\phi(0)=\phi_0$ in order to obtain a specific solution.
We note that 
the asymptotic condition $f \to 1$ for $l\to \infty$ is always satisfied. 
This can be seen from the asymptotic form of Eq.~(14), $f'+f(f-1)/l =0$, 
which has the general solution $f= l/(l+const)$.
This leaves the asymptotic boundary conditions at radial infinity
\begin{equation}
\lim_{l\to \infty}\nu = 0 \,, \hspace{1cm} \lim_{l\to \infty}\phi = 0 \, .
\end{equation}

Wormhole solutions can be found for every value of $\alpha/r_0^2$
below $\alpha/r_0^2 \approx 0.13$. 
This upper bound on $\alpha/r_0^2$ 
translates into a lower bound on the radius of the throat $r_0$
(for a given $\alpha$).
Thus the radius of the throat $r_0$ can be arbitrarily large. 

In Figs.~1a,b and c, we show the metric functions $f(l)$ and $\nu(l)$
and the dilaton function $\phi(l)$, respectively, for an indicative set of
wormhole solutions.
We also exhibit the scaled GB term $\alpha R^2_{\rm GB}$ in Fig.~1d.
Close to the throat the functions show a distinct dependence on both
parameters, $f_0$ and  $\alpha/r_0^2$.
However, for intermediate values of $l$, the functions 
$f(l)$, and likewise the functions $\nu(l)$, corresponding to
the same value of $\alpha/r_0^2$, approach each other and form clusters.
For larger values of $l$, solutions obtained for different values of
$\alpha/r_0^2$ also merge together. The same behaviour is observed for
the GB term whereas for the dilaton function $\phi(l)$ the same tendency
also exists but becomes pronounced at slightly larger values of $l$.

\begin{figure}[t]
\lbfig{f-1}
\begin{center}
\hspace{0.0cm} \hspace{-0.6cm}
\includegraphics[height=.25\textheight, angle =0]{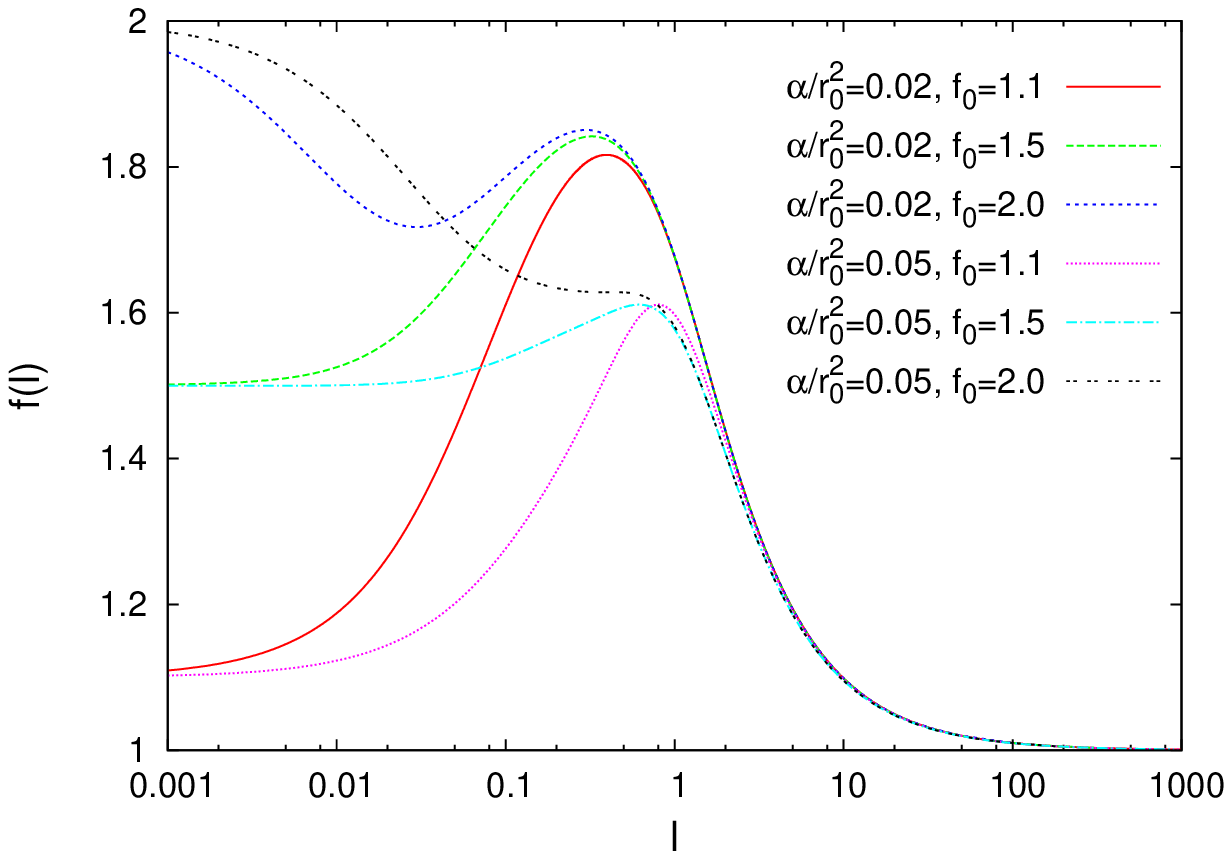}
\hspace{0.5cm} \hspace{-0.6cm}
\includegraphics[height=.25\textheight, angle =0]{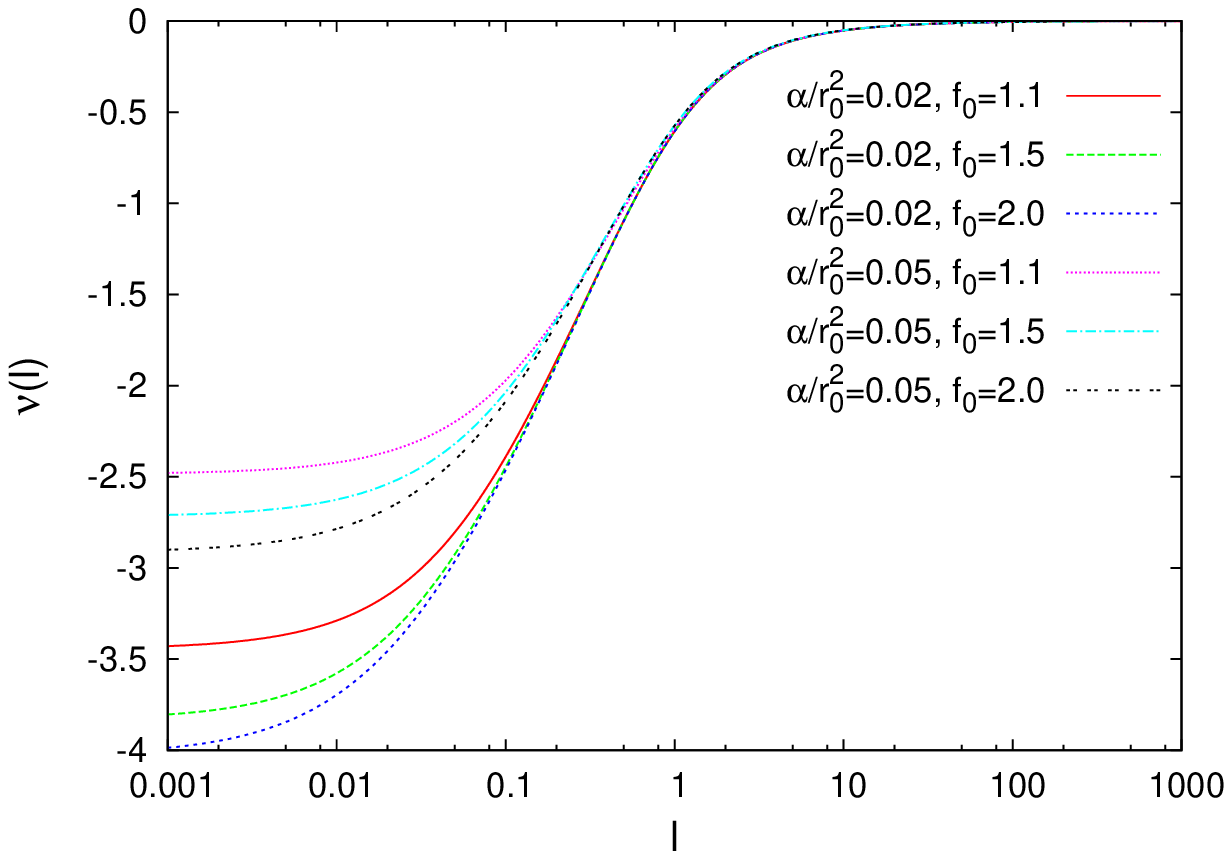}
\\[1mm]
\hspace{0.3cm} {(a)} \hspace{7.5cm} {(b)} \hspace{2cm} \\[1mm]
\hspace{0.0cm} \hspace{-0.6cm}
\includegraphics[height=.25\textheight, angle =0]{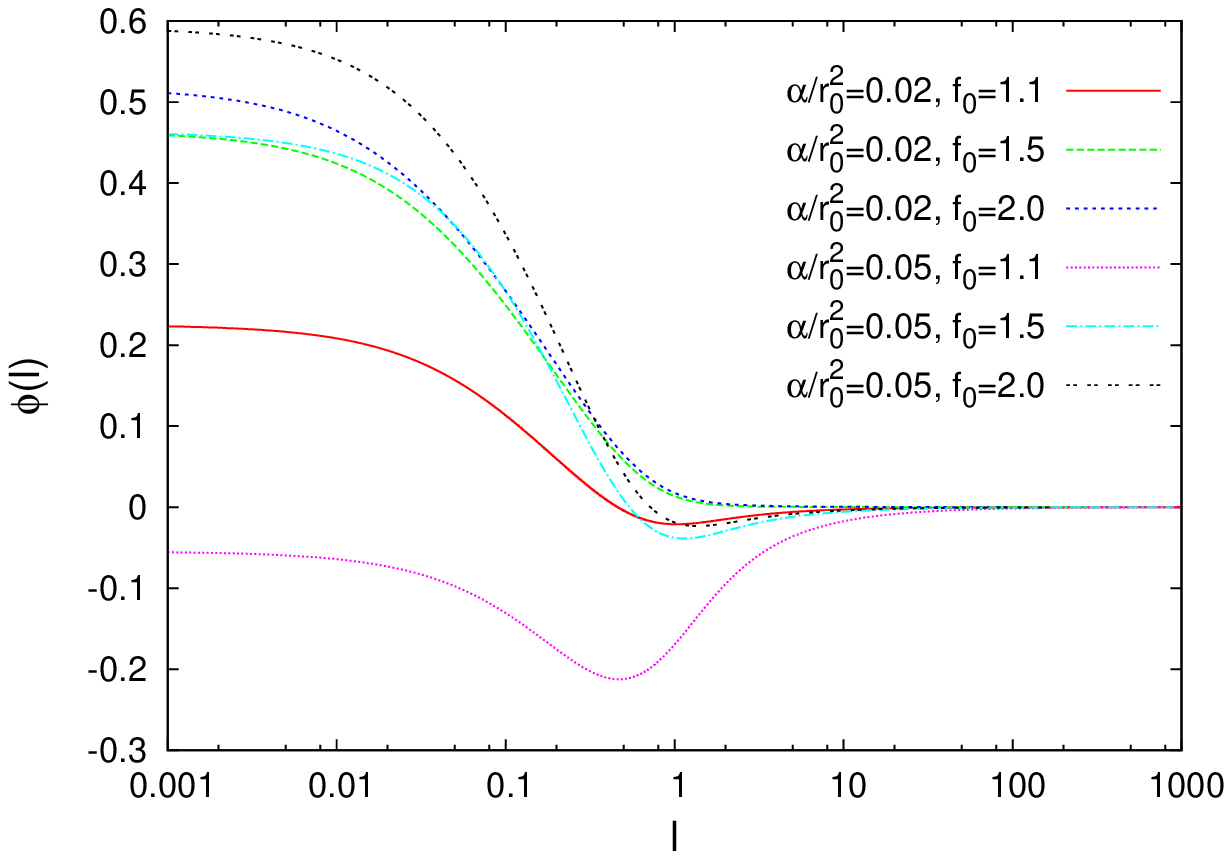}
\hspace{0.5cm} \hspace{-0.6cm}
\includegraphics[height=.25\textheight, angle =0]{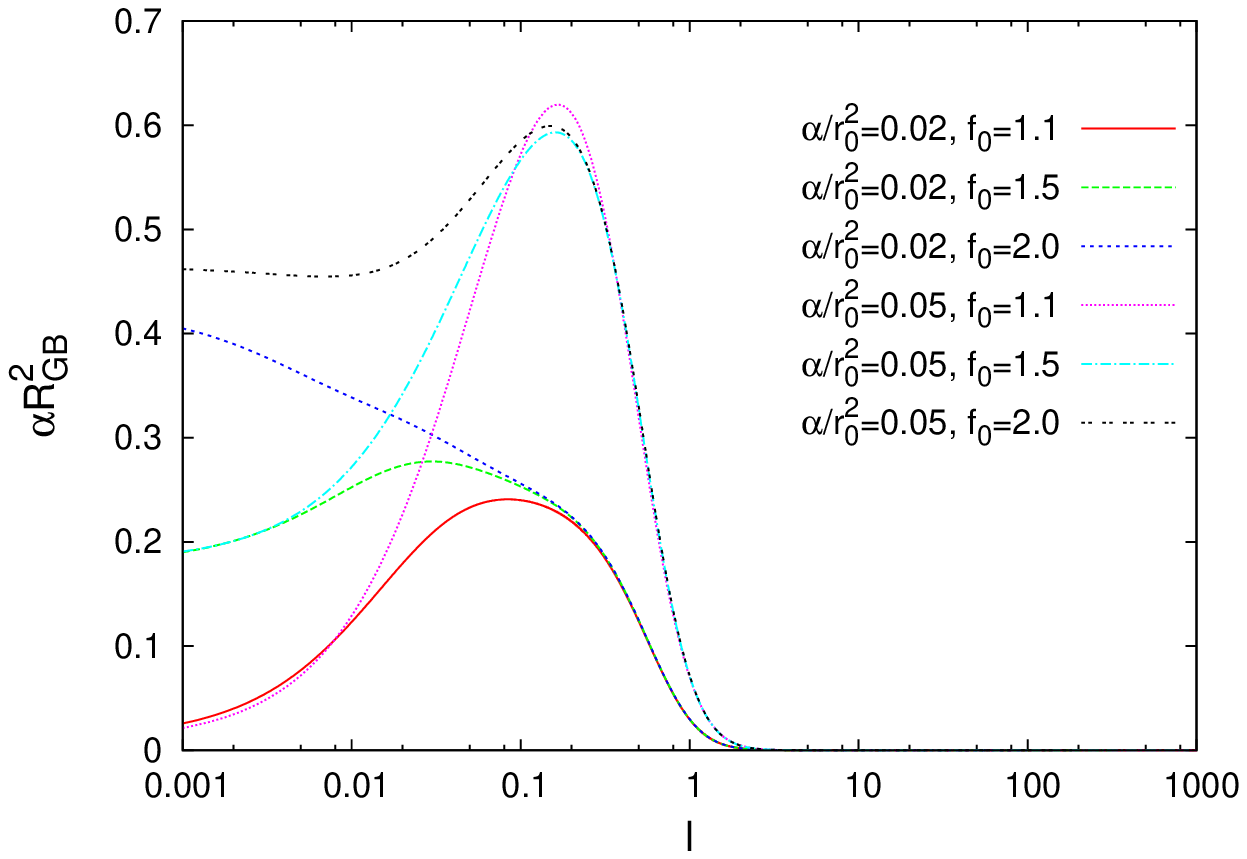}\\
\hspace{0.3cm} {(c)} \hspace{7.5cm} {(d)} \hspace{2cm} 
\end{center}
\caption{
(a) The metric function $f(l)$,  (b) 
the metric function $\nu(l)$,  (c)
the dilaton function $\phi(l)$, 
and  (d) the scaled GB term $\alpha R^2_{\rm GB}$ versus $l$
for wormhole solutions with 
the parameter values $f_0=1.1$, $1.5$, $2.0$ 
and $\alpha/r_0^2=0.02$, $0.05$.}
\end{figure}


In Fig.~\ref{F-2} we visualize the geometry of the wormhole solutions.
In particular, we present as a typical example in Fig.~\ref{F-2}a
the isometric embedding of the solution with $\alpha/r_0^2=0.02$ and $f_0=1.1$
\cite{Kanti:2011jz}.
Here also the curvature radius at the throat, $R_{0} = r_0 f_0$, is shown.
Also for $\alpha/r_0^2=0.02$, the curves $z$ versus $\eta$ near the throat
are shown in Fig.~\ref{F-2}b for several values of $f_0$.

\begin{figure}[t]
\lbfig{F-2}
\begin{center}
\mbox{\includegraphics[height=.27\textheight, angle =0]{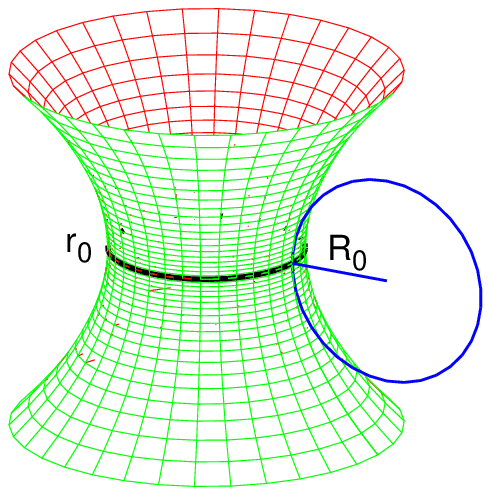}
      \includegraphics[height=.27\textheight, angle =0]{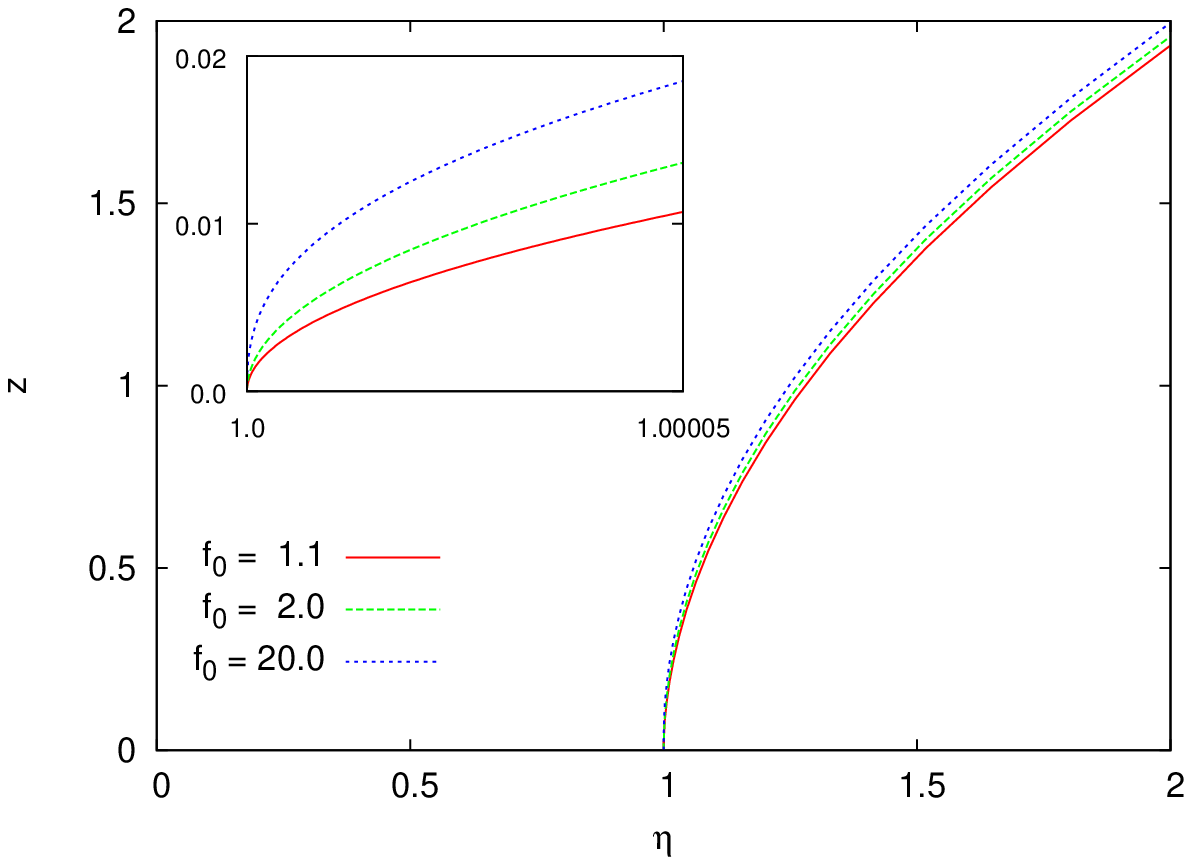}}
\end{center}
\hspace{0.5cm} {(a)} \hspace{8.5cm} {(b)} \hspace{2cm} \\[1mm]
\caption{
(a) Isometric embedding of the wormhole solution for
$\alpha/r_0^2=0.02$ and $f_0=1.1$ (taken from \cite{Kanti:2011jz}),
(b) $z$ versus $\eta$ for $\alpha/r_0^2=0.02$ and $f_0=1.1$, $2.0$, $20.0$.
}
\end{figure}

We may also define the profile function $b(r)$ via the equation
\begin{equation}
e^{-\Lambda(r)} \equiv 1-\frac{b(r)}{r} \ .
\label{bfun}
\end{equation}
We note that, at the throat, the profile function $b/r_0$ goes to one
and, thus, $g^{rr}$ vanishes.
For small values of $f_0$, the value of the metric function 
$\nu$ at the throat keeps decreasing as $f_0 \to 1$.
Therefore, in the limit $f_0 \to 1$, the metric function $g_{tt}$
tends to zero and a horizon emerges, thus recovering the
class of dilatonic black hole solutions \cite{Kanti:1995vq}.


\subsection{Domain of existence}

Let us now explore the domain of existence of these wormhole solutions.
To that end we construct a set of families of wormhole solutions,
where for each family the value of the parameter $\alpha/r_0^2$ is fixed,
while the second parameter $f_0$ varies within a maximal range of
$1 < f_0 < \infty$.
The values of the parameter $\alpha/r_0^2$
cover the range $0 < \alpha/r_0^2 < 0.13$.
For larger values of $\alpha/r_0^2$ no wormhole solutions are found.

In Fig.~\ref{f-3} we present the domain of existence of the wormhole solutions.
Fig.~\ref{f-3}a shows the scaled area of the throat $A/16 \pi M^2$
versus the scaled dilaton charge $D/M$.
The domain of existence is mapped by the families of solutions 
obtained for a representative set of fixed values of $\alpha/r_0^2$. 
We observe that the domain of existence of the wormhole solutions is bounded by 
three curves indicated by asterisks, crosses and dots. 
The boundary indicated by asterisks coincides with the EGBd black hole curve 
\cite{Kanti:1995vq}
and corresponds to the limit $f_0 \to 1$.
The limit $f_0 \to \infty$ on the other hand is indicated by crosses. 
At the third boundary, marked by dots,
solutions are encountered 
that are characterized by a curvature singularity.

\begin{figure}[t]
\lbfig{f-3}
\begin{center}
\mbox{\includegraphics[height=.27\textheight, angle =0]{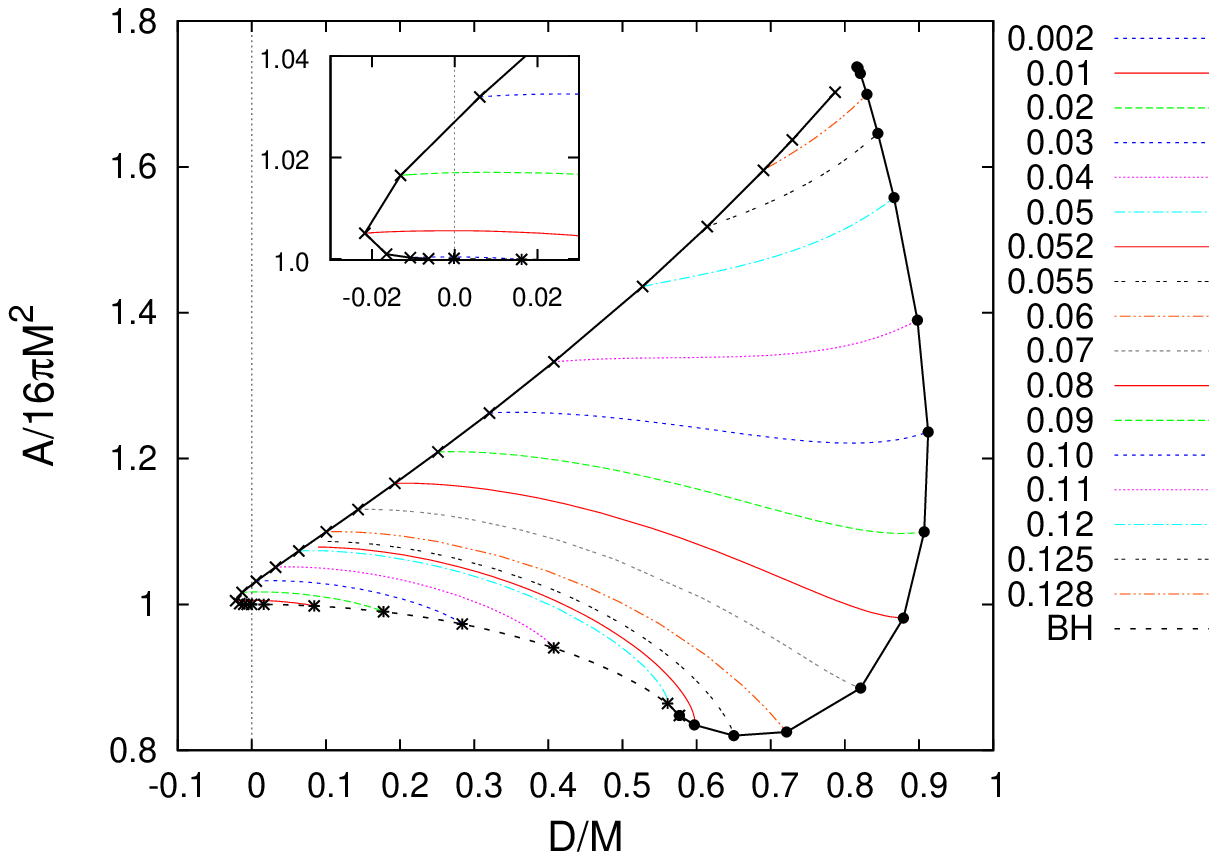}
      \includegraphics[height=.27\textheight, angle =0]{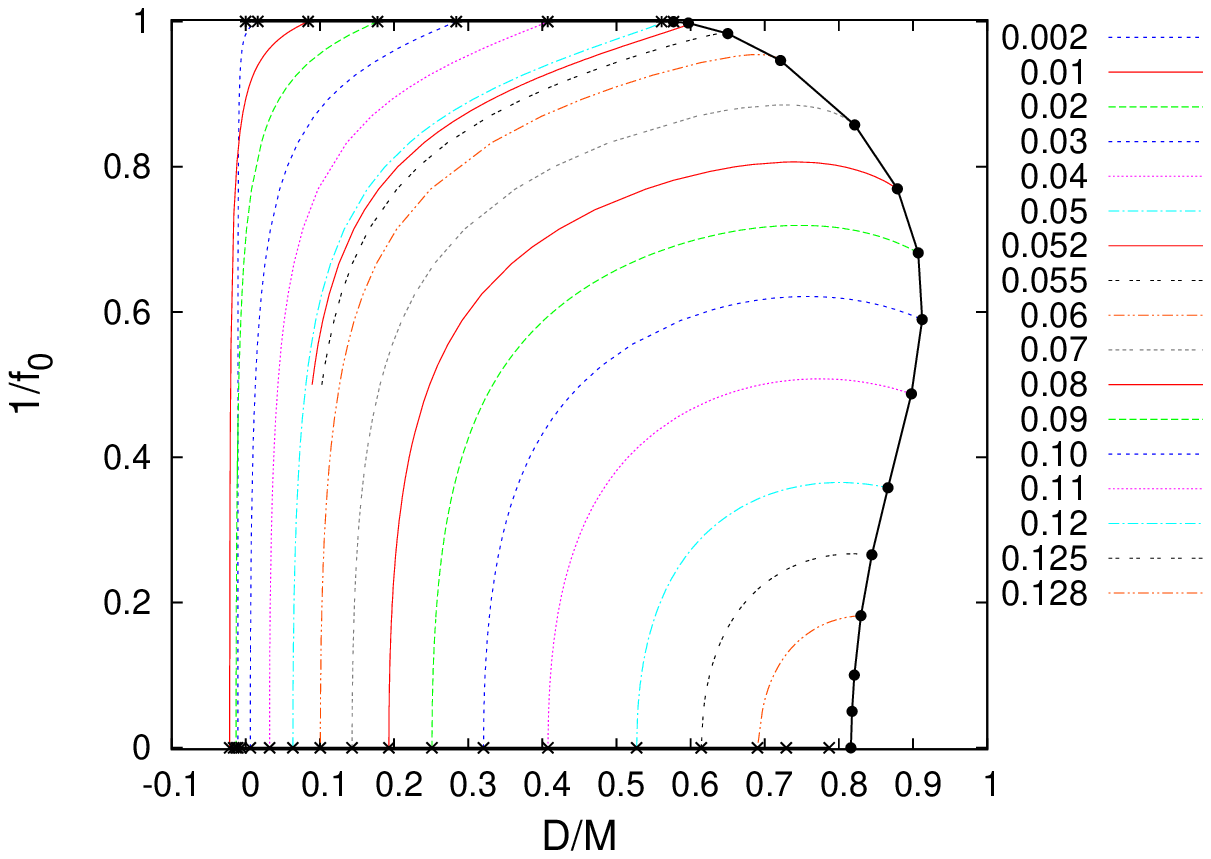}}
\hspace{0.5cm} {(a)} \hspace{8cm} {(b)} \hspace{2cm} \\[1mm]
\mbox{\includegraphics[height=.27\textheight, angle =0]{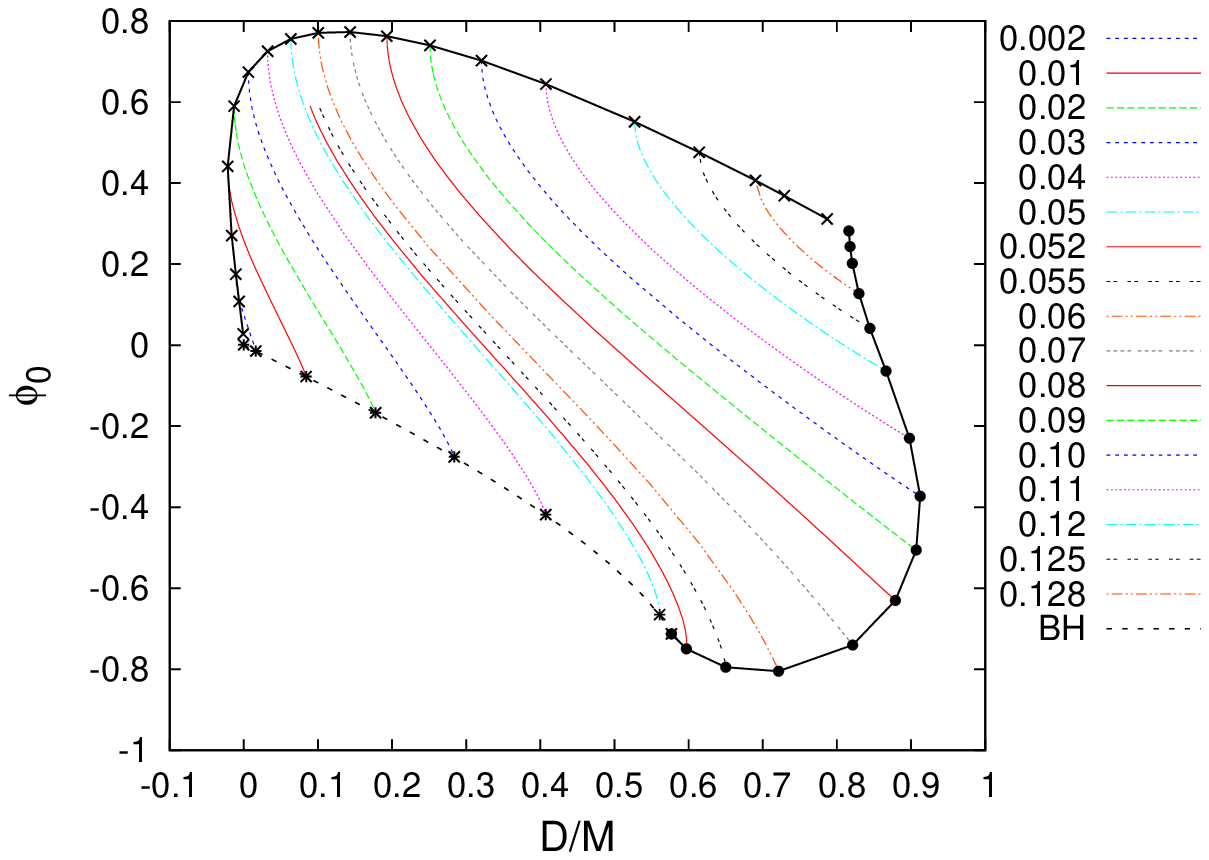}
      \includegraphics[height=.27\textheight, angle =0]{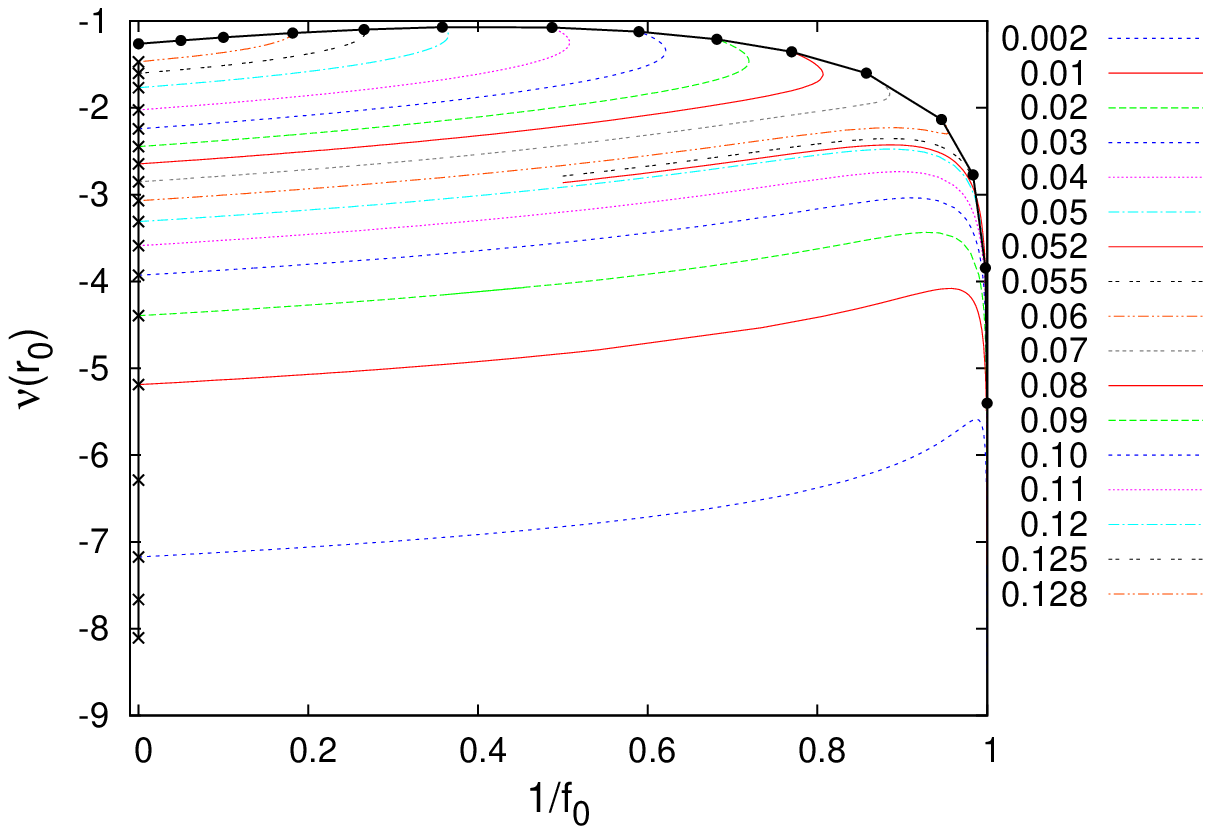}}
\hspace{0.5cm} {(c)} \hspace{8cm} {(d)} \hspace{2cm} \\[1mm]
\end{center}
\caption{
(a) Scaled area $A/16 \pi M^2$ of the throat 
versus the scaled dilaton charge $D/M$
for several values of $\alpha/r_0^2$. 
The boundaries represent EGBd black holes (asterisks),
limiting solutions with $f_0\to \infty$ (crosses)
and solutions with curvature singularities (dots).
(b) Metric function $f_0$ at the throat
(determining the curvature radius $R_0=r_0 f_0$)
versus the scaled dilaton charge $D/M$;
(c) dilaton function $\phi(0)$ at the throat 
versus the scaled dilaton charge $D/M$;
(d) redshift function $\nu(0)$ at the throat
versus $1/f_0$.
}
\end{figure}

Before discussing the three limiting cases in more detail,
let us consider further the boundary conditions at the throat.
We illustrate the metric function $f_0$ at the throat 
for the same set of families of wormhole solutions
in Fig.~\ref{f-3}b.
For fixed values of $\alpha/r_0^2$ below the limiting value for EGBd black holes,
$\left. \alpha/r_0^2 \right|_{\rm bh}=0.0507$
\cite{Kanti:1995vq}, $f_0$ covers the full range $1 < f_0 < \infty$.
Beyond this critical value, however, the minimal value of $f_0$ increases
with increasing $\alpha/r_0^2$.
We note that, for a certain intermediate range of $\alpha/r_0^2$,
$f_0$ is not monotonic.

In Fig.~\ref{f-3}c we show the dilaton field $\phi_0$ at the throat 
versus the scaled dilaton charge $D/M$ for the same set of solutions.
Clearly, the domain of $\phi_0$ is bounded.
Again we observe that in the limit $f_0 \to 1$ the 
black hole values are obtained.
Inspection of the redshift function $\nu_0$ at the throat,
exhibited in Fig.~\ref{f-3}d,
reveals that $-g_{00}(r_0)$ 
tends to zero as $f_0 \to 1$.
Thus a horizon emerges in this limit.

\subsubsection{Black hole limit}

\begin{figure}[h!]
\lbfig{f-5}
\begin{center}
\mbox{\includegraphics[height=.27\textheight, angle =0]{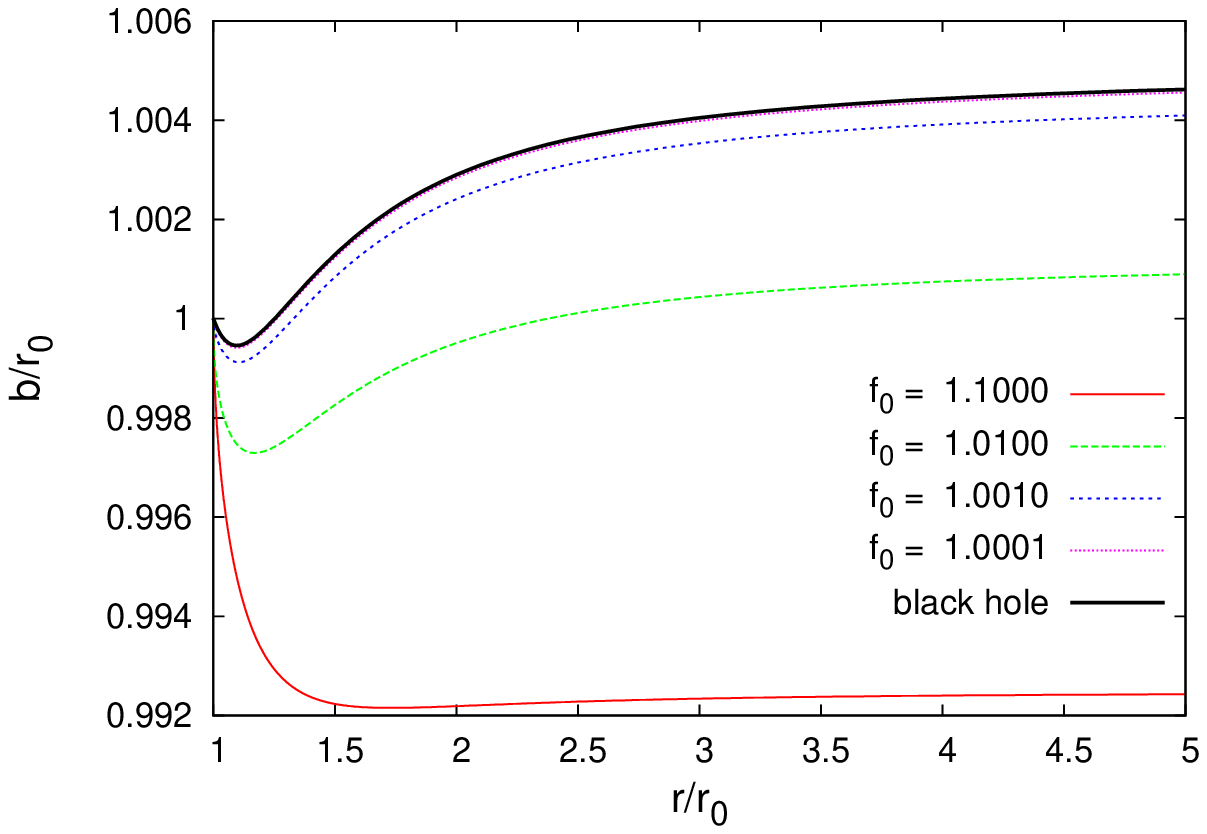}
      \includegraphics[height=.27\textheight, angle =0]{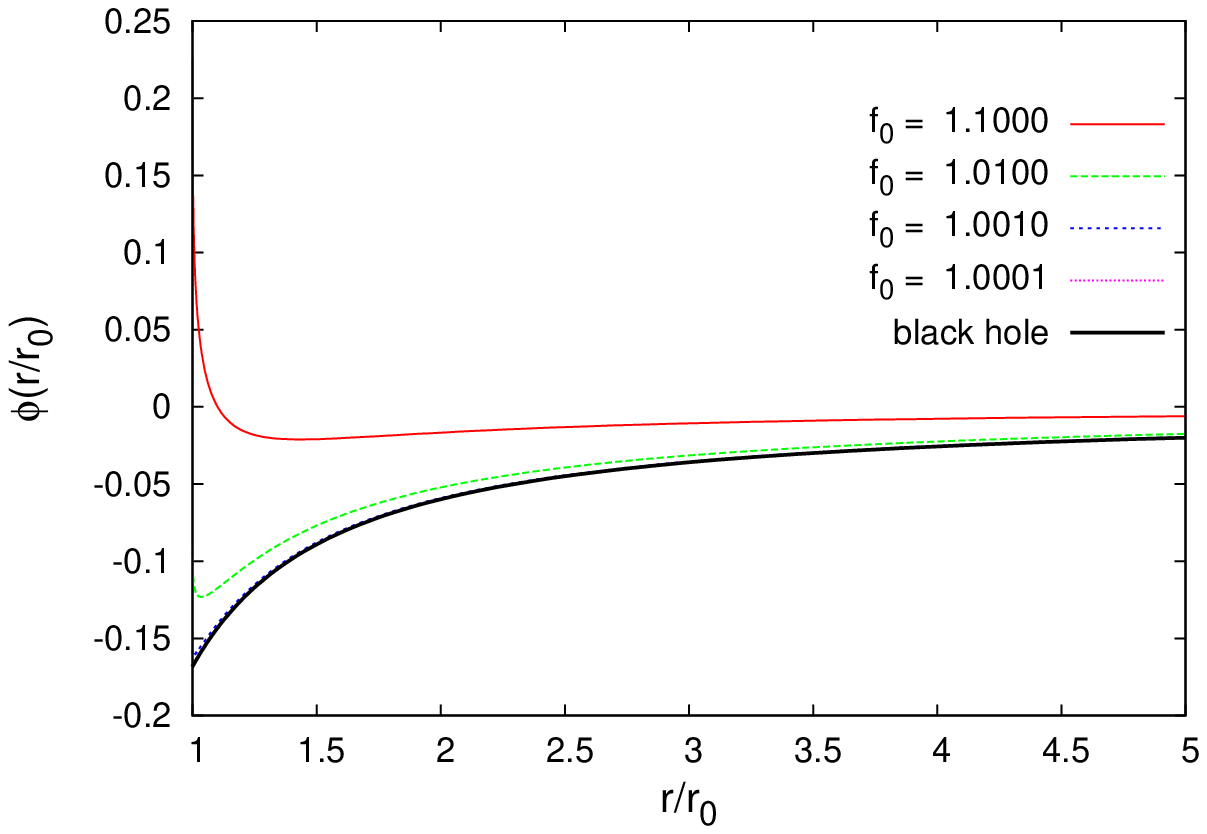}}
\end{center}
\hspace{1.0cm} {(a)} \hspace{8cm} {(b)} \hspace{2cm} \\[1mm]
\caption{
(a) Profile function $b/r_0$ and (b) dilaton function $\phi$ 
versus $r/r_0$ for $f_0=1.1$, $1.01$, $1.001$, $1.0001$ 
and $\alpha/r_0^2=0.02$ together with the limiting black hole functions.}
\end{figure}

To study the black hole limit in more detail 
let us consider a sequence of wormhole solutions
approaching the black hole solution.
We demonstrate this limiting behaviour for solutions 
with $\alpha=0.02$ in Fig.~\ref{f-5}. 
Here, we exhibit a sequence of solutions with values 
of $f_0$ tending to one.
Clearly, the profile function $b(r)$ (Fig.~\ref{f-5}a) and 
the dilaton function $\phi(r)$ (Fig.~\ref{f-5}b)
tend fast to the limiting black hole solutions. 
For $f_0=1.0001$ the wormhole functions
and their black hole counterparts are already very close.
The redshift functions $\nu$ are not distinguishable for this set of solutions
except very close to the throat, where $\nu_0$ diverges in the limit.

\subsubsection{Large $f_0$ limit}

\begin{figure}[h!]
\lbfig{f-6}
\begin{center}
\mbox{\includegraphics[height=.27\textheight, angle =0]{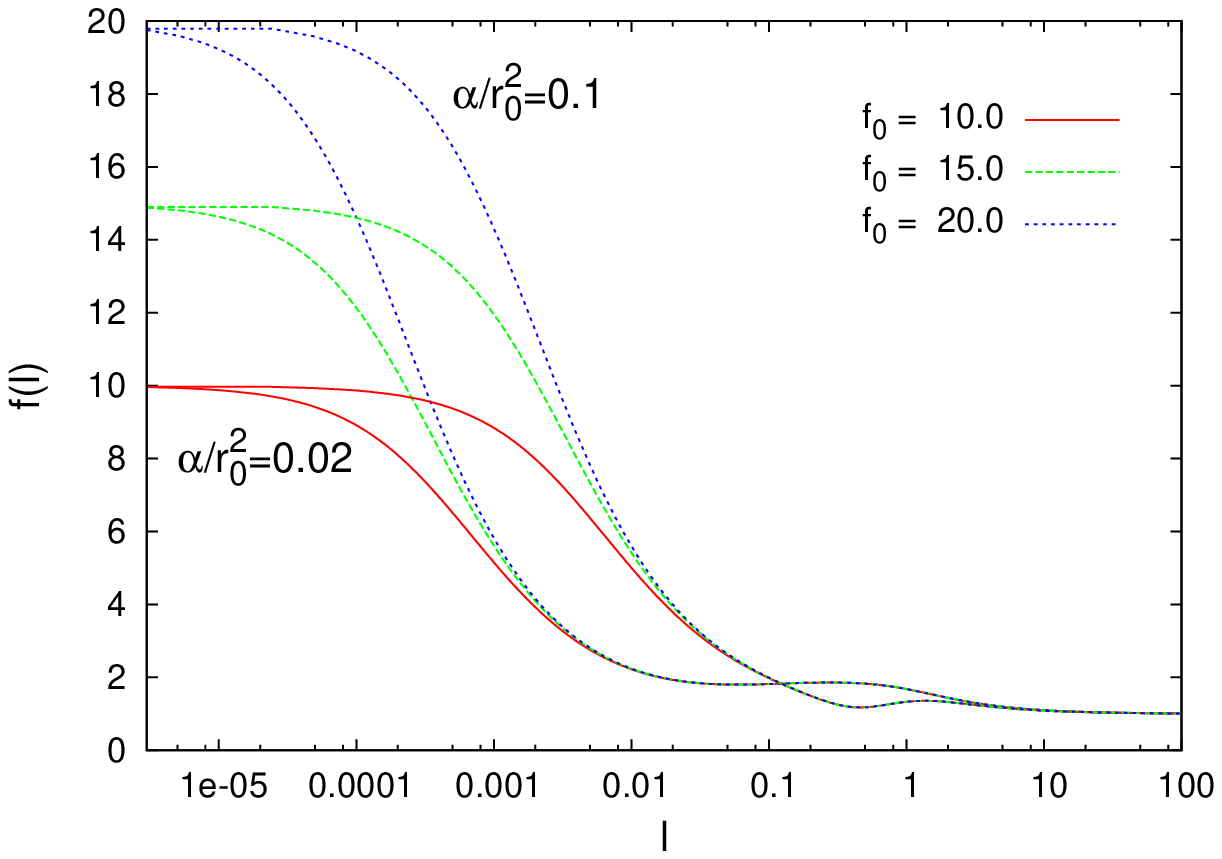}
      \includegraphics[height=.27\textheight, angle =0]{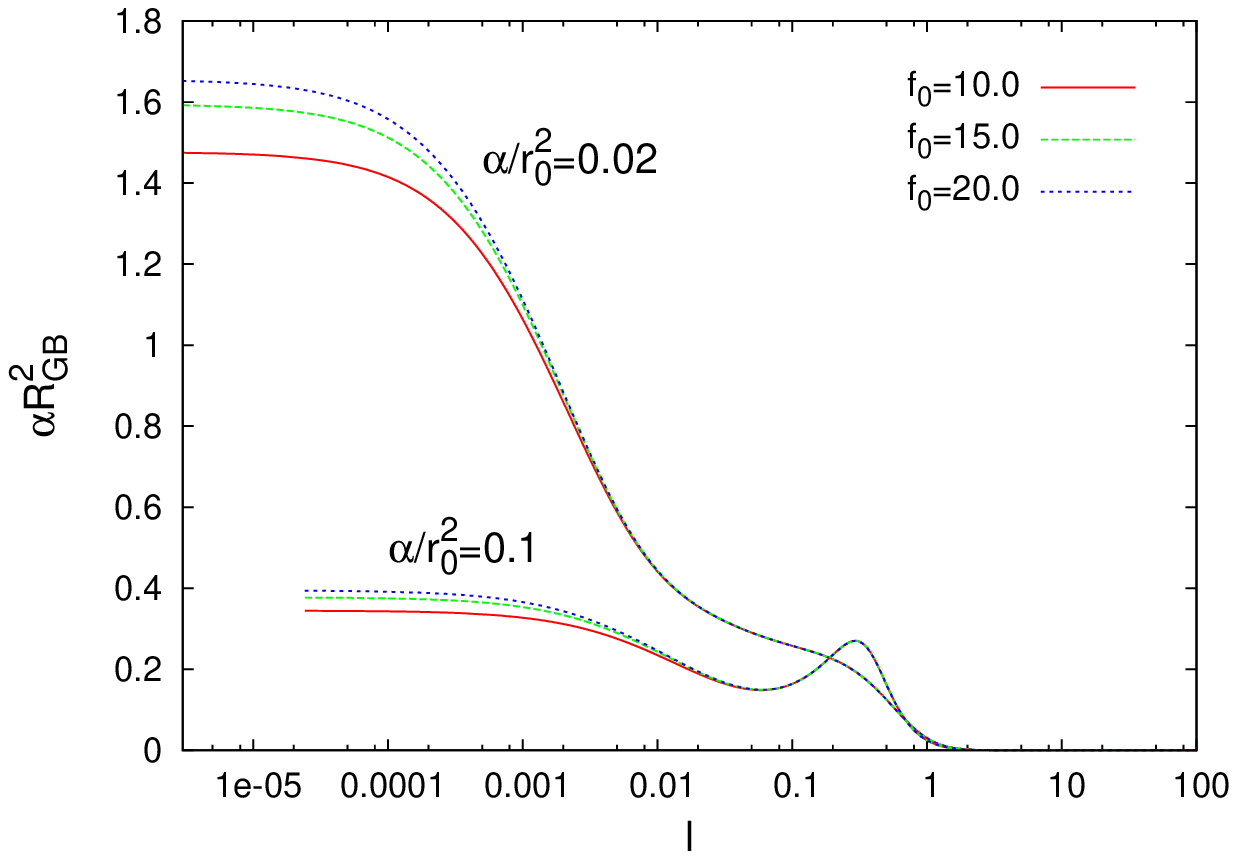}}
\hspace*{0.8cm} {(a)} \hspace{8cm} {(b)} \hspace{2cm} \\[1mm]
\end{center}
\caption{
(a) Metric function $f(l)$ and (b) scaled GB term $\alpha R_{\rm GB}^2$
versus $l$ for
$f_0=10.0$, $15.0$, $20.0$ and $\alpha/r_0^2=0.02$, $0.1$.}
\end{figure}

Next we consider the limit of large $f_0$, 
indicated by crosses in Figs.~\ref{f-3}.
As $f_0 \to \infty$,
the mass $M$ and the dilaton charge $D$ assume finite values.
The same holds for the redshift function $\nu_0$
and the dilaton field $\phi_0$ at the throat,
although the derivative of the dilaton field $\phi_0'$
with respect to the coordinate $l$ diverges like $\sqrt{f_0}$,
as seen from the boundary condition Eq.~(\ref{BC_phi}).

As an example we exhibit the function $f(l)$
in Fig.~\ref{f-6}a for two values of $\alpha/r_0^2$ 
and increasing values of $f_0$.
We observe that for a given value of $\alpha/r_0^2$,
the functions $f(l)$ deviate from each other 
only for small values of $l$ close to the throat, 
but coincide for larger $l$. 
Moreover, the region where the functions coincide increases
with increasing values of $f_0$. 
Thus the solutions approach a limiting solution for $f_0 \to \infty$,
which depends on $\alpha/r_0^2$.

This limiting behaviour is even more pronounced
in the scaled GB term $\alpha R_{\rm GB}^2$,
a curvature invariant which is exhibited
in  Fig.~\ref{f-6}b for the same set of parameters.
Extrapolating the value of $\alpha R_{\rm GB}^2$ at the throat
for large $f_0$ indicates
that $\alpha R_{\rm GB}^2$ remains finite for $f_0 \to \infty$.


\subsubsection{Singularity limit}


\begin{figure}[h!]
\lbfig{f-7}
\begin{center}
\mbox{(a) \includegraphics[height=.27\textheight, angle =0]{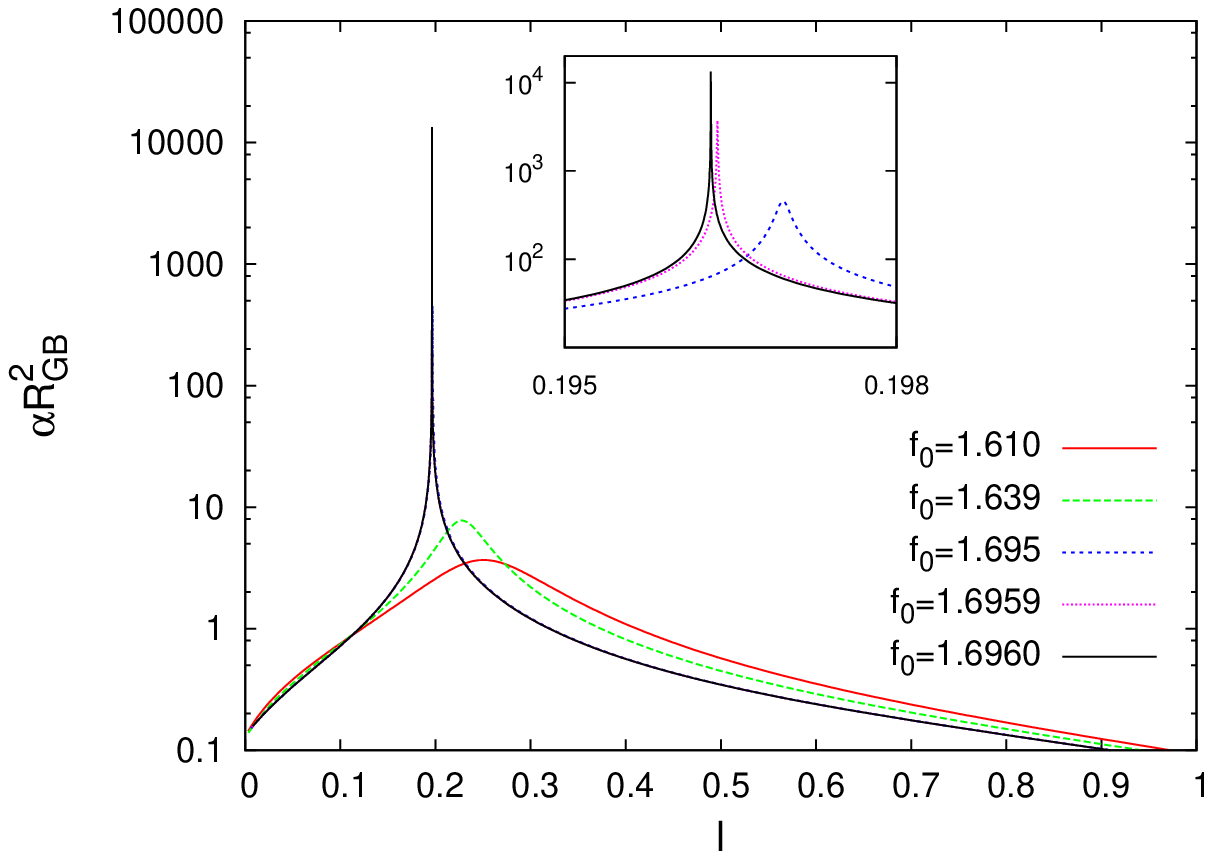}
      (b) \includegraphics[height=.27\textheight, angle =0]{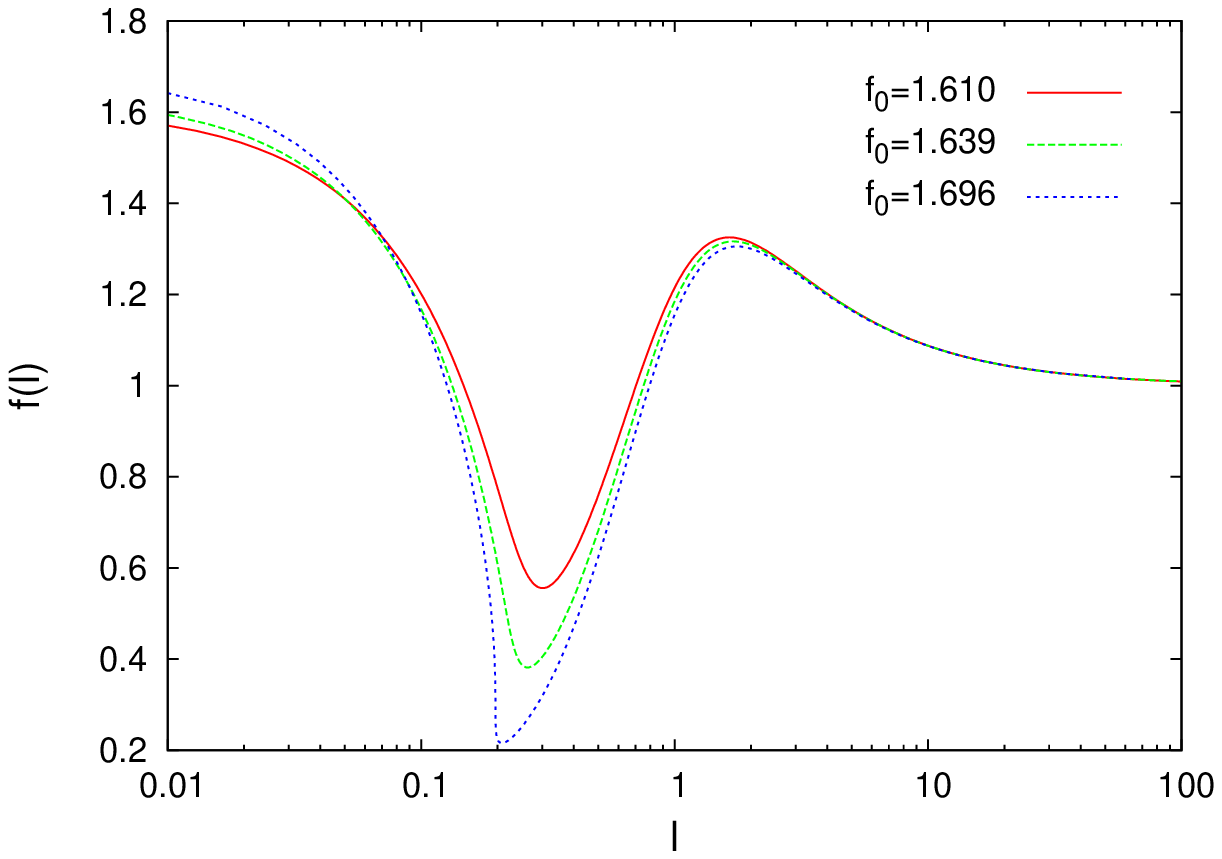}}
\mbox{(c) \includegraphics[height=.27\textheight, angle =0]{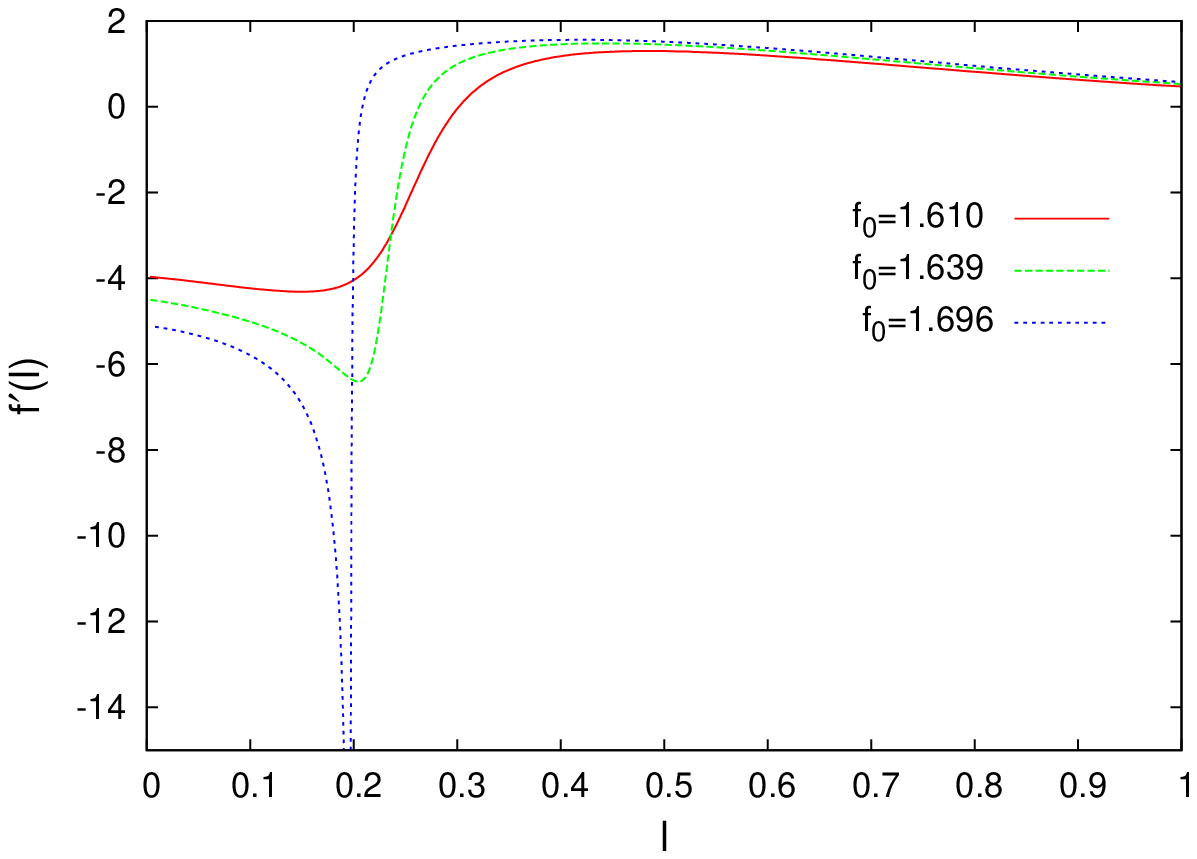}
      (d) \includegraphics[height=.27\textheight, angle =0]{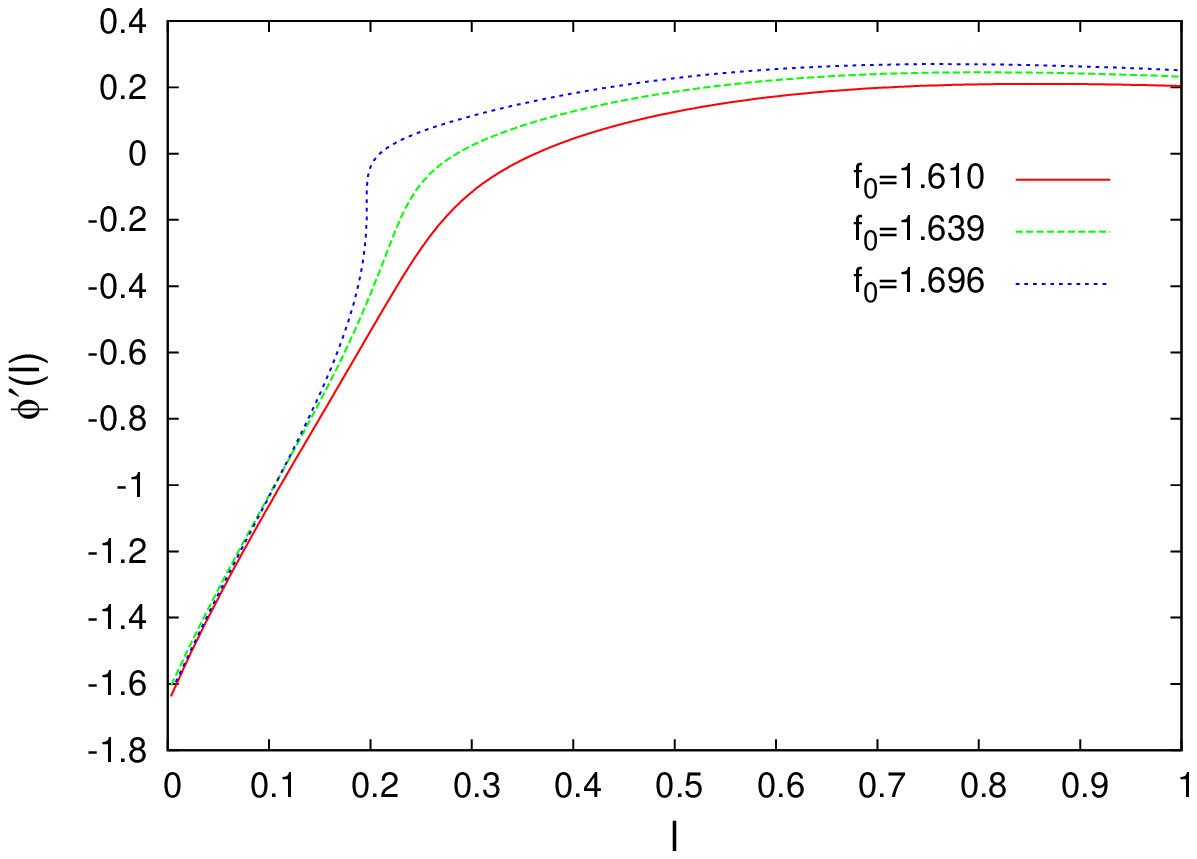}}
\end{center}
\caption{
(a) Scaled GB term $\alpha R_{\rm GB}^2$ versus $l$,
(b) metric function $f(l)$, (c) derivative $f'(l)$,
and (d) derivative of the dilaton function $\phi'(l)$
for several values of $f_0$ and $\alpha/r_0^2=0.1$
demonstrating the emergence of a curvature singularity.
}
\end{figure}

Finally we turn to the third boundary curve, 
indicated by dots in Figs.~\ref{f-3}.
This boundary emerges when branches of solutions 
with fixed $\alpha/r_0^2$ terminate at singular configurations. 
We demonstrate that these singular configurations possess a curvature singularity
at a critical value of $l$, $l_{\rm crit}$, in Fig.~\ref{f-7}a.
Here the scaled GB term $\alpha R_{\rm GB}^2$ is shown versus $l$
for $\alpha/r_0^2=0.1$ for a sequence of solutions approaching the
singular configuration. 
We observe a sharp peak in the vicinity of $l_{\rm crit}$, which increases in
size and diverges as the singular configuration is approached.

To elucidate this emergence of a curvature singularity at some
finite critical value of $l$,
we show in Fig.~\ref{f-7}b the metric function $f(l)$,
in Fig.~\ref{f-7}c its derivative $f'(l)$,
and in Fig.~\ref{f-7}d the derivative of the dilaton function $\phi'(l)$
for a fixed value of $\alpha/r_0^2=0.1$ and increasing
values of $f_0$ approaching the singular configuration. 
We note that all functions are continuous, 
but as the singular configuration is approached, their derivatives develop 
a discontinuity at some point $l_{\rm crit}$ outside the throat. 
In fact the derivative of the function $f$ has a pole at $l_{\rm crit}$.

\subsection{Energy conditions}

We now turn to the energy conditions, Eqs.~(\ref{Nulleng}).
We demonstrate the violation of the null energy condition in Figs.~\ref{f-9}.
Here we show `normalized' quantities in order to emphasize the change of sign.
The normalization factor is 
\begin{equation}
N= \sqrt{(T_0^0)^2+(T_l^l)^2+2(T_\theta^\theta)^2} \ . 
\end{equation} 
This normalization is responsible for the steep rise observed in the figures
for $f_0=1.1$, which occurs when all quantities are close to zero.

The null energy condition is violated 
close to the throat for all wormhole solutions,
as seen explicitly from Eq.~(\ref{eng_pert_cond})
and demonstrated for the set of solutions of Fig.~\ref{f-9}a.
However, for large values of $f_0$ and at the same time
small values of $\alpha/r_0^2$
the dilaton charge $D$ can become negative. 
In this case the null energy condition is violated also
in the asymptotic region, as seen from Eq.~(\ref{term2_as}).
In fact we observe that 
for the larger values of $f_0$ in Fig.~\ref{f-9}b 
the combination $-T_0^0+T_\theta^\theta$ is negative everywhere.
Thus in this case the null energy condition is violated in the whole spacetime.

\begin{figure}[t]
\lbfig{f-9}
\begin{center}
\end{center}
\mbox{\includegraphics[height=.27\textheight, angle =0]{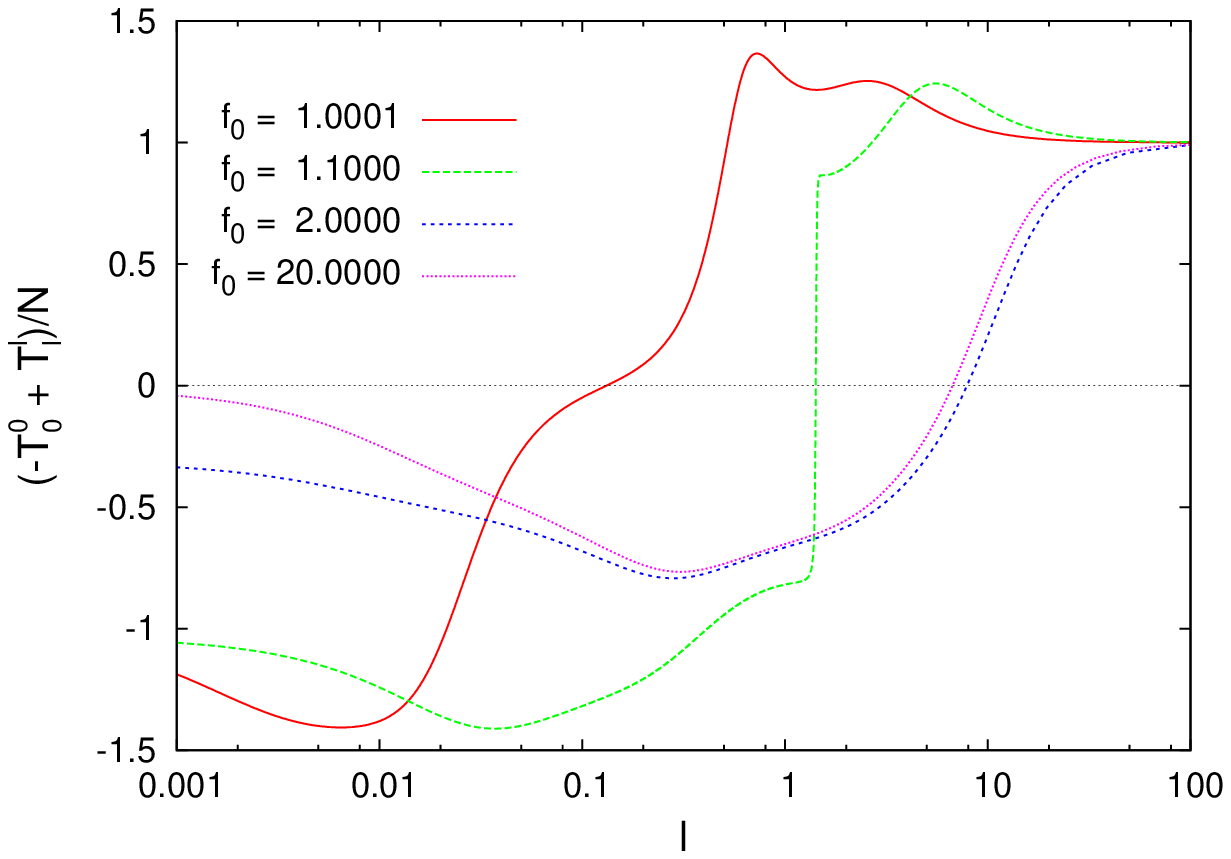}
      \includegraphics[height=.27\textheight, angle =0]{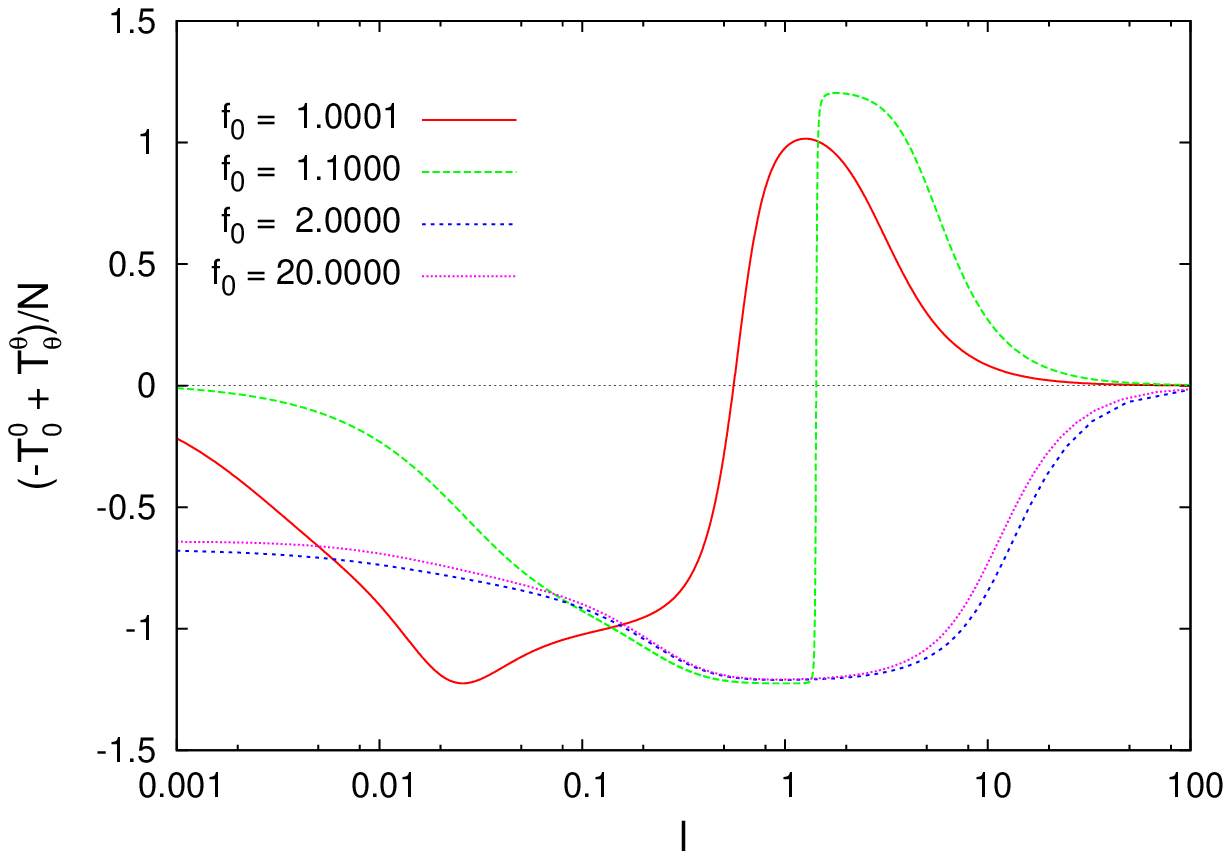}}
\hspace*{1.0cm} {(a)} \hspace{8cm} {(b)} \hspace{2cm} \\[1mm]
\caption{
Demonstration of the violation of the null energy condition for 
$\alpha/r_0^2=0.02$ and several values of $f_0$.}
\end{figure}

\subsection{Smarr relation}

\begin{figure}[t]
\begin{center}
\mbox{\includegraphics[height=.27\textheight, angle =0]{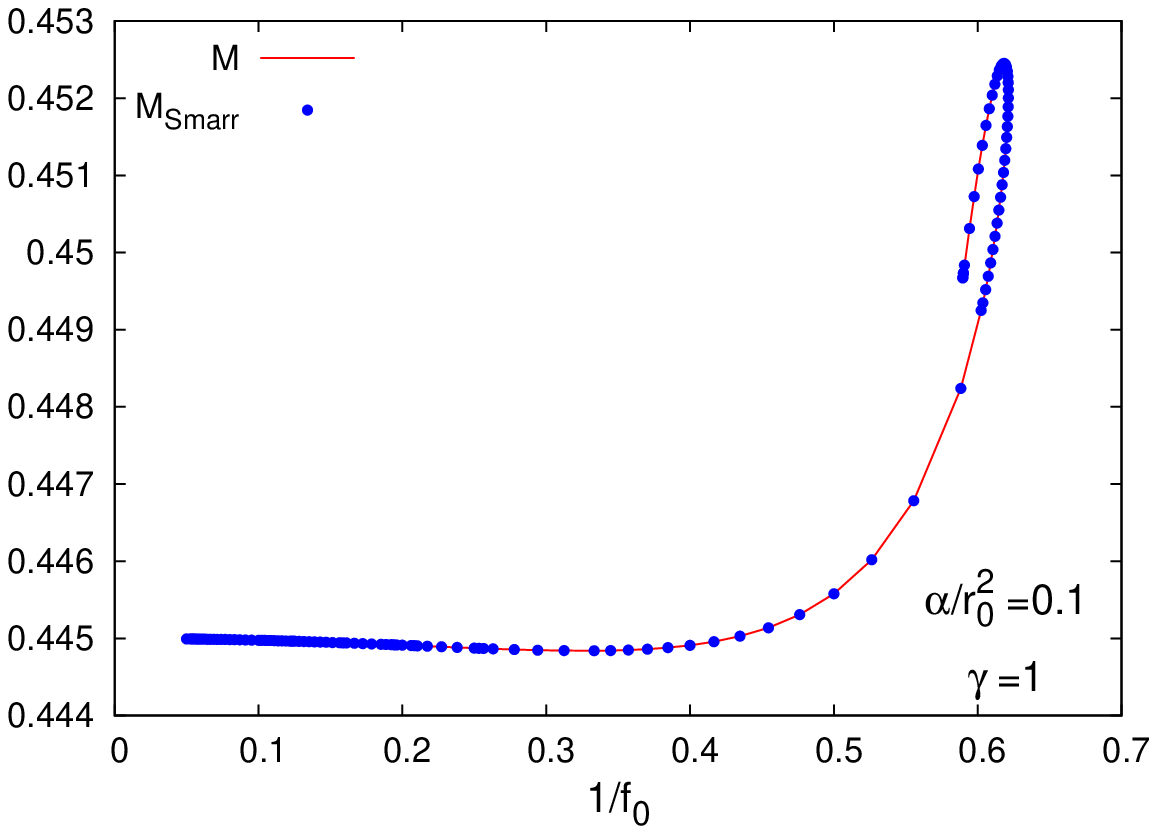}
\includegraphics[height=.27\textheight, angle =0]{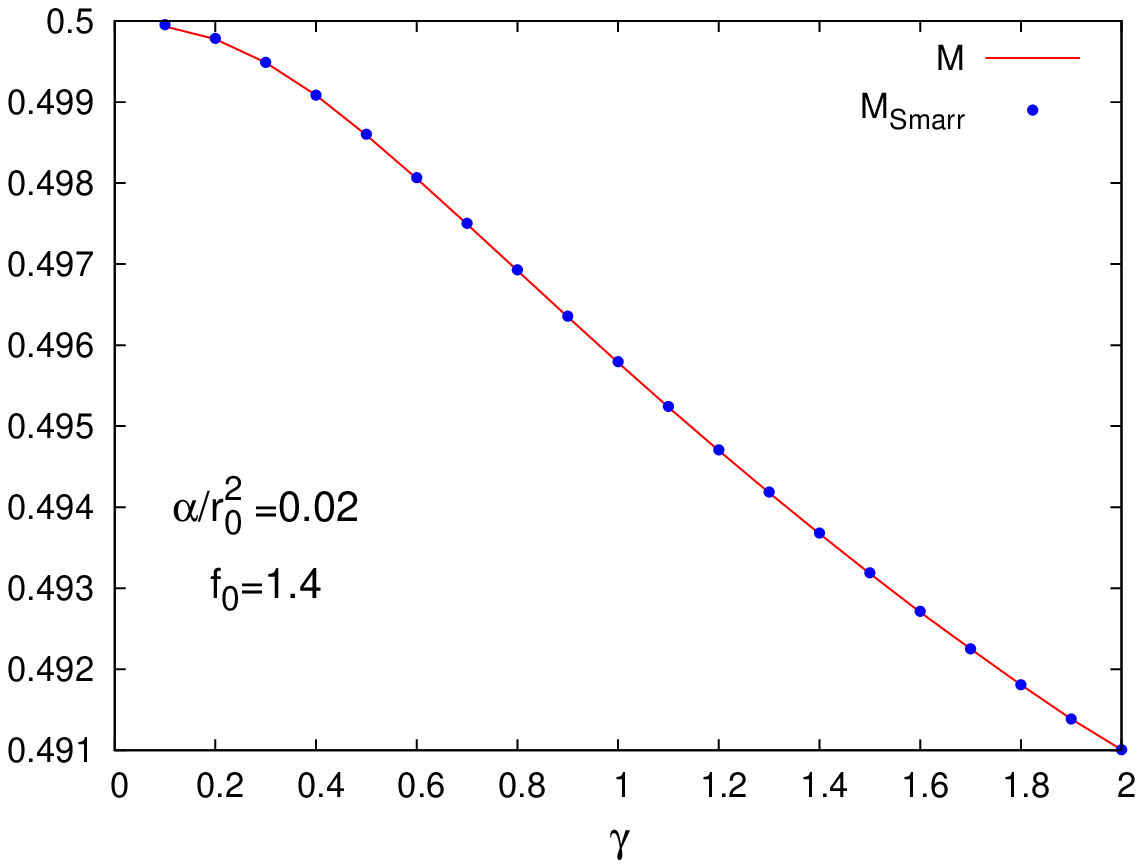}}
\hspace*{1.0cm} {(a)} \hspace{8cm} {(b)} \hspace{2cm} \\[1mm]
\caption{Check of the Smarr relation for several sets of wormhole solutions:
(a) mass $M$ versus $1/f_0$ for $\alpha/r_0^2=0.1$, $\gamma=1$,
(b) mass $M$ versus $\gamma$ for $\alpha/r_0^2=0.02$, $f_0=1.4$.}
\lbfig{smarrfig}
\end{center}
\end{figure}

Let us finally turn to the Smarr-like mass relation,
Eq.~(\ref{smarr}).
This mass relation is a perfect check for the
numerical accuracy and thus for the quality of the solutions.
We demonstrate the Smarr relation in Figs.~\ref{smarrfig}
by comparing the mass $M$ obtained from the asymptotic fall-off of the
metric functions with the mass $M_{\rm Smarr}$ obtained
by evaluating the various terms in the Smarr relation Eq.~(\ref{smarr}).

The Smarr-like formula is well satisfied for all sets of solutions.
In Fig.~\ref{smarrfig}a, we keep $\gamma=1$ as in the main body of the paper
and exhibit the masses $M$ and $M_{\rm Smarr}$ versus $1/f_0$
for $\alpha/r_0^2=0.1$.
The relative error in this case is below $3 \cdot 10^{-5}$
(for $\alpha/r_0^2=0.05$ it is below $10^{-5}$,
and for $\alpha/r_0^2=0.02$ it is below $5 \cdot 10^{-6}$).

In  Fig.~\ref{smarrfig}b we have fixed $\alpha/r_0^2=0.02$, $f_0=1.4$
and varied $\gamma$ to address the $\gamma$-dependence of the
mass formula for a set of solutions.
Also for these solutions the relative error is small
and, in particular, it is below $4 \cdot 10^{-5}$.

\begin{figure}[h!]
\lbfig{f-smarr}
\begin{center}
\mbox{\includegraphics[height=.27\textheight, angle =0]{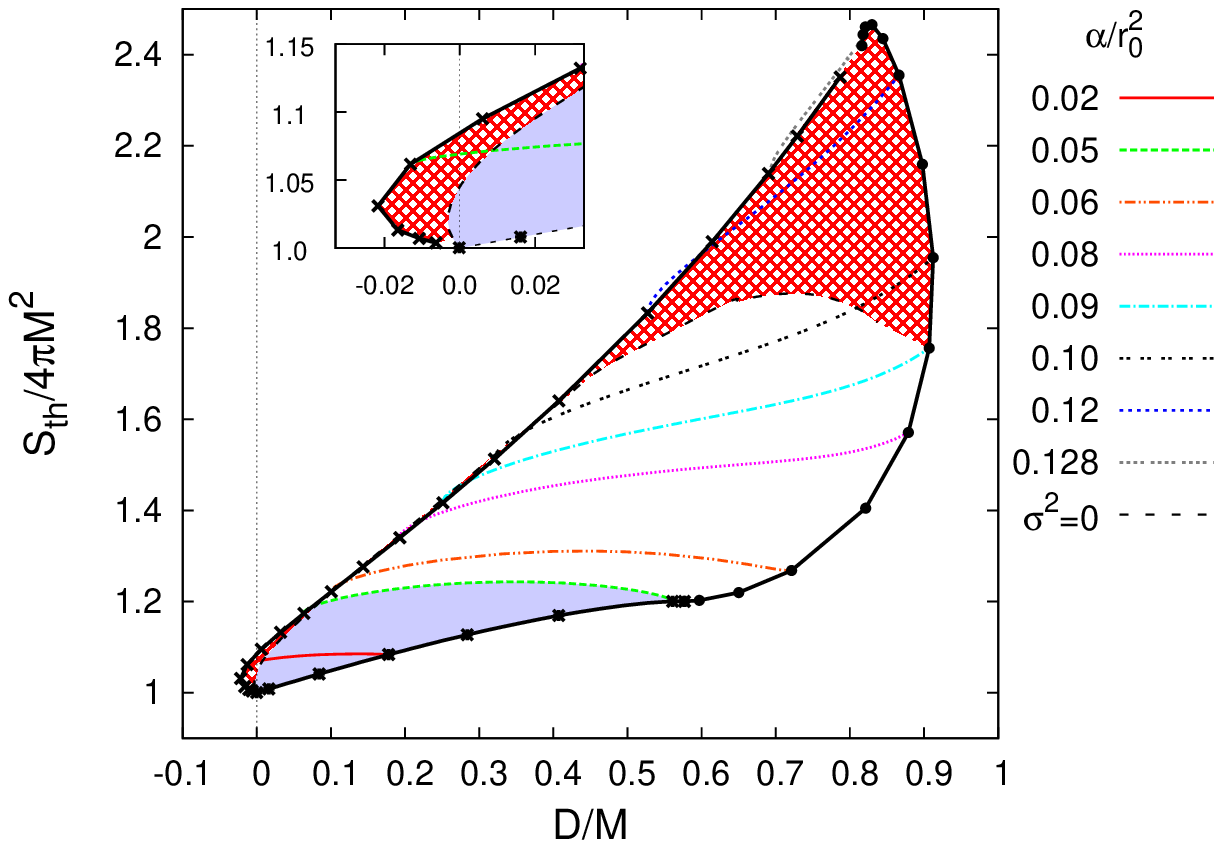}
      \includegraphics[height=.27\textheight, angle =0]{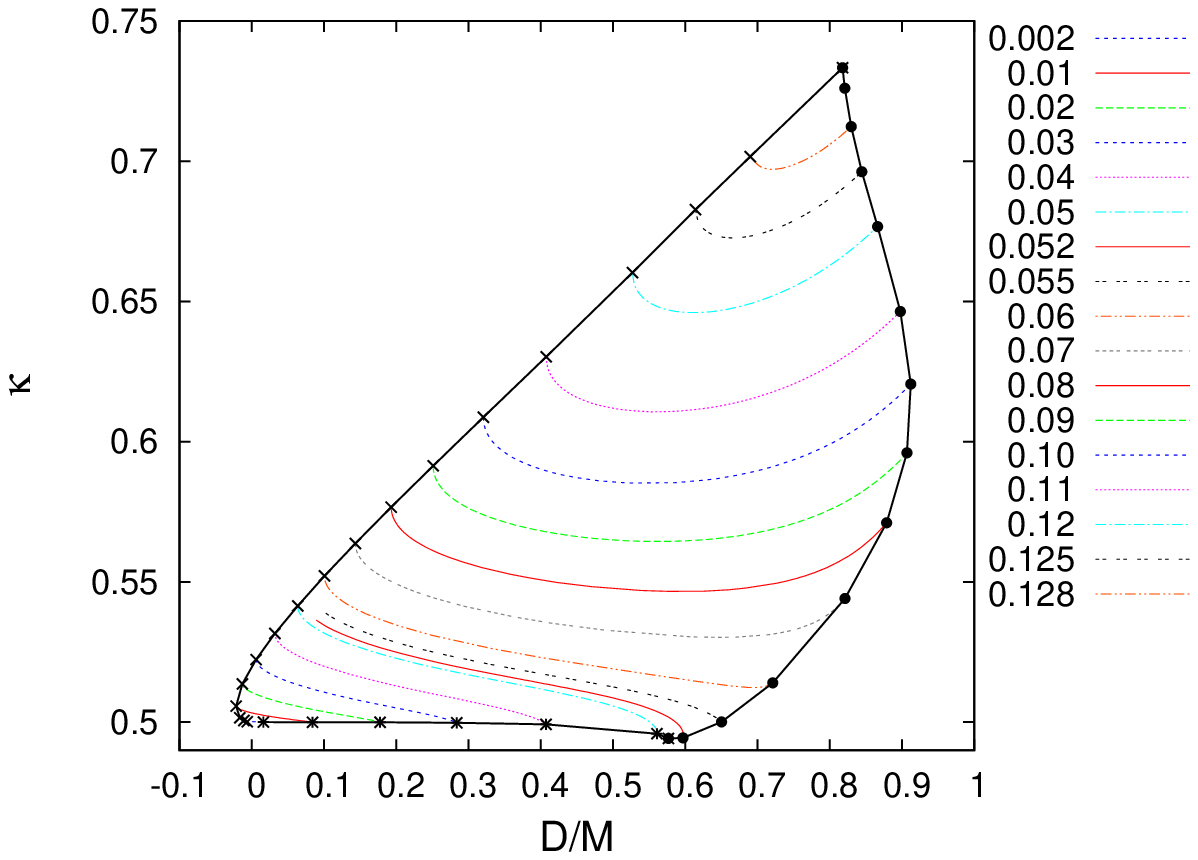}}
\hspace{0.5cm} {(a)} \hspace{8cm} {(b)} \hspace{2cm} \\[1mm]
\mbox{\includegraphics[height=.27\textheight, angle =0]{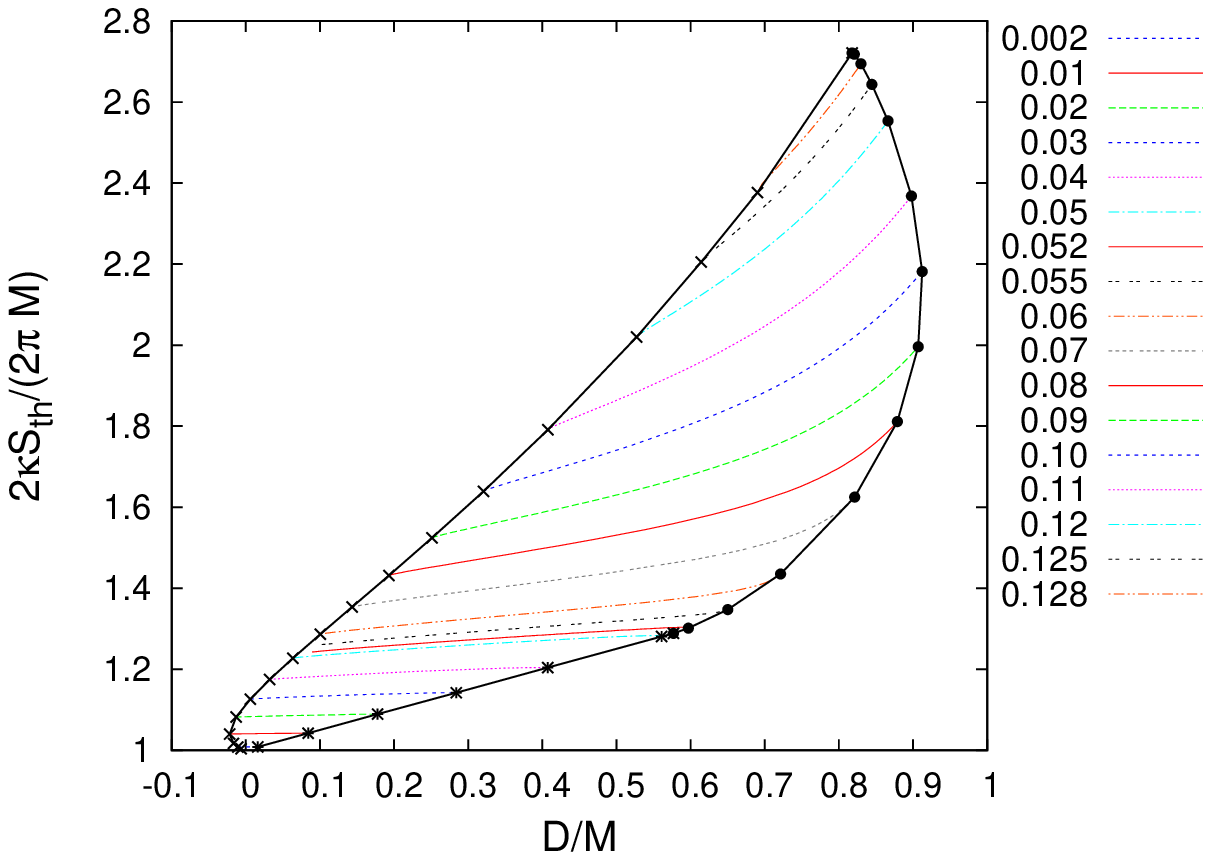}
      \includegraphics[height=.27\textheight, angle =0]{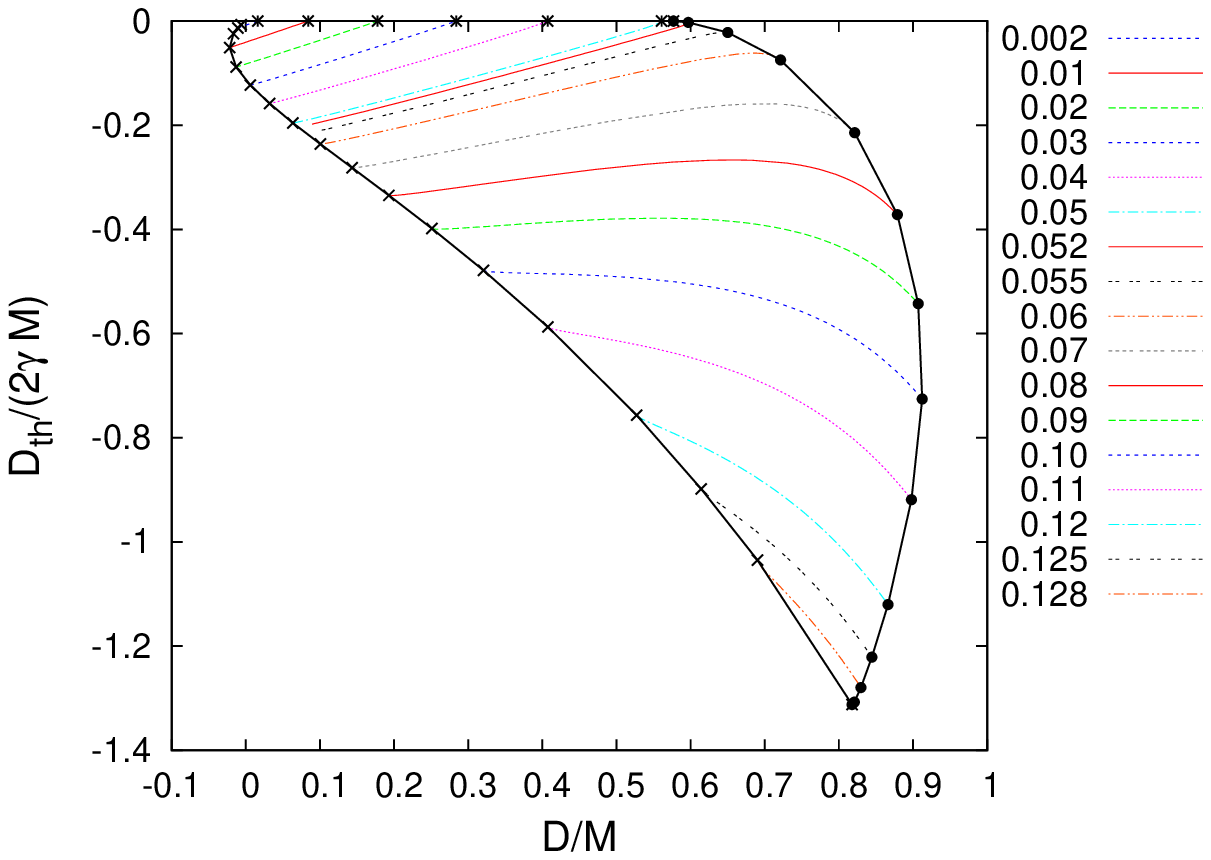}}
\hspace{0.5cm} {(c)} \hspace{8cm} {(d)} \hspace{2cm} \\[1mm]
\end{center}
\caption{
(a) Scaled entropy-analogue throat quantity $S_{\rm th}/4 \pi M^2$ 
versus the scaled dilaton charge $D/M$
for several values of $\alpha/r_0^2$.
The boundaries represent EGBd black holes (asterisks),
limiting solutions with $f_0\to \infty$ (crosses)
and solutions with curvature singularities (dots).
The shaded areas indicate linear stability (lilac or lower), instability (red or upper),
undecided yet (white) w.r.t.~radial perturbations
(Sect.~IV).
(b) Scaled surface gravity $\kappa r_0$ at the throat
versus the scaled dilaton charge $D/M$;
(c) scaled first term of the mass formula;
(d) scaled third term of the mass formula.
}
\end{figure}

To gain a better understanding of the quantities contributing in the
Smarr formula, we now consider the various throat properties.
First, we exhibit in Fig.~\ref{f-smarr}a the scaled throat quantity
$S_{\rm th}/4 \pi M^2$ versus the scaled dilaton charge $D/M$
for the full domain of existence of the wormhole solutions.
$S_{\rm th}/4 \pi M^2$ resembles in its structure the entropy
of black holes, since it contains the same correction term
to the throat area $A_{\rm th}$ as in the black hole case
to the horizon area.
In the black hole limit, $S_{\rm th}$ assumes the meaning
of the black hole entropy.

The scaled entropy-analogue throat quantity $S_{\rm th}/4 \pi M^2$
is always greater than one. Only in the Schwarzschild black hole limit,
where the dilaton vanishes, it assumes precisely the value one. 
A comparison with Fig.~\ref{f-3}a shows, that 
the quantity $S_{\rm th}/16 \pi M^2$ is always larger than
the scaled throat area $A/16 \pi M^2$. This was observed before
for black holes \cite{Kanti:1995vq,Kleihaus:2011tg}.
The color coding of Fig.~\ref{f-smarr}a is related to the stability
of the solution, discussed in the next section.

The scaled surface gravity $\kappa r_0$ at the throat 
of the wormhole solutions is exhibited in Fig.~\ref{f-smarr}b.
It is bounded, with its lower boundary given by the
black hole values, which remain close to the Schwarzschild value of 1/2.
The product of the quantities $S_{\rm th}$ and $\kappa$
represents the first term of the Smarr formula, up to numerical factors.
We exhibit this first term, divided by the mass, in 
Fig.~\ref{f-smarr}c.
Since the first term is always greater than one
(except for the Schwarzschild case),
the remaining pieces of the mass formula must contribute
negatively to cancel this excess.

The second term in the mass formula, $-D/2\gamma$,
is negative, except for the small region with
large $f_0$ and small $\alpha/r_0^2$.
Its contribution to the scaled mass formula can be read off the 
horizontal axis of Fig.~\ref{f-smarr}c.
We exhibit the third term of the scaled mass formula
in Fig.~\ref{f-smarr}d.
It contains the scaled throat dilaton charge, $D_{\rm th}$,
which is modified by the GB term,
analogous to the modification of the area by the GB term.
As expected, this contribution is negative for the wormhole solutions.
It vanishes for the black holes, which represent the upper limit
for $D_{\rm th}$.

\section{Stability}

Our starting point for the study
of the stability is the line element for spherically symmetric
solutions 
\begin{equation}
ds^2 =  -e^{2\tilde{\nu}}dt^2+\tilde{f}dl^2
+(l^2+r_0^2)\left(d\theta^2+\sin^2\theta d\varphi^2 \right) \,.
\label{metricLL} 
\end{equation}
Here we consider only the pulsation modes.
Thus we allow the metric and dilaton functions to depend on the radial
coordinate $l$ and the time coordinate $t$,
\begin{equation}
\tilde{\nu} = \tilde{\nu}(l,t) \ , \ \ \ \ 
\tilde{f} = \tilde{f}(l,t) \ , \ \ \ \ 
\tilde{\phi} = \tilde{\phi}(l,t) \ .
\label{tdepfun} 
\end{equation}

For the study of the stability behaviour of our solutions, we also need 
the time-dependent Einstein and dilaton equations.
These were presented in \cite{Kanti:1998} in the context of the stability analysis
of the dilatonic black hole solutions, and thus we refrain from repeating them here
\footnote{In \cite{Kanti:1998}, the line-element (\ref{metricS}) was used for the
derivation of the time-dependent field equations. 
The new set of equations with respect to the new coordinate $l$
may be derived under the redefinitions
$\Gamma \equiv 2 \nu$, $\Lambda\equiv \ln[r^2 f/l^2]$
and the change of variable $dl/dr=r/l$.
Also, the following changes should be made
due to the different conventions used: $\phi \rightarrow -\phi$ and
$\alpha' e^{\phi}/(4g^2) \rightarrow  \alpha e^{-\gamma \phi}$.}.
Next, we decompose the metric and dilaton functions into the unperturbed 
functions and the perturbations
\begin{eqnarray}
 \tilde{\nu}(l,t) & = & \nu(l) +\epsilon \delta\nu(l) e^{i\sigma t} \ , 
\label{decomp_nu}\\
 \tilde{f}(l,t) & = & f(l) +\epsilon \delta f(l) e^{i\sigma t} \ , 
\label{decomp_f}\\
 \tilde{\phi}(l,t) & = & \phi(l) +\epsilon \delta\phi(l) e^{i\sigma t} \ , 
\label{decomp_phi}
\end{eqnarray}
where we assume a harmonic time dependence of the perturbations 
and $\epsilon$ is considered as small.
Now we substitute the perturbed functions into the Einstein and dilaton 
equations and linearize in $\epsilon$.
This yields a system of linear ODEs for the functions 
$\delta\nu(l)$, $\delta f(l)$ and $\delta\phi(l)$, where the 
coefficients depend on the unperturbed functions and their derivatives.
We use the unperturbed equations to eliminate $\nu'$, $\nu''$,
$f'$ and $\phi''$.

The $tl$ part of the Einstein equations yields $\delta f$ 
in terms of $\delta\phi$ and $\delta\phi'$. Thus $\delta f$ and $\delta f'$ 
can be eliminated from the rest of the ODEs. The dilaton equation and the 
$\theta\theta$ part of the Einstein equations can be diagonalized with 
respect to $\delta \nu''$ and  $\delta \phi''$. Finally 
$\delta \nu'$ can be eliminated from the dilaton equation by adding
the $rr$ part of the Einstein equations with a suitable factor.
Thus we end up with a single second order equation for $\delta \phi$,
\begin{equation}
\delta \phi'' + q_1 \delta\phi' + (q_0+q_\sigma\sigma^2) \delta\phi = 0 \ , 
\label{lineq1}
\end{equation}
where the coefficients $q_1$, $q_0$ and $q_\sigma$ depend on the unperturbed solution. 

The coefficient $q_\sigma$ tends to one asymptotically
and is bounded at $l=0$.
The coefficients $q_1$ and $q_0$ tend to zero asymptotically, however, they
diverge at $l=0$ as $1/l$. Thus, to obtain solutions which are regular at $l=0$
suitable boundary conditions are needed. 
Since the perturbations are assumed to be normalizable, $\delta\phi$
has to vanish asymptotically. In order to obtain a unique solution
we also fix $\delta \phi$ at $l=0$. Thus we are left with three boundary
conditions for a second order ODE, which can be satisfied only for certain
values of the eigenvalue $\sigma^2$.  

We can avoid the singularity of $q_1$ and $q_0$ at $l=0$, by employing
the transformation 
$\delta \phi = F(l) \psi(l)$, where $F(l)$ satisfies 
$F'/F = -q_1(l)/2$. This yields 
\begin{equation}
\psi'' + Q_0\psi +\sigma^2 q_\sigma \psi = 0 \ , 
\label{lineq2}
\end{equation}
where the coefficient $Q_0=-q'_1/2-q_1^2/4+q_0$ is bounded at $l=0$.
We note that $F(l)$ diverges like $1/l$ at $l=0$. Thus for acceptable solutions
we have to impose the boundary condition $\psi=0$ at $l=0$. In addition 
$\psi$ has to vanish asymptotically to ensure normalized solutions.
However, these boundary conditions do not lead to unique solutions, since 
the ODE (\ref{lineq2}) is homogeneous. Therefore, we supplement 
Eq.~(\ref{lineq2}) with the auxiliary ODE $N' = \psi^2$ and impose the 
boundary conditions $N(0)=0$ and $N \to 1$ asymptotically.
These conditions give exactly the normalization
$\int_0^\infty \psi^2 dl = 1$.
Again, solutions exist only for certain values of the eigenvalue $\sigma^2$.

The ODE (\ref{lineq2}) can be written as a Schr\"odinger equation by
introducing the new coordinate $y$ via $dy/dl = \sqrt{q_\sigma}$,
and eliminating the first derivative $d\psi/dy$ as above.
This yields
\begin{equation}
\frac{d^2\chi}{dy} - V_{\rm eff}\chi +\sigma^2  \chi = 0 \ , 
\label{lineq3}
\end{equation}
where
$$
 - V_{\rm eff} = 
 \frac{1}{2\sqrt{q_\sigma}}\left(\frac{1}{\sqrt{q_\sigma}}\right)''
-\frac{1}{4}\left[\left(\frac{1}{\sqrt{q_\sigma}}\right)'\right]^2 
+\frac{Q_0}{q_\sigma} \ ,
$$
and $\chi$ is related to $\psi$. The new potential $V_{\rm eff}$ is 
bounded for all $y$.

Note that the derivation of Eq.~(\ref{lineq3}) holds only if the
function $q_\sigma$ is positive for all $l$. 
However, if $q_\sigma<0$ on some interval,
one cannot rely on the rule that the smallest eigenvalue  
corresponds to the eigenfunction without knots. 
Indeed we found in some cases
two different eigenfunctions $\psi_1$, $\psi_2$ without knots which 
satisfy the orthogonality condition $\int q_\sigma \psi_1 \psi_2 dl =0$.

In order to determine the change of stability 
we consider families of wormhole solutions with fixed $\alpha/r_0^2$. 
The ODE (\ref{lineq1}) is then solved
together with the normalization constraint for varying values of $f_0$. 
In Fig.~\ref{f-pert}(a) we show the eigenvalue $\sigma^2$ versus $f_0$
for several values of $\alpha/r_0^2$. 
We observe that solutions for negative values of $\sigma^2$ 
exist for large values of $f_0$, corresponding to
an instability of the wormhole solutions. 
For these solutions the function $\delta\phi$, 
respectively $\psi$, decays exponentially.
Thus these solutions correspond to bound states of the equivalent
Schr\"odinger equation.
As $f_0$ decreases, the eigenvalue increases 
and tends to zero for some critical value 
$f_0=f_0^{\rm (cr)}$, which depends on $\alpha/r_0^2$.
No solutions were found for $f_0<f_0^{\rm (cr)}$. 
Solutions of Eq.~(\ref{lineq1}) with positive 
$\sigma^2$ would oscillate in the asymptotic region, and are not normalizable.

\begin{figure}[h!]
\lbfig{f-pert}
\begin{center}
\includegraphics[height=.25\textheight, angle =0]{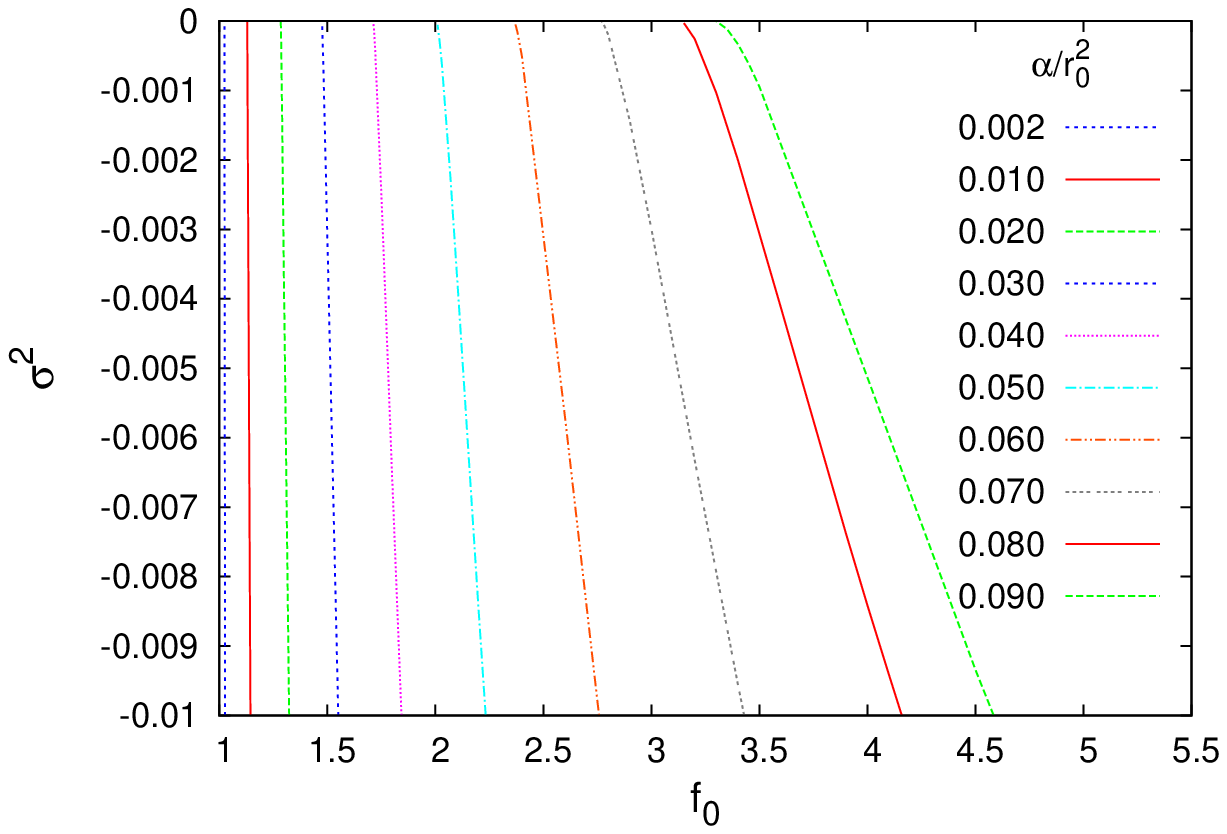}
\includegraphics[height=.25\textheight, angle =0]{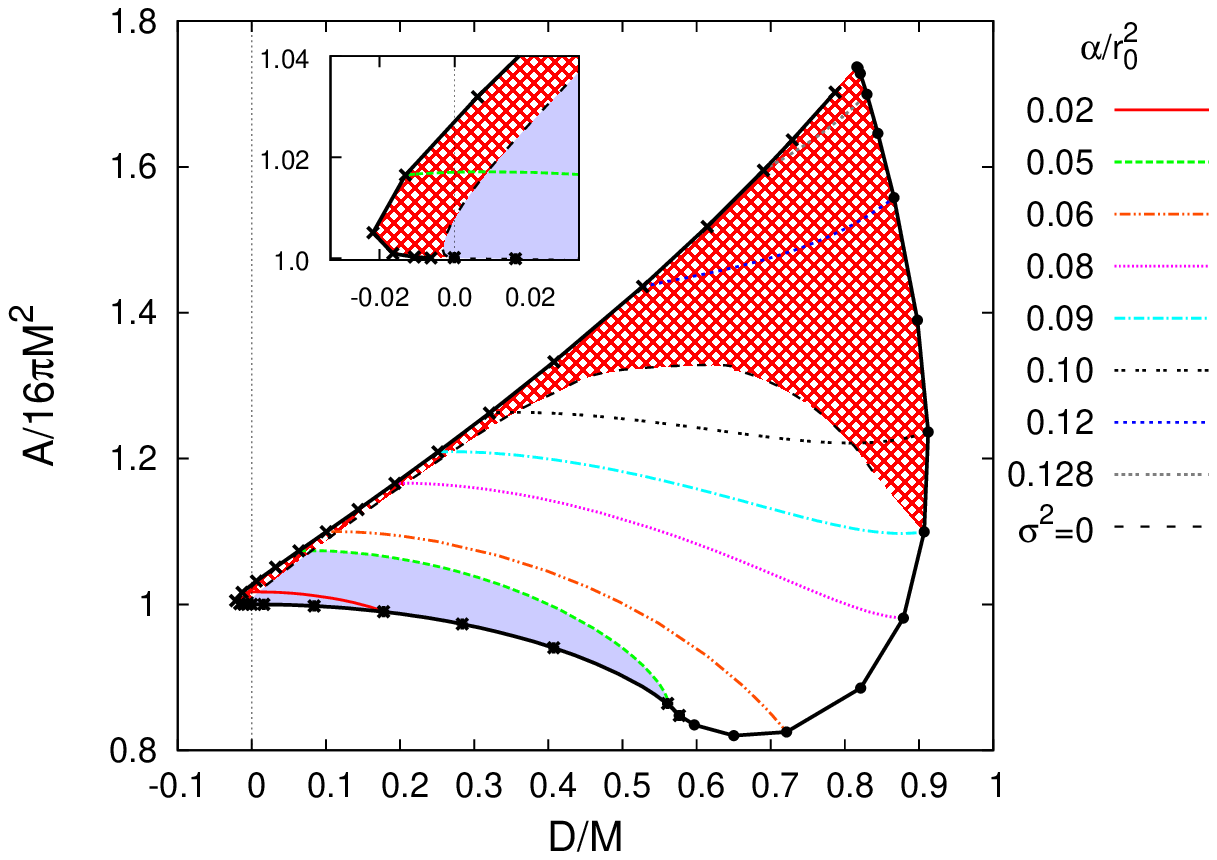}
\end{center}
\hspace*{0.5cm} {(a)} \hspace{8cm} {(b)} \hspace{2cm} \\[1mm]
\caption{
(a) Eigenvalue $\sigma^2$ versus $f_0$ for $\alpha/r_0^2=0.02$.
(b) Domain of existence: scaled area $A/16 \pi M^2$ of the throat
versus the scaled dilaton charge $D/M$
for several values of $\alpha/r_0^2$.
The shaded areas indicate linear stability (lilac or lower), instability (red or upper),
undecided yet (white) w.r.t.~radial perturbations (taken from \cite{Kanti:2011jz}).
}
\end{figure}

The interpretation is that the instability mode vanishes 
at some critical value  $f_0^{\rm (cr)}$.
Hence we conclude that the wormhole solutions are stable for $f_0<f_0^{\rm (cr)}$. 
However, this holds only if $q_\sigma >0$ for all $l$. 
Otherwise one cannot be sure that $\sigma^2$ is the smallest possible eigenvalue. 
Therefore, eigenfunctions with negative eigenvalues
cannot be excluded even if $\sigma^2=0$. 
Consequently, if $q_\sigma$ is not positive for all $l$
the question of stability cannot be decided by considering the 
standard equivalent Schr\"odinger eigenvalue problem.

Numerically, we found that $q_\sigma$ is positive for all wormhole solutions 
with $\alpha/r_0^2 \leq 0.05$;
but wormhole solutions where $q_\sigma$ is negative on some interval exist 
if $\alpha/r_0^2 \geq 0.05$. 
In Fig.~\ref{f-pert}(b) we show the stability region in the domain of existence
\cite{Kanti:2011jz}. 
Also shown is the line where the eigenvalue $\sigma^2$ changes sign. 
The instability region (red or upper) is characterized by negative $\sigma^2$, 
whereas the stability region (lilac or lower) is characterized
by the absence of negative eigenvalues and positive $q_\sigma$. 
In the remaining (white) 
region the question of stability is not yet decided.

\section{Junction Conditions}

Up to this point, we have discussed the behaviour of the metric functions
and of the dilaton field in only half of the wormhole spacetime, i.e.~the part
with $l>0$.  Our solutions should naturally be extended to the second asymptotically
flat part of the manifold ($l \rightarrow -\infty$). If this is performed by
demanding that the derivatives of the metric and dilaton functions are continuous, 
we observe a singular behaviour corresponding to curvature singularities.
This is demonstrated in Fig.~\ref{nonsymmext-1} for the wormhole solution
with parameters $\alpha/r_0^2=0.02$,
$f_0=1.1$ and $\gamma = 1$.

\begin{figure}[h!]
\lbfig{nonsymmext-1}
\begin{center}
\hspace{0.0cm} \hspace{-0.6cm}
\includegraphics[height=.25\textheight, angle =0]{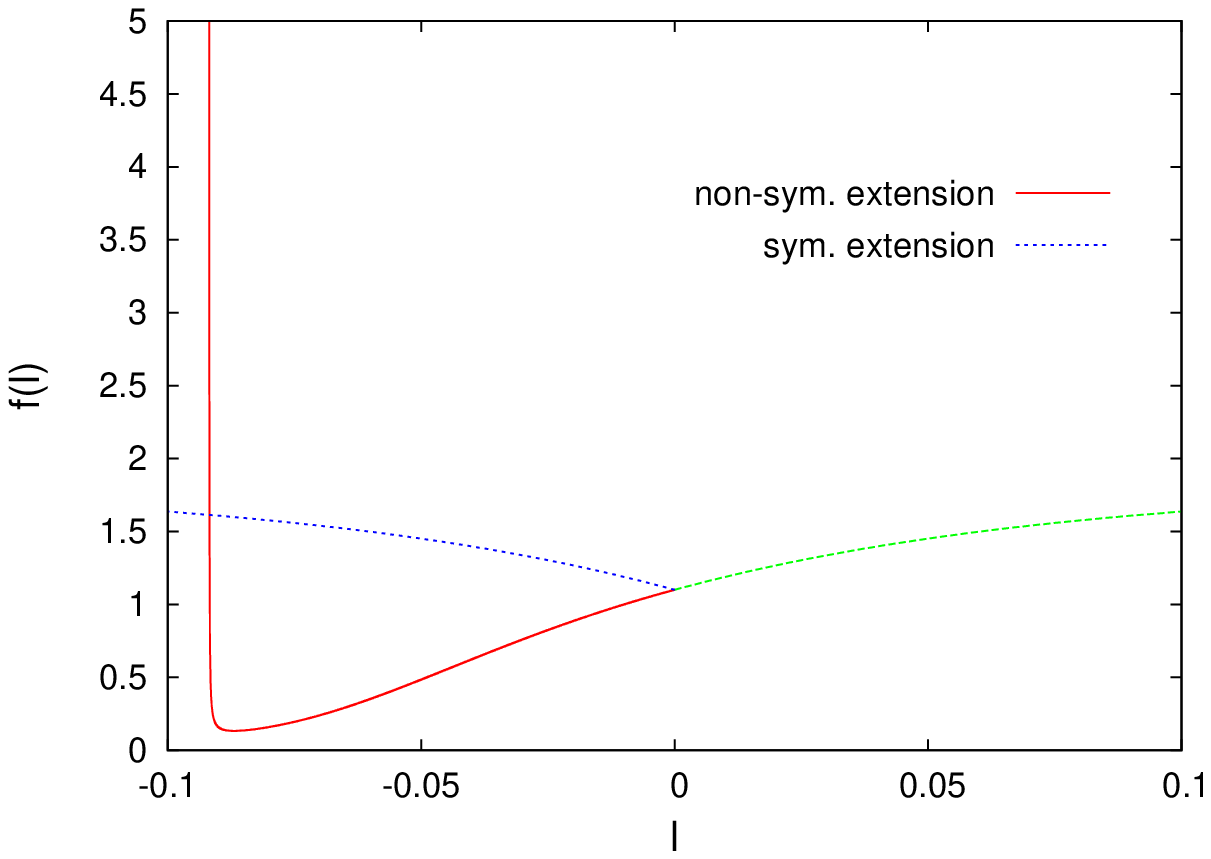}
\hspace{0.5cm} \hspace{-0.6cm}
\includegraphics[height=.25\textheight, angle =0]{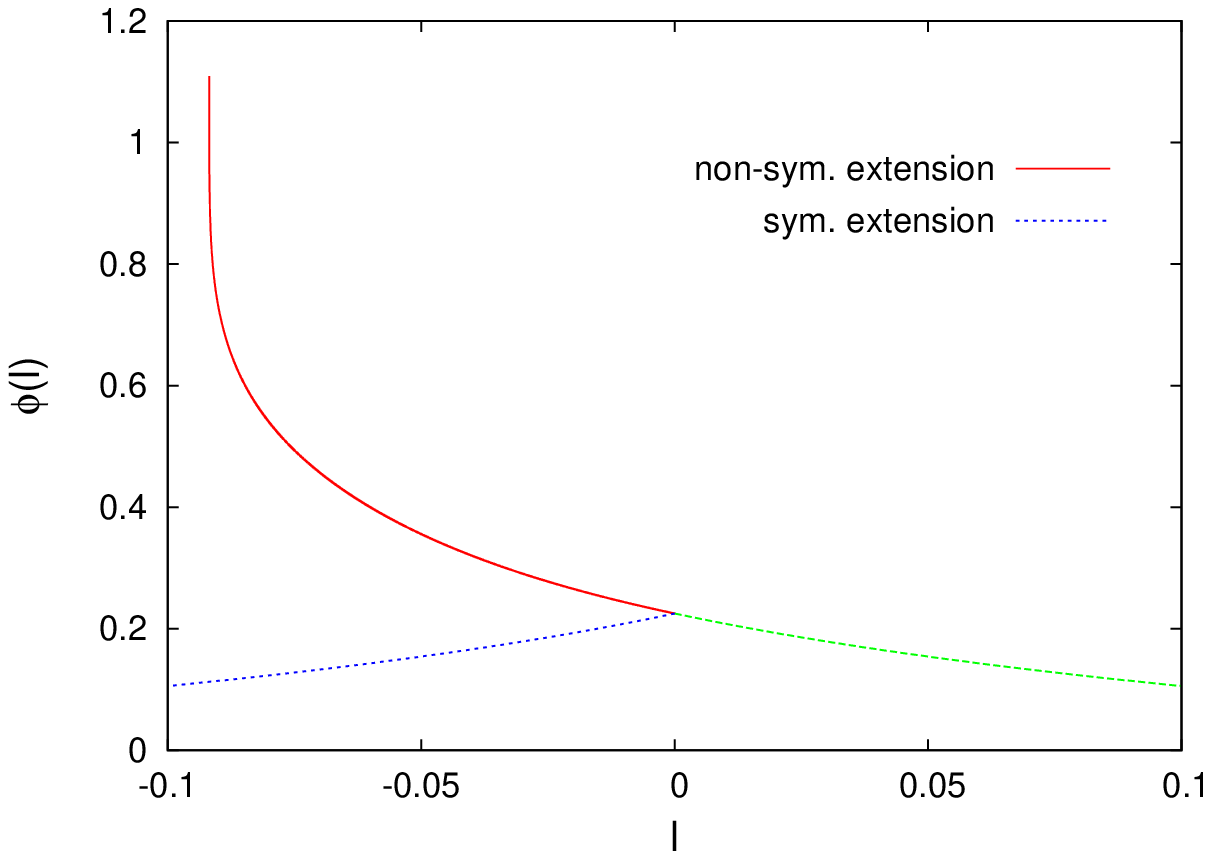}
\end{center}
\hspace*{0.3cm} {(a)} \hspace{7cm} {(b)} \hspace{2cm} \\[1mm]
\caption{
Non-symmetric and symmetric extension of (a)
the metric function $f$ and (b) the dilaton function $\phi$ for
 $\alpha/r_0^2=0.02$ and $f_0=1.1$ versus $l$.}
\end{figure}

However, wormhole solutions without curvature singularities can be constructed
when we extend the wormhole solutions to the second asymptotically
flat part of the manifold in a symmetric way. In this case jumps
appear in the derivatives of the metric and dilaton functions at $l=0$. 
The jumps can be attributed to the presence of matter
located at the throat of the wormhole. 
The corresponding junction conditions are of the form 
\begin{equation}
\langle G_\mu^\nu-T_\mu^\nu \rangle = s_\mu^\nu \  , \ \ \ \ 
\langle \nabla^2 \phi
        -\alpha \gamma e^{-\gamma \phi}R_{GB}^2  \rangle 
= s_{\rm dil} \ ,
\end{equation}
where $s_\mu^\nu$ is the stress-energy tensor of the matter at the throat,
and $s_{\rm dil}$ the corresponding source term of the dilaton field.	
The lhs of the junction conditions can be derived in a standard way
by integrating the Einstein and dilaton equations across the boundary 
$l=0$, i.~e. 
$\langle G_\mu^\nu-T_\mu^\nu \rangle  
= \frac{1}{2}\lim_{L \to 0}\int_{-L}^{L}\left( G_\mu^\nu-T_\mu^\nu\right) dl$.
This yields
\begin{eqnarray}
\langle G_0^0-T_0^0 \rangle 
& = &
 -\frac{8\alpha \gamma e^{-\gamma \phi_0}\phi'_0}{\sqrt{f_0}r_0^2} \ ,
\label{Junc00}\\
\langle G_l^l-T_l^l \rangle 
& = & 0 \ ,
\label{Juncll}\\
\langle G_\theta^\theta-T_\theta^\theta \rangle 
& = & 2\frac{\nu_0'}{\sqrt{f_0}} \ ,
\label{Junctt}\\
\langle G_\vphi^\vphi-T_\vphi^\vphi \rangle 
& = & 
\langle G_\theta^\theta-T_\theta^\theta \rangle  \ ,
\label{Juncpp}\\
\langle \nabla^2 \phi
        -\alpha \gamma e^{-\gamma \phi}R_{GB}^2  \rangle
& = & 
  \frac{\phi_0'}{\sqrt{f_0}}
 +8 \frac{\alpha\gamma e^{-\gamma \phi_0}}{\sqrt{f_0}r_0^2} \nu_0' \ ,
\label{Juncdil}
\end{eqnarray}
where the subscript $0$ indicates evaluation at $l=0$.

Next we assume that the matter at the throat takes the form of
a perfect fluid  with energy density $\rho$ 
and pressure $p$ and a dilaton charge $\rho_{\rm dil}$. 
We also introduce the action 
\begin{equation}
S_\Sigma = \int\left(\lambda_1 +\lambda_0 2 \alpha e^{-\gamma \phi}\bar{R}
           \right)\sqrt{-\bar{h}} d^3x 
\label{S_sig}
\end{equation}
at the throat, 
where $\bar{h}_{ab}$ denotes the (2+1)-dimensional induced metric on the throat, 
$\bar{R}$ the corresponding Ricci scalar, 
and $\lambda_1$, $\lambda_2$ are constants.
Inserting the metric 
this brings the non-trivial junction conditions Eqs.~(\ref{Junc00})-(\ref{Juncdil}) to the form
\begin{eqnarray}
\frac{8 \alpha \gamma e^{-\gamma \phi_0}}{r_0^2}\,\frac{\phi_0'}{\sqrt{f_0}}
&=&
\rho-\lambda_0\,\frac{4 \alpha e^{-\gamma \phi_0}}{r_0^2}-\lambda_1
\,, \\
\frac{2\nu'_0}{\sqrt{f_0}} &=& p+\lambda_1 \,,\\
\left(\phi'_0 +\frac{8 \alpha\gamma  e^{-\gamma \phi_0}}{r_0^2}\,{\nu_0'}\right)\frac{1}{\sqrt{f_0}}
&=& 
\lambda_0\,\frac{4 \alpha\gamma  e^{-\gamma \phi_0}}{r_0^2} +\frac{\rho_{\rm dil}}{2}\,.
\end{eqnarray}
Using these equations $\rho$, $p$ and $\rho_{\rm dil}$ 
can be expressed in terms of 
the metric and dilaton functions and the constants $\lambda_0$, $\lambda_1$. 
In Figs.~\ref{Junction-1} 
we give an example for $\lambda_0=\lambda_1=1$ and $\gamma=1$.
We note that the stable wormhole solutions possess positive energy density $\rho$.
\begin{figure}[h!]
\lbfig{Junction-1}
\begin{center}
\hspace{0.0cm} \hspace{-0.6cm}
\includegraphics[height=.27\textheight, angle =0]{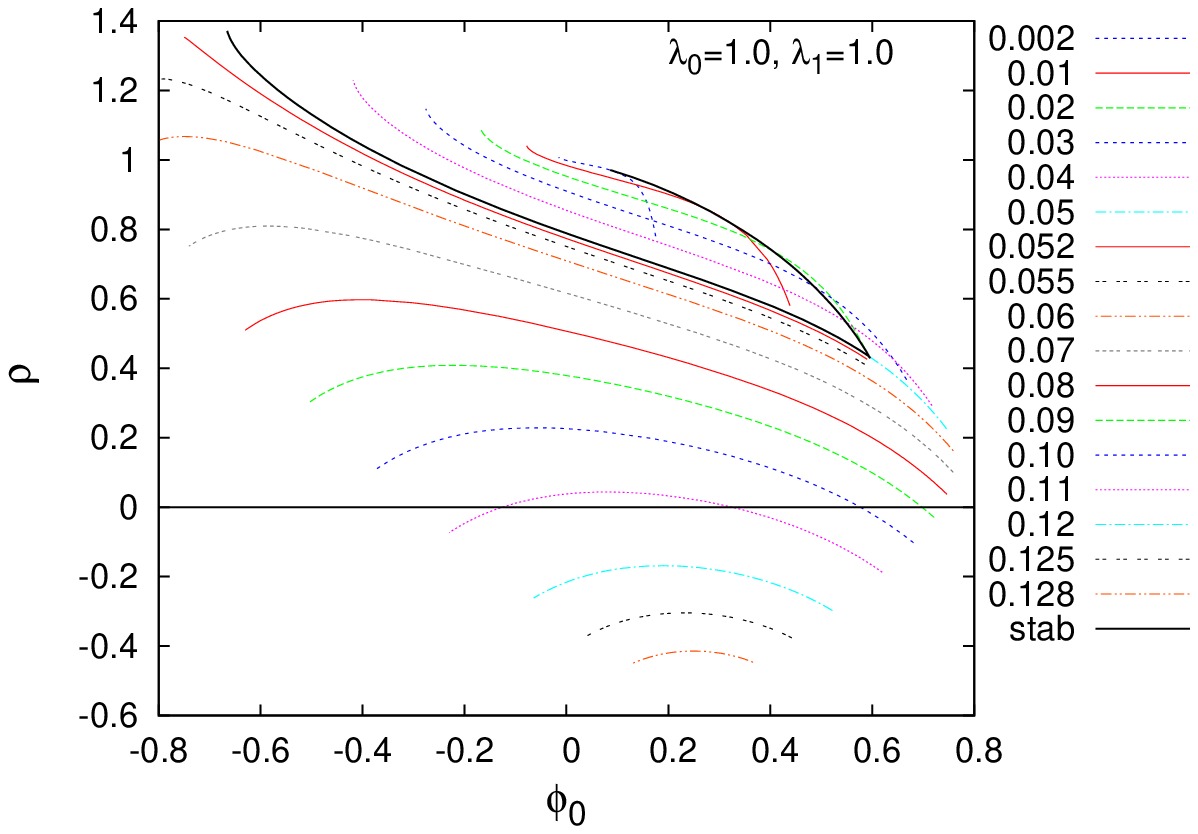}
\hspace{0.5cm} \hspace{-0.6cm}
\includegraphics[height=.27\textheight, angle =0]{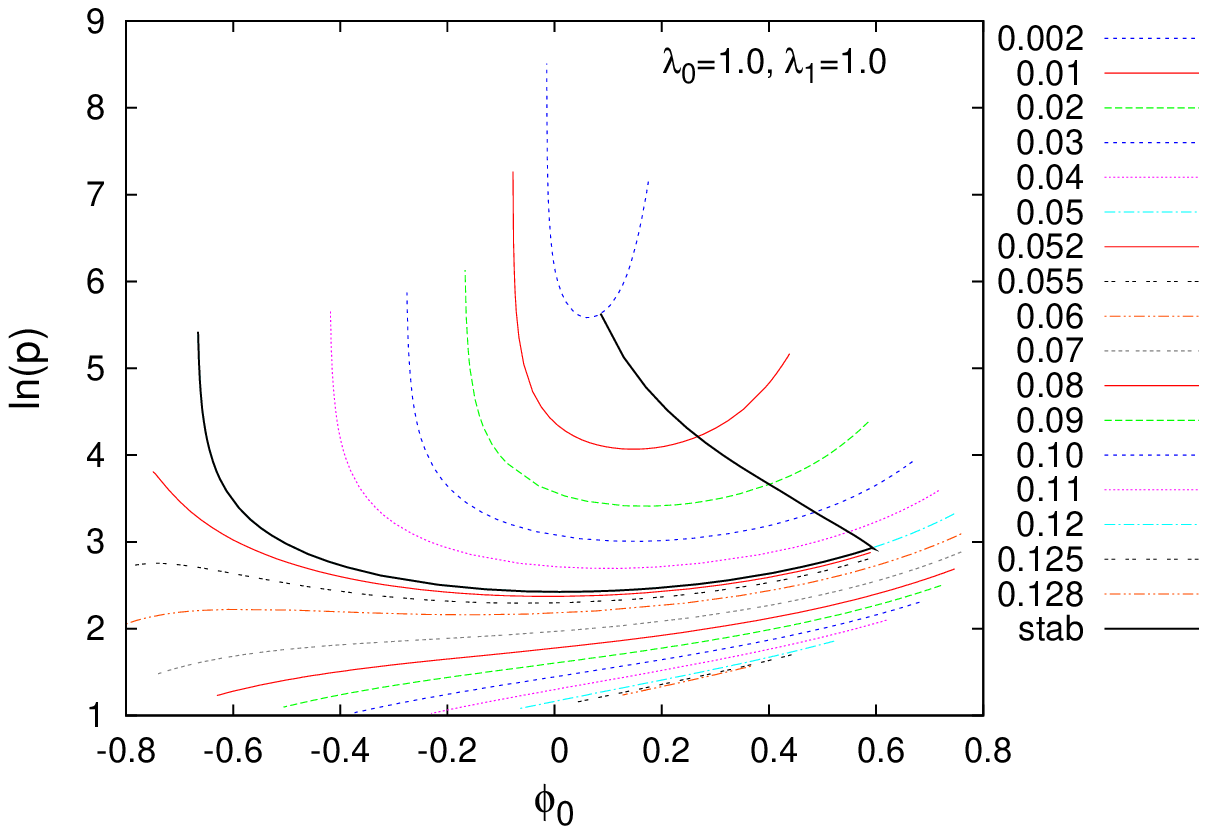}
\\[1mm]
\hspace*{0.0cm} {(a)} \hspace{8cm} {(b)} \hspace{2cm} \\[1mm]
\hspace{0.0cm} \hspace{-0.6cm}
\includegraphics[height=.27\textheight, angle =0]{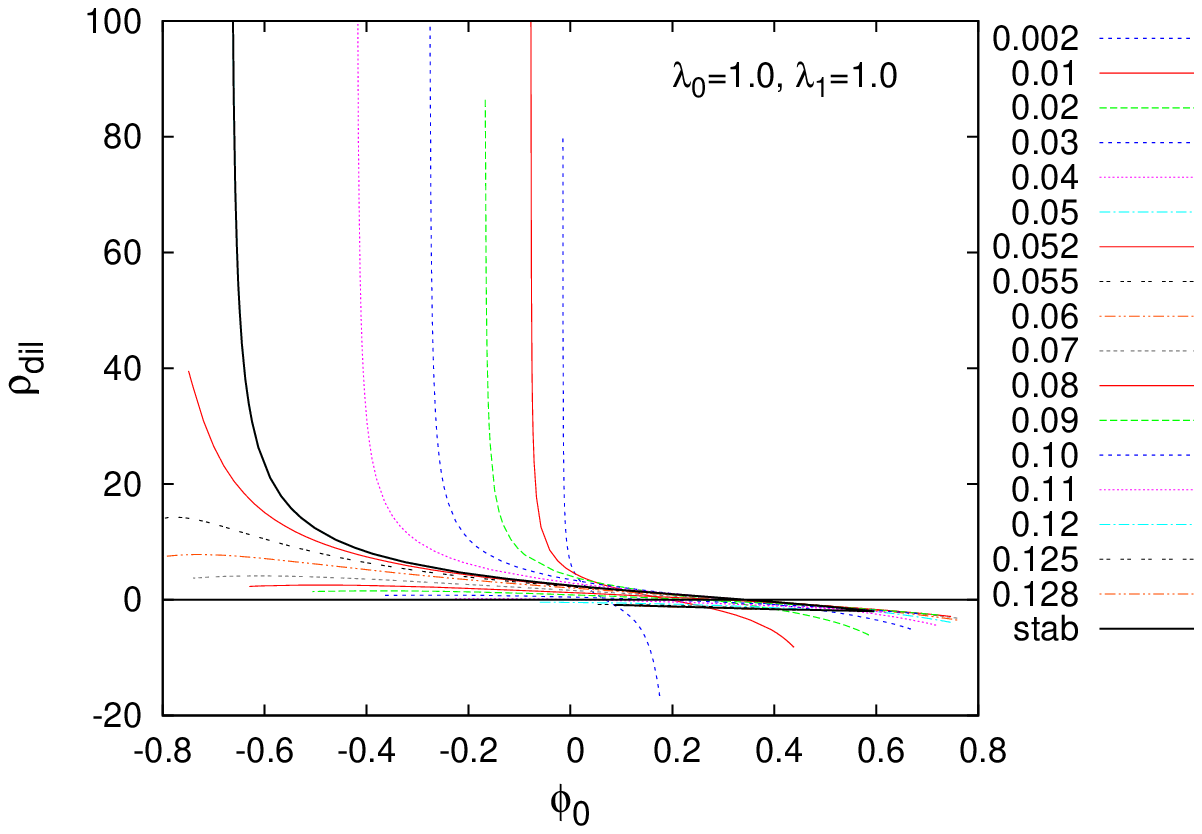}
\hspace{0.5cm} \hspace{-0.6cm}
\includegraphics[height=.27\textheight, angle =0]{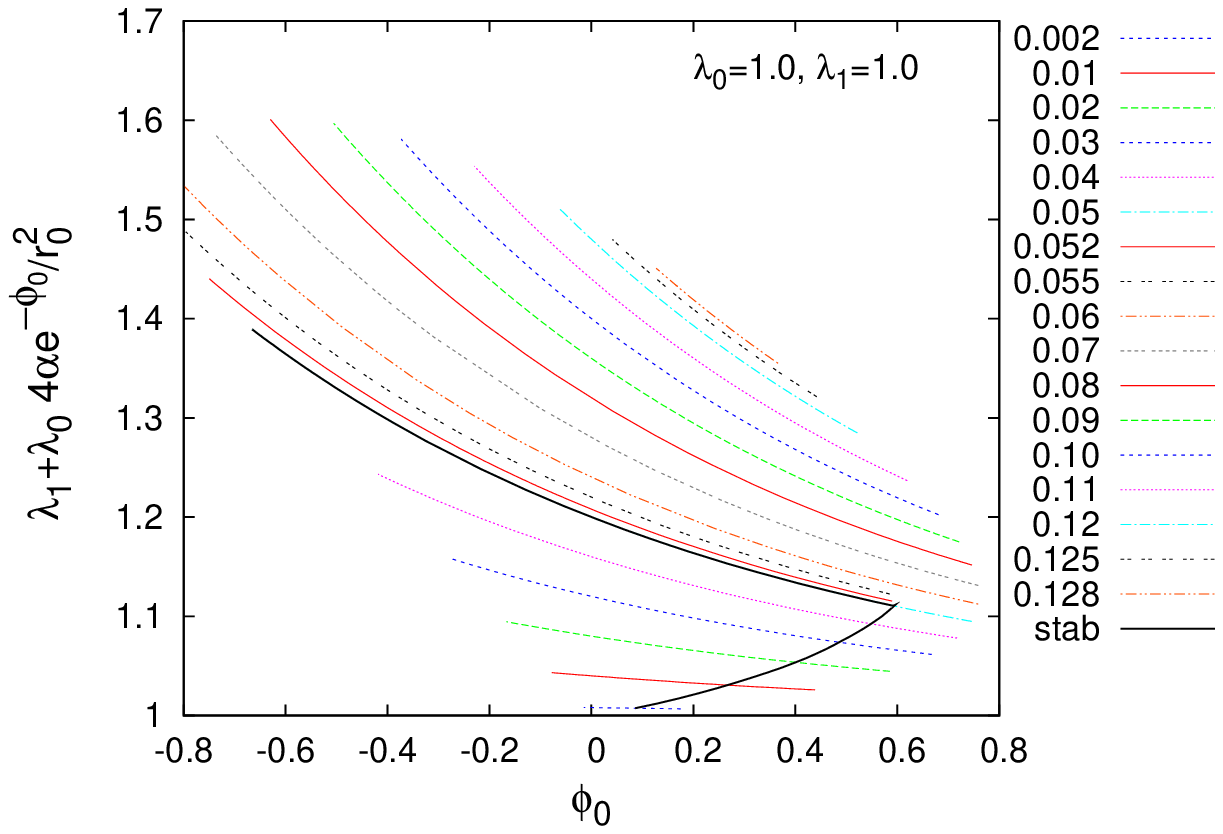}\\
\hspace*{0.0cm} {(c)} \hspace{8cm} {(d)} \hspace{2cm} \\[1mm]
\end{center}
\caption{
(a) Energy density $\rho$ , (b) pressure $p$,
(c) dilaton charge density $\rho_{\rm dil}$ and (d) interaction potential 
versus $\phi_0$.}
\end{figure}

Let us also consider the special case $p=0$ (i.e.~dust) 
and choose the dilaton charge density $\rho_{\rm dil}$
at the throat to be twice the dilaton charge density of the wormhole 
in the `bulk', $\rho_{\rm dil} = 2\phi'_0/\sqrt{f_0}$. 
This yields
\begin{eqnarray}
\lambda_1 & = & \lambda_0  = \frac{2 \nu'_0}{\sqrt{f_0}}
\label{eq_lams}	\\    
\rho  & = & \frac{2 \nu'_0}{\sqrt{f_0}} \left(1+ \frac{4\alpha e^{-\gamma\phi_0}}{r_0^2}\right)
                       +\frac{8 \alpha\gamma e^{-\gamma\phi_0}}{r_0^2} \frac{\phi'_0}{\sqrt{f_0}} 
\nonumber\\		       
 & = & \frac{2 \nu'_0}{\sqrt{f_0}} + \frac{8\alpha e^{-\gamma\phi_0}}{r_0^2}
                        \left(\nu'_0 + \gamma\phi'_0 \right)\frac{1}{\sqrt{f_0}} \ .
\label{eq_rho2}	    
\end{eqnarray}
Interestingly, $\lambda_0=\lambda_1$ and $\rho$ 
are positive for all wormhole solutions,
as shown in Fig~\ref{Junction-3}.

\begin{figure}[h!]
\lbfig{Junction-3}
\centerline{
\includegraphics[height=.28\textheight, angle =0]{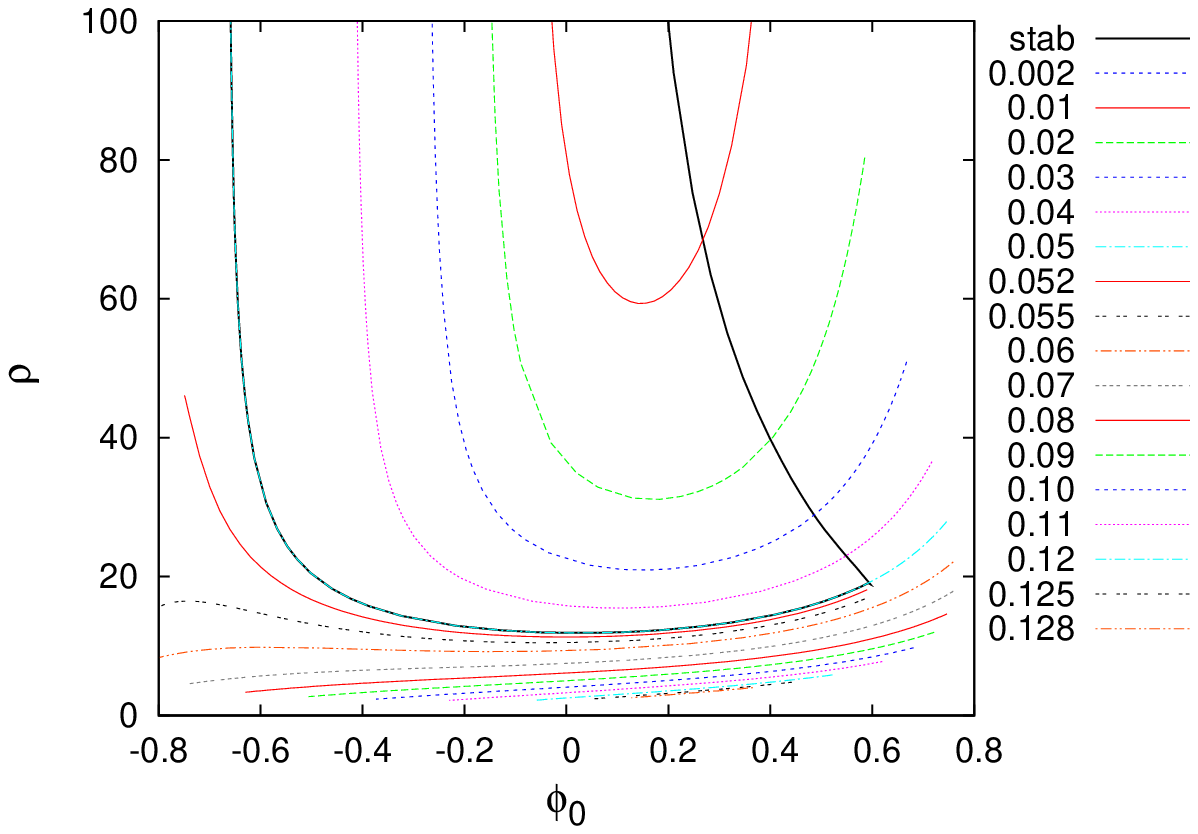}
\includegraphics[height=.28\textheight, angle =0]{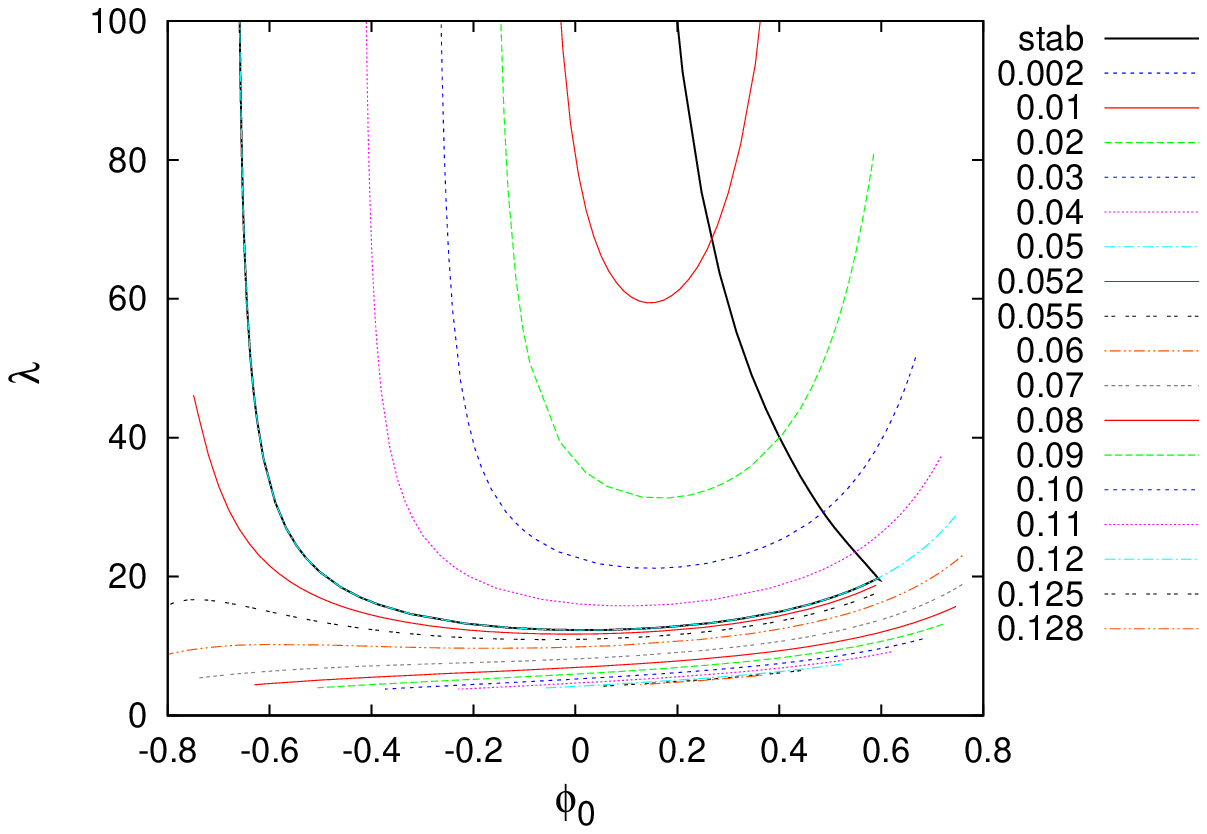}
}
\hspace*{0.0cm} {(a)} \hspace{8cm} {(b)} \hspace{2cm} \\[1mm]
\caption{
(a) Energy density $\rho$ and (b) $\lambda=\lambda_0=\lambda_1$ 
versus $\phi_0$
for dust.}
\end{figure}

Next we consider perturbations of the wormhole solutions in the 
form $\tilde{f}(l,t) = f(l) + \delta f(l) e^{i\sigma t}$, etc. For symmetric wormholes the derivatives
of the perturbations also have a jump at $\Sigma$. Introducing perturbations of the 
energy density $\delta \rho e^{i\sigma t}$, pressure $\delta p e^{i\sigma t}$ and dilaton charge
$\delta \rho_{\rm dil} e^{i\sigma t}$
we find 
\begin{eqnarray}
\delta \rho & = & \frac{4\alpha e^{-\phi_0}}{r_0^2\sqrt{f_0}}
                     \left( \delta \phi'_0 -2\frac{\delta f_0}{f_0} \phi'_0
		            -\left( 2\phi'_0 +\lambda_0\sqrt{f_0} \right)\delta \phi_0\right)		      
\label{eq_dprho} \\
\delta p & = &  2\frac{\delta \nu'_0}{\sqrt{f_0}}
                      -\frac{\delta f_0}{f_0}\frac{\nu'_0}{\sqrt{f_0}}
		      -4\lambda_0 \alpha e^{-\phi_0}\sigma^2 e^{-2\nu_0}\delta \phi_0
\label{eq_dp} \\
\delta \rho_{\rm dil} & = & \delta \phi'_0
        +\frac{8 \alpha e^{-\phi_0} }{r_0^2} \delta \nu'_0
	-\frac{\delta f_0}{2 f_0}\left(\phi'_0 +\frac{8 \alpha e^{-\phi_0}}{r_0^2} \nu'_0\right)
	+\frac{4\alpha e^{-\phi_0}}{r_0^2}\left(\sqrt{f_0}\lambda_0 - 2\nu'_0\right)\delta \phi_0
\label{eq_dphi} 
\end{eqnarray}	

Thus these equations give the perturbations of the energy density, the pressure
and the dilaton charge on the throat. Thus for any perturbation in the `bulk', we 
can find the corresponding perturbation on the throat. Consequently, the stability
of the solutions is not affected by the matter on the throat.

\section{Geodesics}

The study of the orbits of test particles and light
is essential to fully understand the properties of a spacetime.
The motion of test particles in a EGBd wormhole spacetime
with metric $g_{\mu\nu}$, Eq.~(\ref{metricL}), and dilaton field $\phi$
is governed by the Lagrangian
\begin{equation}
{\cal L} = \frac{1}{2} e^{-2\beta \phi} g_{\mu\nu} \dot{x}^\mu \dot{x}^\nu \ ,
\label{lag}
\end{equation}
where the dot denotes the derivative with respect to the affine parameter $\tau$,
and $\beta$ is a constant ($\beta =1/2$ for heterotic string theory).
The conjugate momenta 
$p_\mu = \frac{\partial {\cal L}}{\partial\dot{x}^\mu}$
are found to be
\begin{equation}
p_t = - e^{-2\beta \phi} e^{2\nu}\dot{t}\ , \ \ \ 
p_l =  e^{-2\beta \phi} f\dot{l}\ , \ \ \ 
p_\theta =  e^{-2\beta \phi} (r_0^2+l^2)\dot{\theta}\ , \ \ \ 
p_\vphi =  e^{-2\beta \phi} (r_0^2+l^2)\sin^2\theta\dot{\vphi}\ . 
\label{momenta}
\end{equation}
First integrals of motion are given by
\begin{equation}
p_t = const. = -E \ ,\ \ \  p_\vphi= const. = L \ .
\label{first_int}
\end{equation}
We refer to $E$ as the energy of the test particle 
and to $L$ as its angular momentum.

The affine parameter can be chosen such that $2 {\cal L} = \hat{\kappa}$, 
with $\hat{\kappa}=-1$ for time-like geodesics and $\hat{\kappa}=0$ for null geodesics.
Since we are considering spherically symmetric 
spacetimes we may choose $\theta = \pi/2$, which implies $\dot{\theta}=0$.
Employing Eqs.~(\ref{first_int}) the Lagrangian then reduces to 
\begin{equation}
2{\cal L} = e^{2\beta\phi}e^{-2\nu} 
\left[-E^2 + e^{2\nu}\left(e^{-4\beta\phi}f \dot{l}^2 
                           +\frac{L^2}{r_0^2+l^2}\right)\right]  
 = \hat{\kappa} \ .
\label{lageff}
\end{equation}

Let us first consider time-like geodesics. 
For $\hat{\kappa} = -1$ we obtain 
\begin{equation}
\dot{l}^2 = \frac{e^{4\beta\phi}e^{-2\nu}}{f}
\left[ E^2 -V^2_{\rm eff}(l,L)\right] \ , 
\label{dotl2}
\end{equation}
where we introduced the effective potential 
\begin{eqnarray}
V^2_{\rm eff}(l,L) =
    e^{2\nu}\left(-\hat{\kappa} e^{-2\beta\phi}+\frac{L^2}{r_0^2+l^2}\right) 
\quad
\Longleftrightarrow  
\quad
V^2_{\rm eff}(r,L) =
    e^{2\nu}\left(-\hat{\kappa} e^{-2\beta\phi}+\frac{L^2}{r^2}\right) \ .
\label{veff}    
\end{eqnarray}
The effective potential $V_{\rm eff}$ 
is a suitable quantity to discuss qualitatively
the trajectories of test particles. 
We note that the effective potential
approaches the value of one asymptotically.
Thus it follows from Eq.~(\ref{dotl2})
that unbound trajectories are only possible if $E \ge 1$.
The turning points $r_i$ of trajectories for a given 
energy $E$ and angular momentum $L$ are determined by the condition 
\begin{equation}
E= V_{\rm eff}(r_i,L) \ .
\end{equation}  
We note that, in contrast to black hole spacetimes, the wormhole
spacetimes do not possess an event horizon. Consequently, particles
cannot disappear behind an event horizon. 
Instead they can travel through the throat of the wormhole
from one asymptotically flat part of the manifold to the other.

\begin{figure}[h!]
\lbfig{ff-1}
\begin{center}
\includegraphics[height=.33\textheight, angle =0]{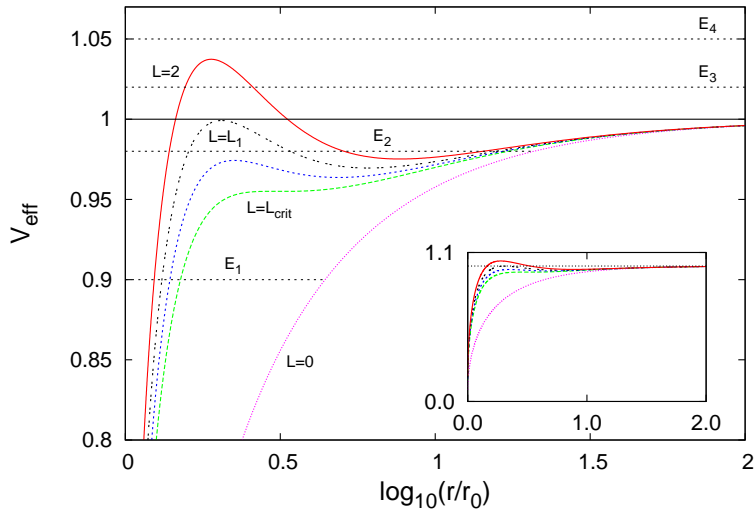}
\end{center}
\caption{
Effective potential $V_{\rm eff}$ versus $\log_{10}(r/r_0)$ for several
values of the angular momentum $L$ for the wormhole solution with 
$\alpha/r_0^2=0.05$ and $f_0=1.1$.
}
\end{figure}

To discuss the various types of trajectories in these
wormhole spacetimes we consider as an example
the wormhole solution with parameters $\alpha=0.05$ and $f_0=1.1$.
We exhibit the corresponding effective potential $V_{\rm eff}(r,L)$
in Fig.~\ref{ff-1} for several values
of the angular momentum $L$.
In particular, we first discuss the dependence of the shape
of the effective potential on the angular momentum $L$ of the test particle,
and next consider the different types of trajectories
with their dependence on the energy $E$ of the test particle.

In the case $L=0$
the effective potential is a monotonically increasing function of $r$, 
assuming its minimum at the throat $r=r_0$ and tending to one 
asymptotically. When the angular momentum is increased, the effective potential 
remains monotonic until it develops a saddle point for some critical
value $L_{\rm crit}$. 
For $L>L_{\rm crit}$ the effective potential is no longer monotonic,
since a local maximum $V_{\rm max}(L)$ 
and a local minimum  $V_{\rm min}(L)$ occur.
If $L$ is larger than a certain value, $L_1$ (say), 
the maximum $V_{\rm max}(L)$ exceeds the asymptotic value of the
effective potential, and thus it becomes the global maximum.
We note that the essential features of the
effective potential $V_{\rm eff}(r,L)$ are
the same for all the wormhole solutions considered.

Concerning the types of orbits of test particles we note that there are
two kinds of trajectories. 
For trajectories of the first kind the particles remain on a single
asymptotically flat manifold, whereas for trajectories of
the second kind the particles travel from one asymptotically flat part 
of the manifold to the other, 
passing through the throat.

Trajectories of the first kind exist only if the effective potential
is non-monotonic, i.e.~for $L>L_{\rm crit}$. 
Bound orbits exist for $V_{\rm min}(L)\leq E<1$, 
e.g.~$E_2$ in Fig.~\ref{ff-1}, 
whereas unbound orbits exist for $1< E <V_{\rm max}(L)$, 
e.g.~$E_3$ in Fig.~\ref{ff-1}.
For unbound orbits we need $L>L_1$, 
thus $V_{\rm max}(L)$ must be the absolute maximum.

Trajectories of the second kind exist for all values
of $E$ above the minimum of the effective potential
$V_{\rm eff}(r_0,L)$. 
The particles move on bound orbits if  
$V_{\rm eff}(r_0,L)\leq E < \max\{1,V_{\rm max}(L)\}$,
e.g.~$E_1$, $E_2$, $E_3$ in Fig.~\ref{ff-1}.
Thus for $E=V_{\rm eff}(r_0,L)$ the particles move on circles at the
throat. Radial trajectories are obtained for $L=0$.
In this case the particles oscillate about the throat.
In the limit $E_0=\sqrt{V_{\rm eff}(r_0,0)}$
they possess the  minimal possible energy
and are at rest at the throat.
On the other hand, if $E>\max\{1,V_{\rm max}(L)\}$, 
e.g.~$E_4$ in Fig.~\ref{ff-1},
the particles travel along unbound orbits, starting at infinity on one 
asymptotically flat part of the manifold, passing through the throat and reaching
infinity on the other asymptotically flat part of the manifold.

In Fig.~\ref{ff-2} we exhibit the minimal possible energy
$E_0=\sqrt{V_{\rm eff}(r_0,0)}$ versus $D/M$ for several values of 
$\alpha/r_0^2$. Note that $E_0$ approaches zero in the black hole limit.
The black line on the left, highlighted in the inset,
indicates the stability change of the wormhole solutions.

\begin{figure}[h!]
\lbfig{ff-2}
\begin{center}
\includegraphics[height=.25\textheight, angle =0]{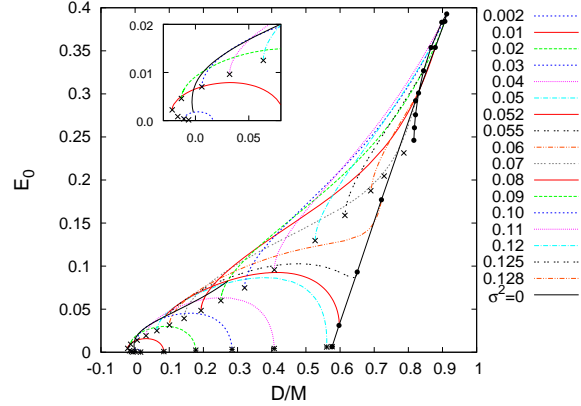}
\end{center}
\caption{
Minimal energy $E_0=\sqrt{V_{\rm eff}(r_0,0)}$ of test particles
versus the scaled dilaton charge $D/M$ 
for several values of $\alpha/r_0^2$.
The black line on the left denoted by $\sigma^2=0$ 
indicates the stability change of the wormhole solutions.
}
\end{figure}

In order to calculate the trajectories we consider
\begin{equation}
\frac{d\vphi}{dl} = \frac{\dot{\vphi}}{\dot{l}} \nonumber \\
= \pm L \frac{e^\nu}{r_0^2+l^2}\sqrt{\frac{f}{E^2 -V^2_{\rm eff}}} \ ,
\nonumber
\end{equation}
from which we find
\begin{equation}
\vphi(l) = \vphi_0 
\pm L \int_{l_0}^l \frac{e^\nu}{r_0^2+l^2}\sqrt{\frac{f}{E^2 -V^2_{\rm eff}}} dl' \ ,
\end{equation}
where $l$ is restricted to intervals where $E^2 -V^2_{\rm eff} \geq 0$ and 
$\vphi_0$ is an integration constant.
The trajectories are then displayed in the $xy$-plane, where
\begin{equation}
x(l) = \sqrt{r_0^2+l^2}\cos(\vphi(l)) \ , \ \ \ \ 
y(l) = \sqrt{r_0^2+l^2}\sin(\vphi(l)) \ .
\end{equation}

In Fig.~\ref{ff-3} we show examples of bound orbits 
of test particles with angular momentum $L=2$
and several values of the energy $E$
for the wormhole solution with $\alpha/r_0^2=0.05$ and $f_0=1.1$.
For an energy of $E=E_1=0.9$ (Fig.~\ref{ff-3}a)
there exists only a bound orbit of the second kind. 
This is highlighted in the figure by using different colors (line styles) for
the two asymptotically flat manifolds, when projecting into the
$xy$-plane.
In contrast, for $E=E_2=0.98$ (Fig.~\ref{ff-3}b)
there exists in addition a bound orbit of the first kind.
Examples for unbound orbits are shown in Fig.~\ref{ff-4}a for 
energies $E=E_3=1.02$ and $E=E_4=1.05$.

\begin{figure}[h!]
\lbfig{ff-3}
\begin{center}
a) \includegraphics[height=.25\textheight, angle =0]{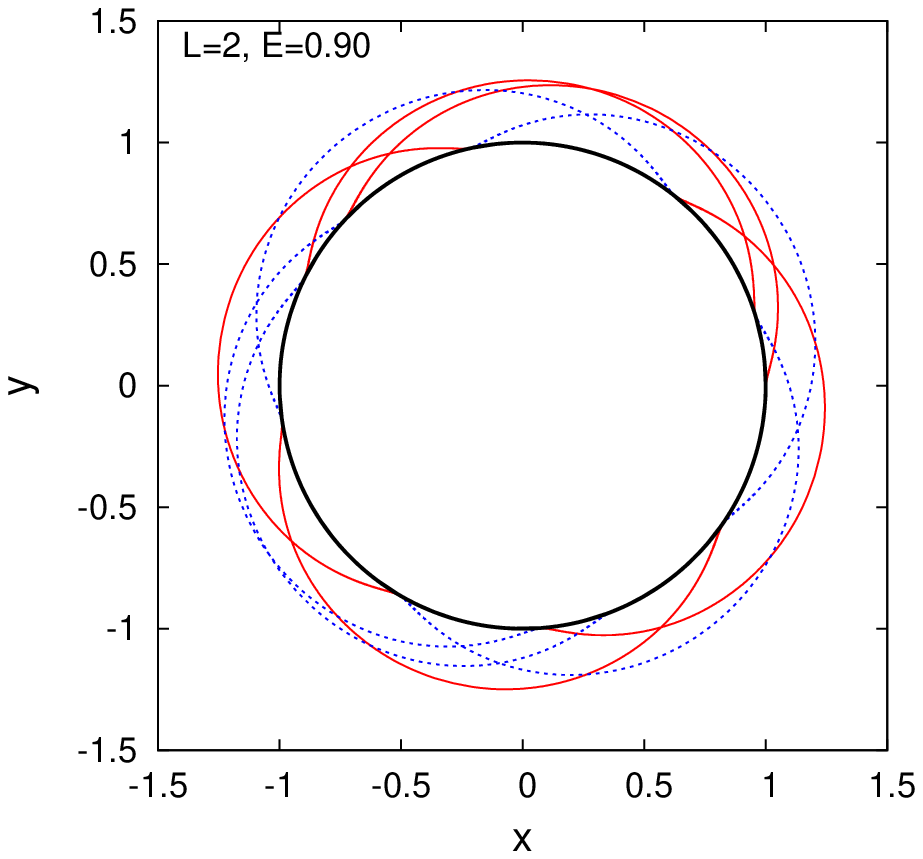}
  \includegraphics[height=.25\textheight, angle =0]{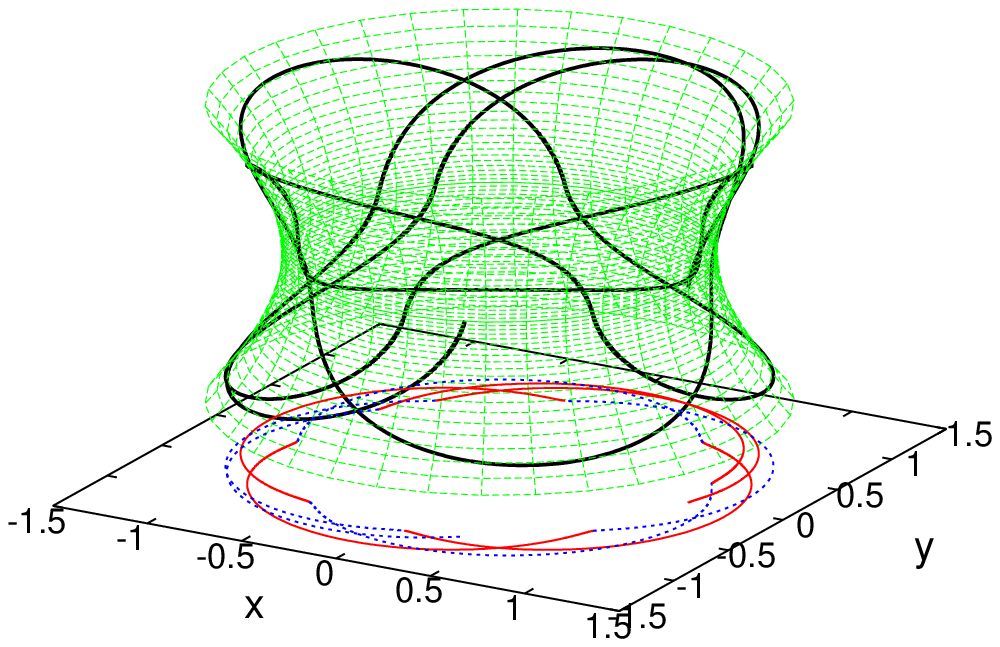}\\
b) \includegraphics[height=.25\textheight, angle =0]{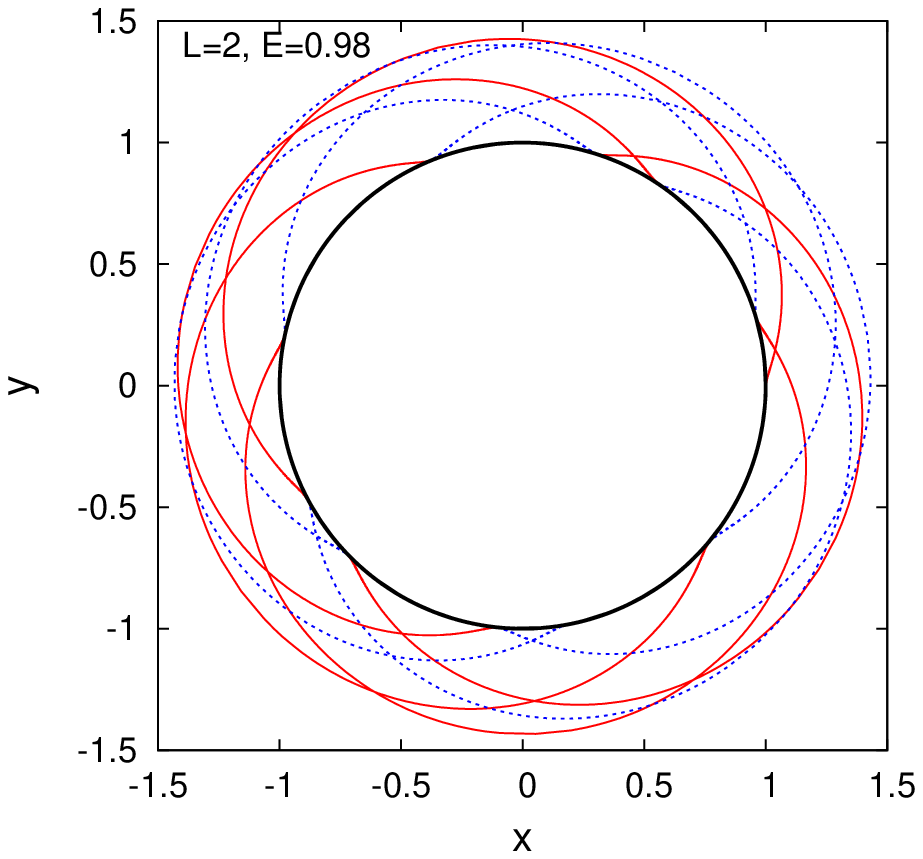}
   \includegraphics[height=.25\textheight, angle =0]{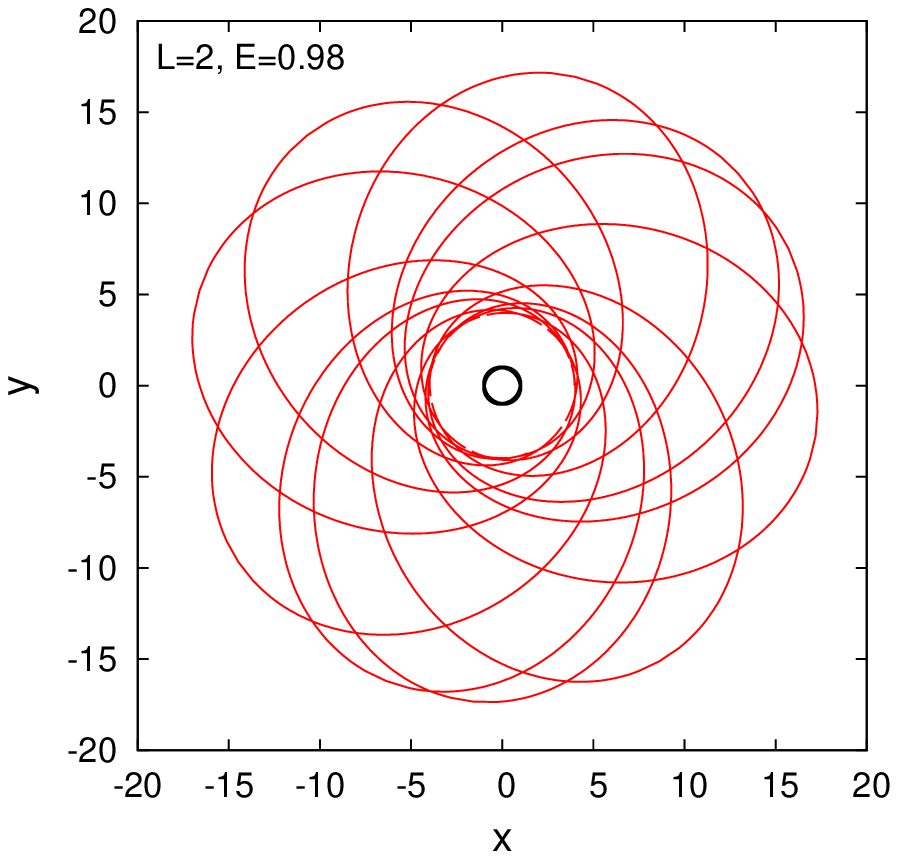}
\end{center}
\caption{
Bound orbits of massive test particles with angular momentum $L=2$
and energy (a) $E=0.9$ and (b) $E=0.98$
in the wormhole spacetime with $\alpha/r_0^2=0.05$ and $f_0=1.1$. 
The colours red (solid) and blue (dotted) indicate 
motion on the first and the second asymptotically flat part of 
the manifold, respectively.
The throat of the wormhole is shown by the black circle.
The right panel of a) shows the orbit on the isometric embedding of the wormhole.
}
\end{figure}

\begin{figure}[h!]
\lbfig{ff-4}
\begin{center}
\includegraphics[height=.25\textheight, angle =0]{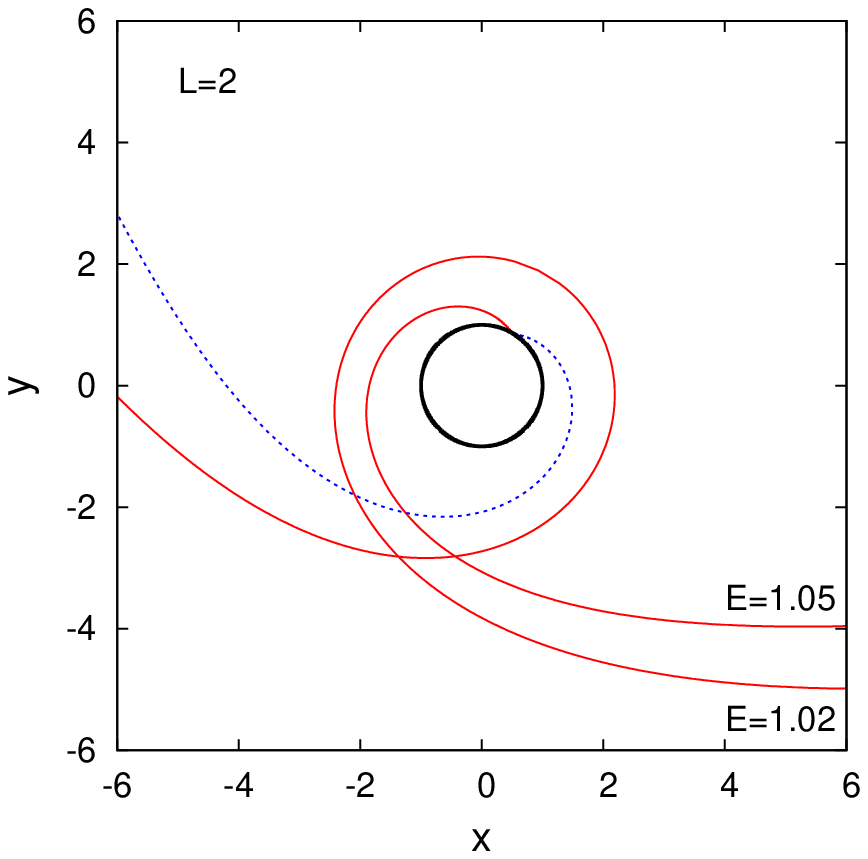}
\includegraphics[height=.25\textheight, angle =0]{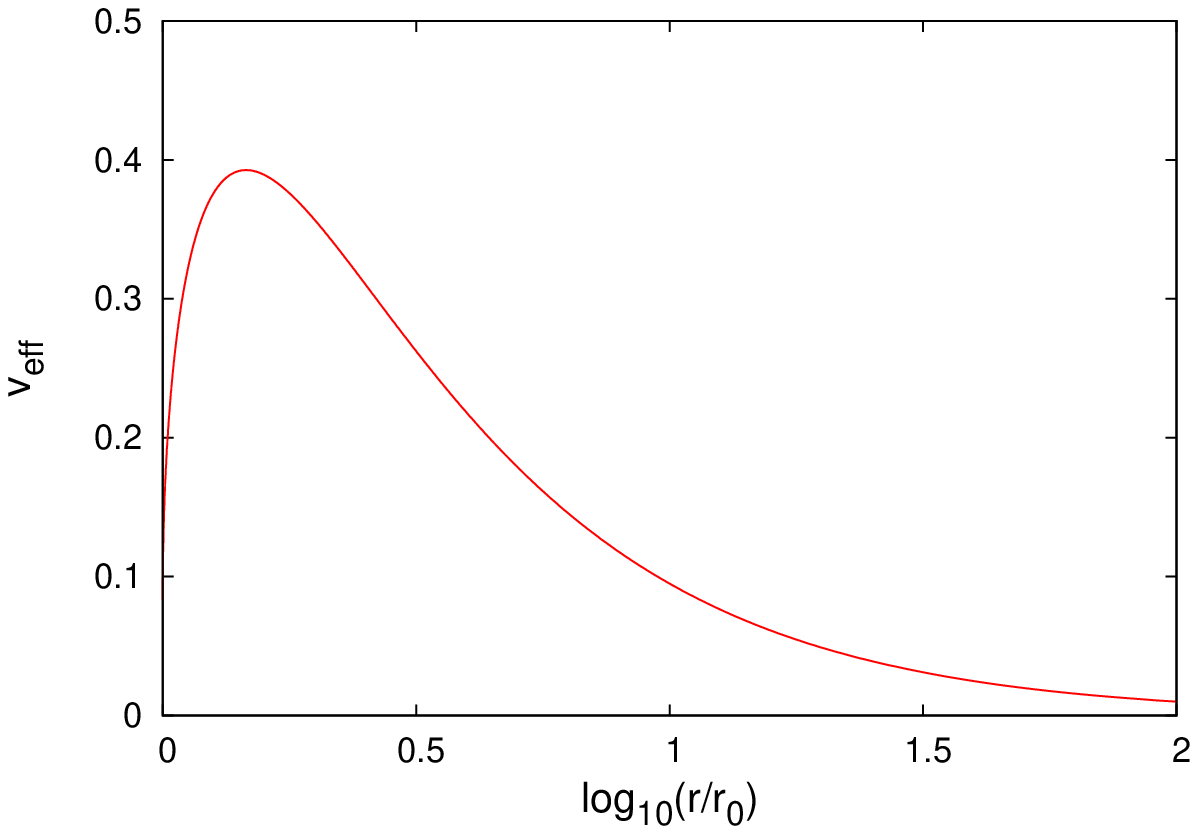}
\end{center}
\hspace*{0.7cm} {(a)} \hspace{7.5cm} {(b)} \hspace{2cm} \\[1mm]
\caption{
(a) Same as Fig.~\ref{ff-3} for unbound orbits of massive test particles
and energies $E=1.02$, $1.05$.
(b) ``Normalized'' effective potential $v_{\rm eff}$ for massless test particles
versus $\log_{10}(r/r_0)$ for the wormhole solution with $\alpha/r_0^2=0.05$ and $f_0=1.1$.
}
\end{figure}

Let us now consider the null geodesics in these wormhole spacetimes. 
Setting $\hat{\kappa}=0$ in Eq.~(\ref{lageff}) we find
\begin{equation}
\dot{l}^2 = \frac{e^{4\beta\phi}e^{-2\nu}}{f}
\left[ E^2 -L^2\frac{e^{2\nu}}{r_0^2+l^2}\right] 
=E^2 \frac{e^{4\beta\phi}e^{-2\nu}}{f}
\left[1  -\left(\frac{L}{E}\right)^2 \frac{e^{2\nu}}{r_0^2+l^2}\right]\ .
\label{dotl2N}
\end{equation}
We define an effective potential $V_{\rm eff}(l,L/E)$ by
\begin{equation} 
 V^2_{\rm eff}(l,L/E) =
   \left(\frac{L}{E}\right)^2 \frac{e^{2\nu}}{r_0^2+l^2} 
   = \left(\frac{L}{E}\right)^2 v^2_{\rm eff}(l)\  .
\label{veffN}    
\end{equation}
We note that for the discussion of the null geodesics it is more convenient
to consider the ``normalized'' effective potential
$v_{\rm eff}(l)=\frac{e^{\nu}}{\sqrt{r_0^2+l^2}}$, since it is independent
of $L/E$. 

As an example, we show the ``normalized'' effective 
potential $v_{\rm eff}$ for the parameter values $\alpha/r_0^2 =0.05$ 
and $f_0=1.1$ in Fig.~\ref{ff-4}.
The ``normalized'' effective potential possesses a 
local minimum $v_{\rm min}$ at the throat, 
a maximum  $v_{\rm max}$ at some distance from the throat 
and tends to zero asymptotically.

As for the timelike geodesics there are two kinds of trajectories:
either the massless test particle remains on one of the asymptotically 
flat parts of the manifold,
or it passes through the throat from one asymptotically flat part to the other. 
The first kind of (unbound) trajectories exists only if the ratio $E/L$ is 
smaller than the maximum of the ``normalized'' effective potential.
The second kind of trajectories includes unbound geodesics if 
$E/L > v_{\rm max}$  and bound geodesics if $v_{\rm min}\leq E/L < v_{\rm max}$.
Circular orbits exist for $E/L=v_{\rm min}=v_{\rm eff}(l=0)$
(and unstable ones for $E/L=v_{\rm max}$).

\section{Acceleration and Tidal Forces}

Following the formalism of \cite{Morris:1988cz}, we will now calculate the magnitude of the
acceleration and tidal forces that a traveler traversing the wormhole would feel.
For this purpose, it is particularly convenient to make two changes of reference
frames: first, starting from the standard reference frame with basis vectors
$({\mathbf e}_t, {\mathbf e}_l, {\mathbf e}_\theta, {\mathbf e}_\varphi)$,
in which the line-element takes the form of Eq.~(\ref{metricL}),
we change to an orthonormal reference frame with
\begin{equation}
{\mathbf e}_{\hat t}=e^{-\nu} {\mathbf e}_t, \quad
{\mathbf e}_{\hat l}=\frac{1}{\sqrt{f}}\,{\mathbf e}_l, \quad
{\mathbf e}_{\hat \theta}=\frac{1}{r}\,{\mathbf e}_\theta, \quad
{\mathbf e}_{\hat \varphi}=\frac{1}{r \sin\theta}\,{\mathbf e}_\varphi
\label{ortho1}
\end{equation}
in terms of which the metric tensor assumes the form: 
$g_{\hat \alpha \hat \beta}={\mathbf e}_{\hat \alpha} \cdot {\mathbf e}_{\hat \beta}=
\eta_{\hat \alpha \hat \beta}$.
Alternatively, we may write the set of equations in (\ref{ortho1}) as 
${\mathbf e}_{\hat \alpha} =L_{\hat \alpha}^{\,\,\mu}\,{\mathbf e}_\mu$, with
\begin{equation}
L_{\hat \alpha}^{\,\,\mu}=\left[\begin{array}{rccc}
e^{-\nu} & 0 & 0 & 0 \\ 0 & 1/\sqrt{f} & 0 & 0\\ 0 & 0 & 1/r & 0\\
0 & 0 & 0 & 1/(r \sin\theta)\end{array} \right]. \label{trans1}
\end{equation}
The above allows us to write the transformation law of the components of the
Riemann tensor as
\begin{equation}
R^{\hat \alpha}_{\,\,\hat \beta \hat \gamma \hat \delta}=L^{\,\,\hat \alpha}_{\mu}
\,L_{\hat \beta}^{\,\,\nu}\,L_{\hat \gamma}^{\,\,\rho}\,L_{\hat \delta}^{\,\,\sigma}\,
R^\mu_{\,\,\nu \rho \sigma}\,,
\label{Riemann-trans1}
\end{equation}
with $L^{\,\,\hat \alpha}_\mu=(L_{\hat \alpha}^{\,\,\mu})^{-1}$. Next, we introduce
the orthonormal reference frame of the traveler which is related to the previous one
by a Lorentz transformation
\begin{equation}
{\mathbf e}_{\tilde t}=\gamma\,{\mathbf e}_{\hat t} \mp \gamma\,\frac{v}{c}\,
{\mathbf e}_{\hat l}\,, \quad
{\mathbf e}_{\tilde l}=\mp\gamma\,{\mathbf e}_{\hat l}+
\gamma\,\frac{v}{c}\,{\mathbf e}_{\hat t}\,, \quad
{\mathbf e}_{\tilde \theta}={\mathbf e}_{\hat \theta}, \quad
{\mathbf e}_{\tilde \varphi}={\mathbf e}_{\hat \varphi}\,,
\label{ortho2}
\end{equation}
with $\gamma=[1-(v/c)^2)]^{-1/2}$ and $v=\mp (\sqrt{f} dl/e^\nu dt)$ the
radial velocity of the traveler at radius $l$ as measured by a static
observer there. As before, we may write 
${\mathbf e}_{\tilde \alpha} =\Lambda_{\tilde \alpha}^{\,\,\hat \mu}\,
{\mathbf e}_{\hat \mu}$, with
\begin{equation}
 \Lambda_{\tilde \alpha}^{\,\,\hat \mu}=\left[\begin{array}{cccc}
\gamma & \mp\gamma (v/c) & 0 & 0 \\ \gamma\,(v/c) & \mp \gamma & 0 & 0\\
0 & 0 & 1 & 0\\ 0 & 0 & 0 & 1\end{array} \right],
\label{trans2}
\end{equation}
so that
\begin{equation}
R^{\tilde \alpha}_{\,\,\tilde \beta \tilde \gamma \tilde \delta}=
\Lambda^{\,\,\tilde \alpha}_{\hat \mu}\,\Lambda_{\tilde \beta}^{\,\,\hat \nu}\,
\Lambda_{\tilde \gamma}^{\,\,\hat\rho}\,\Lambda_{\tilde \delta}^{\,\,\hat\sigma}\,
R^{\hat \mu}_{\,\,\hat \nu \hat \rho \hat \sigma}\,.
\label{Riemann-trans2}
\end{equation}

We note that ${\mathbf e}_{\tilde t} \cdot {\mathbf e}_{\tilde t}=-1$, and thus
${\mathbf e}_{\tilde t}$ can be naturally considered as the traveler's normalized
vector of four-velocity ${\mathbf u}$. For the magnitude of ${\mathbf u}$ to
be fixed in the rest frame of the traveler, the four-acceleration
${\mathbf a}=d{\mathbf u}/d \tau$ should be orthogonal to the four-velocity, 
i.e.~${\mathbf a} \cdot {\mathbf u}=0$, and therefore ${\mathbf a}=(0, a^i)$ --
if we further assume that the traveler moves radially, then
${\mathbf a}= a {\mathbf e}_{\tilde l}$. For a traveler moving along their
world-line in which the tangent vector is 
${\mathbf e}_{\tilde t}={\mathbf u}$, the acceleration is given by the formula
${\mathbf a}={\mathbf \nabla}_{\mathbf u}{\mathbf u}$ or
$a^{\tilde \mu}=u^{\tilde \mu}_{;\tilde a} u^{\tilde a} c^2$.
In order to compute the magnitude of the acceleration that the traveler feels,
we work in the following way: we calculate the $a_t$ component in the
$(t,l,\theta,\varphi)$ coordinate frame first according to the formula
\begin{equation}
\frac{a_t}{c^2}=u_{t;a}u^a=u_{t,a} u^a - \Gamma^\lambda_{ta}u_\lambda u^a
=u_{t,l} u^l-\Gamma^t_{tl} u_t u^l-\Gamma^l_{tt}u_lu^t=
\pm\gamma\,\frac{v}{c}\,\frac{1}{\sqrt{f}}\,(e^\nu \gamma)'\,,
\end{equation} 
where we have assumed that the four-velocity is a function of the radial variable
$l$ and rewritten its expression as 
\begin{equation}
{\mathbf u}=\gamma e^{-\nu} {\mathbf e}_t
\mp \gamma\,\frac{v}{c}\,\frac{1}{\sqrt{f}}\,{\mathbf e}_l \equiv u^t {\mathbf e}_t
+u^l\,{\mathbf e}_l\,.
\end{equation}
However, it also holds that
\begin{equation}
a_t={\mathbf a} \cdot {\mathbf e}_{t}=(a {\mathbf e}_{\tilde l}) \cdot  {\mathbf e}_{t}
= a \gamma \frac{v}{c}\,e^{-\nu} ({\mathbf e}_t \cdot {\mathbf e}_t)=
-a \gamma \frac{v}{c}\,e^{\nu}\,.
\end{equation}
Combining the above two results for $a_t$, we find that the magnitude of the
acceleration is
\begin{equation}
|a|=c^2 \Bigl |e^{-\nu} \frac{1}{\sqrt{f}}\,(e^\nu \gamma)'\Bigr|\,.
\label{con-acc}
\end{equation}
For the wormhole to be traversable, the above quantity must remain always finite
and, if possible, take a small value. 

In \cite{Morris:1988cz}, it was also demanded that the gravitational acceleration
should be small at the location of the stations where the trip starts and ends.
Demanding that the time needed to complete the whole trip should not be too large,
the stations should be fairly close to the throat. Both of these conditions are
satisfied in our case, since an asymptotically flat regime is reached fairly quickly. 

The above also means that the traveler, starting from the stations, does not need
to travel with a relativistic velocity in order to approach the wormhole in a
reasonable time. Then, if  $v \ll c$, $\gamma \simeq 1$, and the acceleration
(\ref{con-acc}) reduces to
\begin{equation}
|a|= c^2 \frac{|\nu'|}{\sqrt{f}}\,. \label{accel-final}
\end{equation}
In Fig. \ref{fig_acc}, we depict the dimensionless acceleration
$\hat{a}=|a|/(c^2/r_0)$ at the throat as a function of $D/M$. We observe that
$\hat{a}$ ranges roughly between 10 and 100 for a set of wormhole
solutions with parameter $\alpha/r_0^2$ ranging between
0.128 and 0.01. In the context of superstring theory, $\alpha \sim  \ell_P^2$,
therefore, for the above solutions $r_0 \sim 10\,\ell_P$ and the magnitude
of the acceleration turns out to be $~(10^{51}-10^{52})\,g_\oplus$, where
$g_\oplus$ is the acceleration of gravity at the surface of the earth.
Since there is no upper limit for $r_0$ in our solutions, one may be tempted
to increase the size of the throat of the wormhole so that $|a|$ is of the order
of $g_\oplus$; this would demand a throat radius of the order of
at least $(10-100)$ light-years. The aforementioned result could perhaps have
been anticipated since our analysis is performed within the context of
superstring theory, a theory whose fundamental scale is tied to the Planck scale.

\begin{figure}[ht]
\lbfig{fig_acc}
\begin{center}
\includegraphics[height=.25\textheight, angle =0]{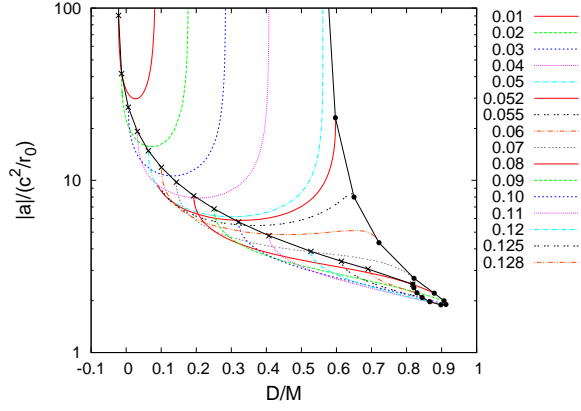}
\end{center}
\caption{
The dimensionless acceleration $\hat{a}=|a|/(c^2/r_0)$ that a traveler would 
feel traversing the wormhole as a function of $D/M$ for a variety of values
of $\alpha/r_0^2$.
}
\end{figure}

Next we turn to the tidal acceleration that a traveler feels between two parts
of her body as the wormhole is crossed. Let ${\mathbf w}$ be the vector separation
between these two parts, which in the reference frame of the traveler is purely
spatial, ${\mathbf w}=(0,w^i)$. Then, the tidal acceleration is given by the expression
\cite{Morris:1988cz}
\begin{equation}
\Delta a^{\tilde \mu}=-c^2 R^{\tilde \mu}_{\,\,\tilde \nu \tilde \rho \tilde\sigma}\,
u^{\tilde \nu} w^{\tilde \rho}\,u^{\tilde \sigma}=
-c^2 R^{\tilde \mu}_{\,\,\tilde t \tilde \kappa \tilde t}\,w^{\tilde \kappa}\,,
\label{tidal}
\end{equation}
where ${\tilde \kappa}$ takes on only spatial values and where we have used that
$u^{\tilde a}=\delta^{\tilde a}_{\tilde t}$. But since the metric is diagonal and the
Riemann tensor is antisymmetric in the first two indices, the superscript ${\tilde \mu}$
should also take only spatial values. Then, the non-vanishing components of the
tidal acceleration are:
\begin{equation} 
\Delta a^{\tilde l}=-c^2 R^{\tilde l}_{\,\,\tilde t \tilde l \tilde t}\,w^{\tilde l}\,,
\qquad 
\Delta a^{\tilde \theta}=-c^2 R^{\tilde \theta}_{\,\,\tilde t \tilde \theta \tilde t}\,
w^{\tilde \theta}\,,
\qquad 
\Delta a^{\tilde \varphi}=-c^2 R^{\tilde \varphi}_{\,\,\tilde t \tilde \varphi \tilde t}\,
w^{\tilde \varphi}\,.
\label{tidal-2}
\end{equation}
The transformation laws (\ref{trans1}) and (\ref{trans2}) allow us to compute the
components of the Riemann tensor in the orthonormal frame of the moving observer
in terms of the ones in the $(t,l,\theta,\varphi)$ coordinates. We then obtain
\begin{eqnarray}
\Delta a^{\tilde l}&=& c^2 w^{\tilde l}\,R^{\hat t}_{\,\,\hat l \hat t \hat l}
= c^2 w^{\tilde l}\,\frac{1}{f}\,R^{t}_{\,\,l t l}=
c^2\,w^{\tilde l}\,\frac{1}{f}\left(\frac{f'\nu'}{2f}-\nu'^2-\nu''\right)\,.
\end{eqnarray}
Similarly, we find:
\begin{eqnarray}
\Delta a^{\tilde \theta}=\Delta a^{\tilde \varphi} &=&
-c^2 w^{\tilde \theta}
\left(\gamma^2\,R^{\hat \theta}_{\,\,\hat t \hat \theta \hat t}+
\gamma^2\,\frac{v^2}{c^2}\,
R^{\hat \theta}_{\,\,\hat l \hat \theta \hat l}\right) =
-c^2\,w^{\tilde \theta}\,\gamma^2\left(e^{-2\nu}
R^{\theta}_{\,\, t \theta t} + \frac{v^2}{c^2 f}\,
R^{\theta}_{\,\, l \theta l}\right) \nonumber \\
&=& 
-\frac{c^2\,w^{\tilde \theta}\,\gamma^2}{r_0^2+l^2}
\left[\frac{l\nu'}{f}+\frac{v^2}{c^2 f}\,\left(\frac{l f'}{2f}-
\frac{r_0^2}{r_0^2+l^2}\right)\right]\,.
\end{eqnarray}
As in the case of the acceleration, we should demand that the magnitude of the
above components should be small, i.e. of the order of the acceleration of gravity
at the surface of the earth. Using the fact that $|{\mathbf w}| \simeq 2\,{\rm m}$,
the above constraints may be written as
\begin{eqnarray}
\frac{1}{f}\left|\frac{f'\nu'}{2f}-\nu'^2-\nu''\right|
&\leq & \frac{g_\oplus}{c^2 (2\,{\rm m})}=\frac{1}{(10^{10} {\rm cm})^2}\,,
\\[2mm]
\frac{\gamma^2}{r_0^2+l^2}
\left|\frac{l\nu'}{f}+\frac{v^2}{c^2 f}\,\left(\frac{l f'}{2f}-
\frac{r_0^2}{r_0^2+l^2}\right)\right| 
&\leq & \frac{g_\oplus}{c^2 (2\,{\rm m})}=\frac{1}{(10^{10} {\rm cm})^2}\,.
\end{eqnarray}
The second inequality involves the velocity with which the traveler moves and thus
may be considered as a constraint on this quantity. The first inequality restricts
again the profile of the metric functions - a similar analysis to the one above leads to
the same results regarding the magnitude of the tidal forces and the necessary
size of the throat in order to bring these down to a reasonable value.

\section{Conclusions}

The existence of traversable wormholes in the context of General Relativity 
relies on the presence of some form of exotic matter. However, in the
framework of a string-inspired generalized theory of gravity, the situation
may be completely different. 
Here we have investigated wormhole solutions in EGBd theory, 
which corresponds to a simplified action
that is motivated by the low-energy heterotic string theory.
Indeed, as we have demonstrated, EGBd theory allows for stable,
traversable wormhole solutions, without the need of introducing
any form of exotic matter. The violation of the energy conditions,
that is essential for the existence of the wormhole solutions
\footnote{According to recent studies \cite{Houndjo},
wormhole solutions arise also in the context of $f(T)$ gravitational
theories, where $T$ is the torsion, without the energy conditions being violated.},
is realized via the presence of an effective energy-momentum tensor
generated by the quadratic-in-curvature Gauss-Bonnet term.

We have determined the domain of existence of these wormhole solutions
and shown that it is bounded by three sets of limiting solutions.
The first boundary consists of the EGBd black hole solutions
of \cite{Kanti:1995vq}.
The second boundary is approached asymptotically,
when the curvature radius at the throat of the wormhole diverges
(the $f_0 \to \infty$ limit).
Finally, at the third boundary, solutions with a curvature singularity
at a finite distance from the throat are encountered.

We have investigated the properties of these EGBd wormholes
and derived a Smarr-like mass relation for them. In this,
the horizon properties in the black hole case
are replaced by the corresponding throat properties of the wormholes,
thus the area and surface gravity here refer to the ones at the throat.
Moreover, as is well-known for black holes in EGB theories,
their entropy does not correspond merely to their horizon area but it
receives a GB correction term. Similarly, we find that the 
mass formula for the wormhole solutions includes an analogous GB
correction term; in addition, another term, that vanishes in the
black hole case, appears that represents the GB corrected dilaton
charge at the throat. We have demonstrated that the Smarr relation is
satisfied very well by the numerical solutions.

We have also investigated the stability of the solutions.
We have shown that a subset of our wormhole solutions, the one that
lies close to the border with the linearly stable dilatonic black holes
in the domain of existence, is also linearly stable
with respect to radial perturbations.
While we have also shown that another subset is unstable,
we have concluded that the study of the standard equivalent Schr\"odinger equation
cannot determine the stability for the full domain of existence
We hope to resume the question of the existence of an alternative
method for the study of the stability of all of our wormhole
solutions at some future work.

When the wormhole solutions are extended to the second
asymptotically flat region,
this extension must be made in a symmetric way,
since otherwise a singularity is encountered.
As a consequence of this symmetric extension,
the derivatives of the metric and dilaton functions 
become discontinuous at the throat. This discontinuity
demands the introduction of some matter distribution
at the throat. We have shown that this may be realized
by the introduction of a perfect fluid at the throat whose
energy density is positive for the subset of stable wormhole solutions.

Next, we have studied and classified the geodesics
of massive and massless test particles in these wormhole spacetimes.
Depending on the respective effective potential,
there are two general kinds of trajectories for these particles.
The particles may remain on bound or escape orbits
within a single asymptotically flat part of the spacetime,
or they may travel from one asymptotically flat part
to the other on escape orbits, and travel back and forth
on bound orbits.

In addition, we have calculated the acceleration and tidal forces
which travelers traversing the wormhole would feel. We find that
their magnitude may be small for fairly large values of the size
of the throat of the wormhole. According to our findings, the radius
of the wormhole throat is bounded from below only, therefore, the
wormholes can be indeed arbitrarily large. Astrophysical consequences
will be addressed in a forthcoming paper as well as the existence
of stationary rotating wormhole solutions in the EGBd theory.

\section*{Acknowledgments}
We gratefully acknowledge discussions with Eugen Radu.
B.K. acknowledges support by the DFG.

\end{document}